\RequirePackage{fix-cm} 
\documentclass[natbib,smallextended]{svjour3}       
\usepackage{graphicx}
\usepackage{amsmath}
\usepackage{amssymb}
\usepackage{color}
\usepackage{ulem}		

\newcommand{\beq}{\begin{equation}}
\newcommand{\eeq}{\end{equation}}

\newcommand{\rrho}{\rm R_{\rho}}
\newcommand{\rrhomax}{\rm R_{\rho,\rm max}}
\begin{document}

\title{Layer formation in double-diffusive convection over resting and moving heated plates}

\titlerunning{Layer formation over heated plates}        

\author{Florian Zaussinger \and Friedrich Kupka}


\institute{F. Zaussinger \at
              Brandenburg University of Technology Cottbus-Senftenberg \\
              Department of Aerodynamics and Fluid Mechanics \\
              Siemens-Halske-Ring 14\\
              D-03046, Cottbus, Germany \\
              Corresponding author: florian.zaussinger@b-tu.de 
                            \and
              F. Kupka \at
              Universit\"at G\"ottingen \\
              Institut f\"ur Astrophysik \\
              Friedrich-Hund-Platz 1 \\
              D-37077 G\"ottingen \\
              \\
               MPI for Solar System Research \\
              Justus-von-Liebig-Weg 3 \\
              D-37077 G\"ottingen               
                    }

\date{Received: \today}

\maketitle

\begin{abstract}
We present a numerical study of double-diffusive convection 
characterized by a stratification unstable to thermal convection while at the
same time a mean molecular weight (or solute concentration) difference between top 
and bottom counteracts this instability. Convective zones can form in this case either 
by the stratification being locally unstable to the combined action of both temperature
and solute gradients or by another process, the oscillatory double-diffusive
convective instability, which is triggered by the faster molecular diffusivity of heat 
in comparison with that one of the solute. We discuss successive layer 
formation for this problem in the case of an instantaneously heated bottom (plate)
which forms a first layer with an interface that becomes temporarily unstable and 
triggers the formation of further, secondary layers. We consider both the case of
a Prandtl number typical for water (oceanographic scenario) and of a low Prandtl
number (giant planet scenario). We discuss the impact of a Couette like shear on 
the flow and in particular on layer formation for different shear rates. Additional layers
form due to the oscillatory double-diffusive convective instability, as is observed for 
some cases. We also test the physical model underlying our numerical experiments by 
recovering experimental results of layer formation obtained in laboratory setups.
\end{abstract}

\section{Introduction} \label{sec:introduction}

Various situations are found in nature and in engineering problems where a thermally unstably stratified fluid 
column is in principle stabilized by a counteracting gradient although the resulting stratification may still be unstable 
to small perturbations. The stability then depends on the ratios of the gradients on the one hand and on fluid properties on 
the other. A particular type of the more general double-diffusive instability (DD) belongs to these scenarios. As the name 
suggests, two different diffusivities are needed to describe this instability, which operates in coexistence with 
counteracting gradients in the stratification of the fluid: the diffusivity $\kappa_{\rm T}$ of a fast diffusing component, 
e.g., temperature $\rm T$, and the diffusivity $\kappa_{\rm S}$ of, say, a slowly diffusing solute such as salt in water or 
a component of a gas with mean molecular weight different from the other component(s) of the gas, such as helium 
in comparison to hydrogen in element mixtures found in stars. The ratio of both diffusivities is known as the Lewis 
number ${\rm Le}$ and for all cases of interest here, ${\rm Le =\kappa_S/\kappa_T}\ll 1$. 
Note that the definition of the Lewis number is community dependent and particularly in oceanography
this name is also used for the inverse value $\kappa_{\rm T}/\kappa_{\rm S}$, contrary to the convention followed in 
this study. The case where a temperature stratification unstable against small perturbations is stabilized by the solute 
stratification but can in the end become destabilized due to heat diffusing faster than the solute is known as 
the \textit{diffusive regime}. In astrophysics it is called \textit{semi-convection}. Following work devoted to the 
stability analysis of this scenario \citep{Walin_1964, Kato_1966, Spiegel_1969} it is also known 
as \textit{oscillatory double-diffusive convection}. If instead the solute is unstably stratified and the temperature 
gradient is stably stratified along the vertical direction, so-called \textit{salt-fingers} can form, an instability, 
which again results from heat diffusing faster than the solute. This case is also known as \textit{thermohaline convection} 
or \textit{fingering convection}.

Both instabilities are first of all described by the ratio of the thermal and the solute Brunt-V\"ais\"al\"a 
frequencies, $\rm N_T^2=-g\alpha(dT/dy)$ and $\rm N_S^2=-g\beta(dS/dy)$, where g is the gravitational acceleration,
$\alpha$ and $\beta$ are  positive thermal/solute expansion coefficients for (potential) temperature T and salinity S, respectively.
The vertical direction is denoted here with $y$.
The stability ratio $\rm R_{\rho}$ compares the impact of the solute stratification on the 
thermal buoyancy. For the diffusive regime it is defined as $\rm R_{\rho}= N_S^2/N_T^2$, \citep{Spruit_2013} and see also equations 164, 181, 185 in \citep{Canuto_1999}. To take advantage of 
symmetries in the stability analyses of these configurations it is common to define $\rm R_{\rho}$ for the salt-finger 
regime instead as $\rm R_{\rho}= N_T^2/N_S^2$. For both scenarios the value of $\rm R_{\rho}$ relative to unity is important: 
$\rm R_{\rho}>1$ implies that the stratification is much more stable than in the opposite case of $\rm R_{\rho}<1$. Indeed, the latter
characterizes flow states which rapidly mix a fluid as in the case of pure thermal convection (where $\rm N_S^2=0$). In the
astrophysical literature the case $\rrho<1$ is known as \textit{Ledoux unstable} (\citealt{Ledoux_1947}, see 
also \citealt{Canuto_1999}).

However, even in the Ledoux stable case $\rrho>1$ a fluid can become unstable to small perturbations. 
This is the regime of the DD and the flow states developing in these cases are prone to layering processes 
which lead to so-called \textit{staircases}.
The latter are well mixed layers separated by sharp interfaces with large gradients of $\rm T$ and solute
concentration $\rm S$. A criterion for their formation was proposed by \cite{Radko_2003}, originally for the 
case of salt-fingers, which for symmetry reasons of the underlying equations also applies to the diffusive 
regime (\citealt{Mirouh_2012}). The criterion has been derived as part of a stability analysis of DD 
convection. An alternative model for the diffusive regime which suggests an upper limit 
$\rrhomax$ for layer formation was proposed by \cite{Spruit_1992,Spruit_2013}.
The common challenge of modeling layered DD processes is to calculate the thermal and the solute flux 
over these interfaces. The fluxes in turn are needed to calculate the effective diffusivities, which can be 
used to estimate merging timescales and hence the heat and solute fluxes through as well as the
overall stability of such a stack. 
During the last 50 years a lot of research has been devoted to the DD and especially to layering 
processes, both in geophysics and astrophysics. An overview on these developments
can be found in \citet{Radko_2013}. For convenience of the reader and to explain the motivation for our work 
we first provide an overview on relevant activities in both research areas and separately on relevant 
numerical simulation studies before we summarize the main goals and the structure of this paper.

\subsection{Motivation: Diffusive convection in geophysics}

In the early 1960s \citet{Armitage_1962} discovered that Lake Boney and Lake Vanda (both in Antarctica) 
exhibit a double diffusive character, resulting in a natural \textit{solar pond}. This type of lake is characterized 
by a bottom heated (by solar radiation, e.g.) and surface cooled temperature field, and a stabilizing salt gradient. 
Due to these specific conditions a DD ground layer stores the thermal energy, which is isolated by the
overlaying fluid. This principle is advanced in artificial solar ponds, where the hot water of the ground layer 
is used to power glasshouses and office buildings. In spite of the simplicity of solar ponds, this technology 
seems to be a candidate for a seasonal storage system. Recent developments by \citet{Suarez_2010} 
confirm the importance of solar ponds, although more work is needed especially concerning questions 
about their long-time stability.

First DD experiments under controlled laboratory conditions were done by \citet{Turner_1964}, 
who heated an isothermal, but salt-stratified water column from below. Within 90 minutes three layers evolved 
in the experiment. \citet{Turner_1964} demonstrated the importance of the stability ratio on the ratio of the turbulent flux 
and the salt flux. However, the measured constant values of this ratio for $\rrho>2$ could not be explained 
and it is now known to be in a \textit{linearly stable} region. Based on their observation that DD can trigger layer 
formation, numerous investigations in lakes and in the ocean found stratifications arranged as staircases 
in temperature and salinity. For a comprehensive review on early experiments and field observations 
see \citet{Huppert_1981}.

\citet{Descy_2012} studied microstructure profiles in Lake Kivu in 2010 and 2011, which shows 
DD staircases due to dissolved gases. Below a depth of 100~m up to 300 mixed layers with a thickness of 0.3--0.6~m, 
separated by thin interfaces, were found. Especially the interface thickness could be reproduced satisfactorily 
with numerical simulations. However, theoretically estimated fluxes do not coincide fully with measurements and 
numerical calculations. This points to certain discrepancies between model assumptions and experiments. On the 
other hand, the numerical resolution of $2000 \times 1000$ points captured all relevant scales and is hence 
trustworthy as a DNS.

Parallel to laboratory and field experiments stability analyses were done for the DD scenario including the 
case of diffusive convection. \citet{Walin_1964}, \citet{Veronis_1965}, \citet{Stern_1960}, and subsequently 
in greater details \citet{Baines_1969} made pioneering work in the field of DD stability analysis. 
For a summary on these and some more recent results see \citet{Garaud_2018}. Whereas 
deep insight is gained with the linear stability analysis, the layering process itself eludes a mathematical 
description with this approach. A main result is an upper limit for the DD instability, which reads 
$1 < \rrho <({\rm Pr}+1)/({\rm Pr}+{\rm Le})\approx 1+{\rm Pr}^{-1}$ for ${\rm Le} \ll {\rm Pr}$ in the diffusive
convection regime. However, this inequality is \textit{only valid for initially linearly stratified temperature 
and salinity fields}. This scenario is violated in many applications, since local steep gradients 
in $\rm T$ and $\rm S$ can be induced by external forcing, for instance, by solar radiation on lakes and ponds 
or in case of laboratory experiments by a heating source. 

\subsection{Motivation: Diffusive convection in astrophysics}

Diffusive convection plays an important role in astrophysics. As mentioned before, it is known as 
semi-convection in this field. \citet{Tayler_1956} was uncertain, whether the semi-convective zone should be treated 
as fully mixed (i.e., assuming $\rrho=0$) or treated with the Ledoux criterion (mixing takes
place where $\rrho < 1$ which leads to smaller convective cores). Different approaches led to diverging stellar 
evolutionary tracks (see \citealt{Langer_1985}, \citealt{Aguirre_2011}, and \citealt{Maeder_2013}, e.g.). Two branches 
have established themselves to deal with semi-convection. The theoretical approach, which focusses on the 
underlying physics, is often based on the mixing length theory and the resulting mixing properties. On the 
other hand, numerical simulations of diffusive convection have recently become popular.

More recent models of semi-convection in astrophysics include the following. \citet{Grossman_1996} 
developed a non-local mixing-length theory. For stellar models of 15 to 30~M$_{\odot}$, with 
M$_{\odot}$ denoting the mass of the Sun, they confirmed earlier assumptions that semi-convection 
leads to compositional changes on relevant time scales but not fast enough such that instantaneous 
readjustment is appropriate. \citet{Spruit_1992} estimated the thermal flux in a semi-convection zone 
and in an extension of his model \citep{Spruit_2013} he derived a maximum $\rrhomax < {\rm Le}^{-1/2}$
for which layering can occur. A different model was suggested originally in an oceanographic context: 
the $\gamma$-instability proposed by \citet{Radko_2003} is based on a mean-field theory and it can be 
applied to salt-fingers as well in the diffusive convection regime (see \citealt{Radko_2010} and 
\citealt{Mirouh_2012}). This model has been successfully tested for ${\rm Pr, Le} \ge 10^{-1}$ by 
triple-periodic numerical simulations with the same linear background stratification
of temperature and molecular weight (salinity). It turned out that the thermal and solute fluxes
can be calculated in terms of the thermal and solute Nusselt numbers with an expression similar
to the model of \cite{Spruit_2013} but with different  numerical coefficients and scaling factors as derived
from the simulations \citep{Wood_2013}. An application of that model to semi-convection above the 
core of intermediate mass (F- and late A-type) stars in the mass range of 1.2 to 1.7~M$_{\odot}$ by 
\cite{Moore_2016} resulted in the conclusion that the semi-convective layers mix so efficiently that 
the results are the same as if instantaneous mixing were assumed (i.e., ignore the stabilizing 
composition gradient and apply the Schwarzschild criterion of instability to thermal convection). 
This appears to be at variance with \citet{Grossman_1996}, but the latter 
have considered high mass stars where the convective zone in their core evolves quite differently. 
Each of those models mentioned thus far in the end aims at modeling semi-convection as a gradient
diffusion process. \citet{Xiong_1986} was the first to develop and apply a non-local Reynolds stress model
to semi-convective core convection in massive stars (7 to 60~M$_{\odot}$). This model allows combining
the effects of both semi-convection and overshooting (turbulent mixing into layers stably stratified with
respect to both thermal and composition gradient). He found the temperature gradient in the
overshooting region to be more closely to that one obtained from using the simple Schwarzschild criterion of 
convective instability. The most complete model of this class has been proposed by \citet{Canuto_1999}.
A more accessible variant of it has been published in \citet{Canuto_2011a,Canuto_2011b}.
It allows a step-by-step increase of the completeness and thus also of the complexity of the physical 
model. The interaction between overshooting and semi-convection with a Reynolds stress model was recently 
investigated by \citet{Ding_2014} for massive stars. They found almost opposing tendencies for both processes.

Diffusive convection also plays a role in research fields bridging astrophysics and geophysics. 
\citet{Stevenson_1985} proposed layered semi-convection to operate in giant planets to explain 
anomalies of the mass and radius development as a function of time for the case of Saturn in comparison 
with Jupiter, for which in turn plain adiabatic convection led to consistent models. Alternative models for 
those problems were proposed later on until the situation became a lot more complex with the huge amount 
of data on the ever increasing numbers of currently known giant exoplanets. \citet{Chabrier_2007} renewed 
the idea of layered semi-convection to play an important role for the structure and evolution of giant (exo-) planets. 
They also pointed out that $10^{-3} \lesssim {\rm Le} < {\rm Pr} \lesssim 1$ which is more accessible to direct 
numerical simulations than the stellar case. In follow-up work, \citet{Leconte_2012} analyzed the consequences 
of a model of layered semi-convection on the structure of Jupiter and Saturn and discussed cases of up to 
$10^5 - 10^6$ layers in these objects. \citet{Leconte_2013} provided further evidence for this idea. 

\subsection{Previous numerical simulations of the diffusive case}

Numerical simulations of semi-convection and double-diffusive layering exist since the early 1990s. Due to 
low Lewis numbers and high Rayleigh numbers, it is difficult or even impossible to capture all relevant scales 
by direct numerical simulations (DNS). Additionally, the evolution of layers occurs on thermal time scales 
$\rm t_{\rm therm}$, because either thermal diffusion is the dominant energy transport process in part of the domain 
or it is at least characteristic for the time evolution and transport processes around interfaces between convective 
layers. On the other hand, the time steps of the simulations are limited by the time scales of advective transport, 
$\rm t_{\rm adv}$, on which the solution changes locally within the convective regions of the domain (see also 
\citealt{Kupka_2017}). This entails a very large number of time steps $\rm N_t \propto t_{\rm therm} /t_{\rm adv}$ 
that need to be covered by numerical simulations of DD. Full DNS for realistic stellar 
parameters remains impossible for decades to come (see also \citealt{Kupka_2017}).

The first numerical simulations of diffusive convection have been performed by \cite{Beckermann_1991}, 
followed by \citet{Merryfield_1995} 
who found coherent structures, but 
no layer formation.
\citet{Biello_2001} investigated semi-convection in the same parameter space 
 and observed layering. The effect of sidewalls was investigated in \cite{Young_2000}.
\citet{Bascoul_2007} found semi-convective layering for both regimes, the astrophysical one and 
the geophysical one.

 Two different types of numerical simulations of DD convection
have become common. The first class assumes triple-periodic 
boundary conditions and uses spectral methods to simulate instantaneous layer formation.
Examples include \citet{Radko_2003}, \citet{Rosenblum_2011}, \citet{Mirouh_2012}, and \citet{Wood_2013},
with the first one focussing on the salt-finger case, but providing theoretical grounds for the latter. 
The advantage of this setup is that it directly links to linear stability analysis of the DD. 
A key result from these simulations is that layers form for $1 < \rrho < \rrhomax < ({\rm Pr}+1)/({\rm Pr}+{\rm Le})$ 
where $\rrhomax$ is a consequence of the $\gamma$-instability to layer formation described 
by \citet{Radko_2003}. A large parameter space ($10^{-1} \leqslant {\rm Le} \leqslant {\rm Pr} \leqslant 10$)
has been investigated this way.

The investigation of the formation of layers from steep initial gradients or jumps or even just
a layer that has formed during the simulation and exposes now a sharp interface to a region 
without flow requires a different class of codes. Such scenarios are of interest to understand 
layer formation above planet core regions \citep{Leconte_2012} or for the evolution of thermohaline 
staircases in lakes \citep{Carpenter_2012a} or comparison with laboratory data such as \citet{Turner_1964}. 
Simulations of DD with non-periodic vertical boundary conditions and with extrapolations to 
astrophysical cases have been performed and discussed in
\citet{Zaussinger_2011}, \citet{Zaussinger_2013a}, \citet{Zaussinger_2013b}, \citet{Kupka_2015},
and independently thereof in a geophysical context by \citet{Carpenter_2012a,Carpenter_2012b}
and \citet{Sommer_2014}, among others. 
\citet{Carpenter_2012a,Carpenter_2012b} and \citet{Sommer_2014} demonstrated that the linear stability
analysis does not necessarily apply to the interface between two convective layers of a DD convection zone. 
They explained how values of $\rrhomax \sim 6$ can be found unstable at DD staircases in water.
This is in contrast with the simple linear stability criterion which predicts $\rrhomax \sim 1.14$.
\citet{Sommer_2014} were able to recover the stratification found for Lake Kivu by numerical simulations, 
for which high values of $\rm R_{\rho}$ in the range of 2 to 6 are considered physically relevant and layer
formation has been proven to occur by measurements.

An alternative process possibly responsible for layer formation in water (or rather, at ${\rm Pr} = 10$) 
in diffusive convection has recently been studied by \cite{Radko_2016}. Assuming a linear initial stratification 
in $\rm T$ and $\rm S$ which made the problem accessible to linear stability analysis and numerical simulations 
with double- and triple-periodic boundary conditions for the two- and three-dimensional case, respectively, 
he found the parameter regime for layer formation in terms of $\rrho$ drastically enlarged in case of shear 
acting on the flow. The mean shear profile was assumed sinusoidal, which again is highly suitable for the 
analysis and simulation techniques used. The instability is more pronounced for large Peclet number (i.e., for 
large thermal Rayleigh number at given Prandtl number) with a weak dependence on Le as long as it is clearly 
less than 1. Simulations and stability analyses agreed quite well with respect to the formations of layers and 
it was also confirmed both on theoretical and simulation grounds that the destabilizing effect of shear on 
diffusive convection (with both separate scenarios being linearly stable, while linearly unstable when put together) 
is primarily a two-dimensional effect and simulations in 2D hence already give the right answer with respect
to the stability or instability of the flow when DD occurs concurrently with shear. This conclusion also
agrees with a numerical simulation study of diffusive convection without shear for ${\rm Pr} = 13$ and ${\rm Le} = 0.005$ 
by \citet{Flanagan_2013} who concluded important deviations of heat fluxes through interfaces to occur
only for $\rrho \lesssim 2$. Thus, expensive 3D simulations can be avoided and larger parameter regimes 
studied for $\rrho \gtrsim 2$.

\subsection{Main goals and structure of the paper}

In this paper we present results from an extensive study with 2D direct numerical simulations for
the diffusive regime of the double-diffusive instability and parameters $\rrho>1$. We focus on cases 
where layering is observed, both with and without an additional, Couette flow-like shear that is applied
to the system in some of the numerical experiments. The investigated numerical model is based on 
a modified experimental setup inspired by \cite{Turner_1968}. Thus, we assume a linear and stable initial 
stratification for the salinity $\rm S$ whereas the temperature field $\rm T$ has an initially constant layering from 
the top (cold) downwards with a temperature jump (towards hot) at the bottom and is hence unstably stratified.
With this setup we investigate evolutionary aspects of the formation of double-diffusive staircases over 
a large fraction of the thermal time scale, i.e., the spatial formation (and partial remerging) of layers that 
form above the initial seed layer. The latter is triggered by the initial temperature contrast at the bottom of the 
simulation box. We also study thermal and solute fluxes occurring across the interfaces between those layers.

We first consider the geophysically relevant case of ${\rm Pr}=7$ and ${\rm Le}=10^{-1}$, but also explore
the astrophysically relevant parameter space of ${\rm Pr}=0.1$ and ${\rm Le}=10^{-1}$ which is of direct 
interest to research on giant planets. To close the gap between similarly constructed laboratory experiments 
of layer formation and numerical simulations we also discuss a special setup that aimed at reproducing
the initial and boundary conditions of the laboratory experiments as closely as possible. We demonstrate
being able to recover the main results of those experiments. This serves as a reference problem for the 
investigations presented in the main part of the paper.

The remainder of the paper is structured as follows. Section~\ref{sec:methods} gives a brief 
overview of the governing equations, the numerical setup, the model assumptions, and some basics 
about the validation of the simulations. Results concerning multi-layer formation, their temporal behavior, 
and stability are presented in Section~\ref{sec:results}. We discuss our results and present an
outlook on further work in Section~\ref{sec:discussion}. Our comparison between numerical 
simulations and laboratory experiments is discussed in Appendix~\ref{append:lab_validation}.

\section{Methods and setup of the simulations} \label{sec:methods}
\subsection{Governing equations and simulations}  \label{sec:goveq}

The double-diffusive system is described by the mass conservation equation, the Navier-Stokes equation, 
the temperature equation, and the solute equation. Fluid velocities are typically small compared to the 
sound speed. Hence, the density is assumed to be constant, apart from its effects on the buoyancy term. 
Additionally, the physical fluid properties are assumed to be constant, too. This restriction is valid for small 
temperature differences as they occur in the investigated systems. Finally, the domain considered is small
compared to the local pressure scale height. These assumptions justify the Boussinesq approximation. In the 
following, $\bf u(\bf x,\rm t)=(u(\bf x,\rm t),v(\bf x,\rm t),w(\bf x,\rm t))$ is the velocity where $\rm u$ and $\rm w$ 
are the horizontal velocity components and $\rm v$ is the vertical component, respectively. In the following, 
$\rm T(\bf x,\rm t)$ is the temperature, $\rm S(\bf x,\rm \rm t)$ is the solute concentration, $\rm p(\bf x,\rm t)$ 
is the pressure, and $\bf g=\rm (0,-g,0)$ is the gravitational acceleration. The positive $x$-axis of the Cartesian 
grid points from the origin (bottom left in simulations) towards the right direction. The positive $z$-axis points 
horizontally backwards and is only considered in three-dimensional cases (not considered for the numerical 
simulations presented here). The positive $y$-axis points upwards, in the opposite direction of the gravitational acceleration.
The fluid properties are represented by the Prandtl number ${\rm Pr}=\nu/\kappa_{\rm T}$ and the Lewis number 
${\rm Le}=\kappa_{\rm S}/\kappa_{\rm T}$, where $\kappa_{\rm T}$ is the thermal diffusion coefficient, 
$\kappa_{\rm S}$ is the solute diffusion coefficient, and $\nu$ is the kinematic viscosity. 
We recall again here that the definition of the Lewis number is community dependent and this name is 
also used for the inverse value $\kappa_{\rm T}/\kappa_{\rm S}$, which is not used in this study.
Finally, $\rho(\bf x,\rm t)$ is the density and $\rho_0$ is the constant reference density.
 
The governing equations are hence based on the Boussinesq approximation which have been derived, e.g., 
in \cite{Lesieur_2008} (Eq. 2.108, Eq. 2.111). Consequently, the governing equations can be written as
\begin{equation}  \label{eq:basic}
\begin{aligned}
\nabla \cdot \bf u &= 0, \\
\frac{\partial \,{\bf u}}{\partial\,\rm t} + {\bf u} \cdot \nabla {\bf u} &= - \frac{1}{\rho_0}\nabla \rm p + \nu \nabla^2 {\bf u} + \rm \frac{\rho}{\rho_0} \bf g, \\
\frac{\partial \,\rm T}{\partial\,\rm t} + {\bf u} \cdot \nabla \rm T &=   \kappa_{\rm T} \nabla^2  \rm T, \\
\frac{\partial \,\rm S}{\partial\,\rm t} + {\bf u} \cdot \nabla \rm S  &=   \kappa_{\rm S} \nabla^2 \rm S. \\
\end{aligned}
\end{equation}  
The buoyancy is modeled in terms of the extended Boussinesq approximation, where
the third term of the right hand side describes changes of the density related to the solute field,
given by the linearized relation
\begin{equation}
\rm  \rho=\rho_0(1-\alpha(T-T_0)+\beta(S-S_0)).
\label{eq:BA_extended}
\end{equation}
Here, $\rm T_0$ and $\rm S_0$ are pre-defined reference values, $\alpha$ is the thermal expansion coefficient and  
$\beta$ is the solute expansion coefficient.
The expansion coefficients are assumed to be small compared to thermal or 
solute differences, $\rm A=\alpha \Delta T\ll1$ and $\rm B=\beta \Delta S\ll1$ (cf. \citealt{Mutabazi_2016}) which 
justifies Eq.~(\ref{eq:BA_extended}), too. 
The governing equations are scaled by the thermal diffusion time $\rm H^2/\kappa_T$, the height of the fluid column H,
the thermal diffusion velocity $\rm \kappa_{\rm T}/{\rm H}$, and the 
thermal/solute boundary values given at the lower boundaries, i.e., $\rm \Delta$T and $\rm \Delta$S:
\begin{equation}  \label{eq:entdim}
\begin{aligned}
\rm  \bf{ x}&=\rm \bf{x}^* {\rm H},\\
\rm  t&=\rm \tau \tfrac{\rm H^2}{\kappa_{\rm T}},\\
\rm  {\bf u}&={\bf u}^* \tfrac{\kappa_{\rm T}}{\rm H},\\
\rm  p&=\rm p^*  \tfrac{\rho_0 \kappa_{\rm T}\nu}{\rm H^2},\\
\rm  T&=\rm T_0 + \Delta \rm T T^*,\\
\rm  S&=\rm S_0 + \Delta \rm S S^*.
\end{aligned}
\end{equation}  
Applying Eq.~(\ref{eq:entdim}) and omitting the asterisk we obtain,

\begin{equation}  \label{eq:nondim}
\begin{aligned}
\nabla \cdot \bf u &= 0, \\
\frac{\partial \,{\bf u}}{\partial\,\rm \tau} + {\bf u} \cdot \nabla {\bf u}  &= 
                   - \nabla \rm p + {\rm Pr} \nabla^2 {\bf u} + {\rm Pr}\,Ra_T \, T\, \bf{\hat j} - {\rm Pr}\,\rm Ra_S \, S \,\bf{\hat j} , \\
\frac{\partial \,\rm T}{\partial\,\rm \tau} + {\bf u} \cdot \nabla \rm  T &=   \rm \nabla^2 \rm T, \\
\frac{\partial \,\rm S}{\partial\,\rm \tau} + {\bf u} \cdot \nabla \rm S  &=  \rm Le \nabla^2 S, \\
\end{aligned}
\end{equation}  
where $\rm Ra_T=\tfrac{\alpha \rm g \Delta \rm T H^3}{\kappa_{\rm T} \nu}$ is the thermal Rayleigh number,
 $ \rm Ra_S=\tfrac{\beta \rm g \Delta \rm S H^3}{\kappa_{\rm T} \nu}$ is the solute Rayleigh number and $\bf{\hat j}$ is the unit vector pointing upwards. 
The mechanical stability of a double-diffusive fluid column is parametrized by the stability parameter,
$\rm R_{\rho}:=\frac{\rm Ra_S}{\rm Ra_T}$. The physical time t is scaled by the thermal 
diffusion time, which defines one dimensionless temporal unit.

We set the height of the box to one, where the  position $\rm y=0$ 
is the bottom and the position $\rm y=1$ is the top of the box. 
Additionally, the temperature and solute differences, $\rm \Delta T=1$ and $\rm \Delta S=1$, are set by letting
$\rm T_0=0$ and $\rm S_0=0$.

\subsection{Choice of Pr, Le, Ra$_{\rm T}$, Ra$_{\rm S}$, $\rrho$, and Ri}

Layer formation is expected in an interval of $\rm 1<R_{\rho}<R_{\rho,\rm max}$, where for cases without shear 
the latter may be estimated from the upper limit $\rm R_{\rho,\rm max} \leqslant \rm Le^{-1/2}$ argued by \citet{Spruit_2013}. 
For our specific choice of parameters we guide ourselves by a preliminary study of direct numerical simulations 
in two spatial dimensions at low spatial resolution presented by \citet{Zaussinger_2017}. The entire study included 135 simulations 
with a linear initial stratification in $\rm S$ and a constant stratification with a jump at the bottom in $\rm T$ as well as 
90 simulations with a linear stratification in both variables. No shear was included in those simulations. The focus of the
report in \citet{Zaussinger_2017} had been only on a subset of the doubly linearly stratified case, so we summarize the main 
findings for the cases with a temperature stratification with a jump in $\rm T$, as in the present work. We use 
$\rm Ra^*=Ra_{\rm T} Pr$ to describe the parameter space in the following
(the asterisk here is unrelated to non-dimensionalization but rather follows the notation of \citealt{Spruit_1992}).

For the case of ${\rm Pr}=7$ and ${\rm Le}=10^{-2}$ the simulations revealed multiple layer formation for 
$4.3\cdot 10^6 \lesssim {\rm Ra}^* \lesssim 2.2\cdot 10^8$ and $2 \lesssim \rm R_{\rho} \lesssim 6$. The interval below 
2 was not investigated except for the case $\rm R_{\rho}=1$ which always led to a convectively fully mixed simulation box. 
This is in remarkable agreement with the stability analysis, numerical simulations, and evidence from the Lake Kivu 
system presented in \citet{Carpenter_2012a,Carpenter_2012b} and \cite{Sommer_2014}. For $\rm R_{\rho} = 7$, 
no layer formation was found in this parameter range, which also ceased at even lower $\rm R_{\rho}$ for the 
lower range of ${\rm Ra}^*$ investigated. 

Likewise, for  ${\rm Pr}=0.1$ and ${\rm Le}=10^{-2}$  the simulations revealed multiple layer formation for 
$5\cdot 10^5 \lesssim {\rm Ra}^* \lesssim 5\cdot 10^7$ and $2 \lesssim \rm R_{\rho} \lesssim 5$.
Again the interval below 2 was not investigated except for the control case $\rm R_{\rho}=1$ which always led to 
a convectively fully mixed simulation box. Similar to the case of ${\rm Pr}=7$ layer formation ceased
at large $\rm R_{\rho}$ for the lower range of values of ${\rm Ra}^*$.

Thus, for the present study we have confined the parameter space to cases where all aspects of layer formations 
can be studied, including the first layer and multiple on-top layers. This includes the 
oceanographic case of ${\rm Pr}=7$, ${\rm Ra}^*=3.5\cdot 10^7$, ${\rm Le}=10^{-2}$, for which the stability parameter 
has been set to $\rm R_{\rho}=\{2,4\}$. Layer formation in the giant planet scenario is represented by ${\rm Pr}=10^{-1}$, 
${\rm Ra}^*=7.0\cdot 10^7$, ${\rm Le}=10^{-2}$, and by $\rm R_{\rho}=\{2,3\}$. The influence of shear is parameterized by the
Richardson number, where $\rm Ri=\infty$ represents cases without shear and $\rm Ri=\{10,1,10^{-1}\}$ cases with shear
(we explain the computation of the latter below). Hence, the present study covers 16 simulations and an additional
control simulation. The output rate was set to increments of $\rm \Delta \tau=4\cdot 10^{-5}$ to capture all relevant dynamical 
processes. This results in 25.000 snap shots for a thermal time scale, which, in contrast to Rayleigh-B\'enard convection, 
is the dynamical time scale of layer formation. We recall that ${\rm Ra_S} = \rm R_{\rho}\, {\rm Ra_T}$,
which is hence specified, too.

\subsection{Numerical simulations}

The governing equations are calculated with the ANTARES software suite \citep{Muthsam_2010}, 
which treats advective terms with a 5th-order WENO type scheme on a rectangular grid. Incompressibility
is ensured by an operator splitting technique which requires the solution of a Poisson equation during each 
time stage and which is parallelized using the Schur complement (in \citealt{Happenhofer_2013} the excellent,
strong scaling of this parallelization method is demonstrated for a solver for the generalized Poisson equation).
A comprehensive overview on the numerical details of solving Eq.~(\ref{eq:basic}) and also Eq.~(\ref{eq:nondim})
with ANTARES is presented in \cite{Zaussinger_2011} and \cite{Zaussinger_2013a}. The diffusion terms are approximated 
by 4th-order dissipative stencils compatible with the WENO method  \citep{Happenhofer_2013}. 

The interior domain of staircases consists of `pealed off' layers from the boundary, which determines the resolution. 
Thermal and solute boundary layers are resolved with at least three points to guarantee correctly calculated fluxes 
and Nusselt numbers for very steep gradients. The minimum numerical resolution is estimated via the solute boundary 
layer which scales with the Lewis number and the thermal boundary layer 
(see \citealt{Huppert_1976}),
\begin{equation}  \label{eq:Sscale}
\delta_{\rm S}=\sqrt{\rm Le} \, \delta_{\rm T}.
\end{equation}
Further, the extent of the thermal boundary layer $\delta_{\rm T}$ is expressed in terms of 
$\rm Ra^*$, \cite{Spruit_1992},
\begin{equation}   \label{eq:Tscale}
\frac{\delta_{\rm T}}{\rm H}\approx \sqrt[4]{\frac{1}{\rm Ra^*}}.
\end{equation}
We note that for choosing the grid sizes for our simulations there is little difference if we had used alternative 
scalings of the Nusselt number as a function of $\rm Ra^*$ (e.g., \citealt{Spruit_2013} or \citealt{Wood_2013}), 
which is the basis for Eq.~(\ref{eq:Tscale}). Those would encourage less restrictive grid sizes, by factors of 2 and 3, 
respectively. The sizes of thermal and solute boundary layers found in our simulations, however, confirm that 
the numerical resolution should be chosen according to the less optimistic estimate Eq.~(\ref{eq:Tscale}).
Finally, the solute boundary thickness is calculated according to $\max(\rm Ra^*)=5\cdot 10^7$ and $\rm Le=10^{-2}$. 
This results in a solute boundary thickness of $\delta_{\rm S}=0.006$. Assuming that  $\delta_{\rm S}$ 
is resolved with at least 3 points we obtain a final resolution of $500\times 500$. This incorporates that the thermal boundary 
layer is resolved with at least 30 points. This estimate is independent of ${\rm Pr}$ and applicable to both, the 
oceanographic case (${\rm Pr}=7$) and the astrophysically relevant case (${\rm Pr}=0.1$), since ${\rm Le} \ll {\rm Pr}$
and ${\rm Le} \ll 1$, whence $\kappa_{\rm S} \ll \kappa_{\rm T}$ and $\kappa_{\rm S} \ll \nu$.

\subsection{Boundary conditions and shear}    \label{sec:bc_and_shear}

The boundary conditions are chosen to be compatible with a Rayleigh-B\'enard experiment. Thus,
the top and bottom boundary plates are held at constant values of $\rm T$ and $\rm S$. In addition, they
can either remain at rest (u=v=w=0) or move at a constant velocity ($\rm u \neq 0$ and v=w=0) to apply a constant mean horizontal shear 
on the system as in experiments for plain Couette flow. There are no sideward walls along the vertical (y-) direction. Rather,
the domain is continued periodically with a fixed aspect ratio of one.
The horizontal velocity component $\rm u_{y=0}$ at the bottom (and optionally at the 
top $\rm u_{y=H}$) is calculated from a fixed value of Ri, which is held constant during each simulation: we define 
$\rm Ri={\rm N^2_T}\,({\rm d\,u}/{\rm d\, y})^{-2}={\rm N^2_T}\,({(\rm u_{y=H}-u_{y=0})}/{H})^{-2}$, where $\rm N^2_T$ is the thermal
Brunt-V\"ais\"al\"a frequency and $\rm y$ denotes the vertical coordinate. In the following, the definition of Ri incorporating the 
difference quotient of the horizontal velocity component u over H is used.
Moreover, $\rm Ra^*=\tau^2_{\rm therm} N^2_T$ and 
$\tau_{\rm therm}=\rm H^2/{\kappa_T}$ is the thermal time scale used as temporal scaling factor in Eq.~(\ref{eq:entdim}) to 
express time in dimensionless units. Rearranging Ri and setting  $\rm u|_{y=H}=0$ we obtain
\begin{equation}   \label{eq:shear}
\rm u|_{y=0}=\sqrt{\frac{\rm Pr\,Ra_{\rm T}}{Ri}} \frac{\kappa_T}{H}
\end{equation}
for a \textit{constant average shear rate} exerted through moving the bottom boundary at a constant
horizontal velocity whereas the top boundary remains at rest. 
The dimensionless representation of the shear rate is given through Eq.~(\ref{eq:entdim}),
\begin{equation}   \label{eq:shea_dim-less}
\rm u^*|_{y^*=0}=\sqrt{\frac{\rm Pr\,Ra_{\rm T}}{Ri}},
\end{equation}
where $\rm u^*$ and $\rm y^*$ with their asterisks (omitted in the following) emphasize their dimensionless nature.
As in laboratory experiments, $\bf u=0$ inside the domain at the beginning. In the case of a viscous, 
pure (Couette-type) shear flow such initial and 
boundary conditions give rise to a characteristic horizontal velocity profile that evolves towards a linear variation 
with depth (see \citealt{Batchelor_2000}). In a control experiment we have also equally distributed the
shearing between both plates, i.e.,
$\rm u|_{y=0} \equiv \sqrt{{\rm Ra^*} / {Ri}}  / 2
\equiv -\rm u|_{y=1}$ and we recall the definition $\rm Ra^* = \rm Pr\,Ra_{\rm T}$.
The (potential) temperature is set according to Dirichlet boundary conditions, 
whence $\rm T|_{y=0}=1$ and $\rm T|_{\rm y=1}$=0. The same holds for the 
salinity, $\rm S|_{y=0}=1$ and $\rm S|_{\rm y=1}$=0.
\subsection{Some remarks on model parameters and box geometry}
This study covers two cases. First, the geophysically relevant case of saltwater, 
where ${\rm Pr}=7$ and ${\rm Le}=10^{-2}$. These values are also known from rift lakes, solar ponds 
and oceans, \cite{Carpenter_2012b}.
We point out here that in realistic settings the value of Pr for water varies as a function of 
temperature and salinity. Since we do not intend to compute detailed oceanographic cases in the following, it 
suffices to consider the standard value of $\rm Pr = 7$ even though for some cases this value might vary
by up to a factor of 2.
The second case represents the giant planet case of ${\rm Pr}=10^{-1}$ 
and ${\rm Le}=10^{-2}$, see \cite{Chabrier_2007}. The height of the initial layer does not depend on Pr or Le, 
but on the imposed heat flux, i.e., the temperature difference, and the salinity gradient. 
However, a crucial point for the simulations is the overall height and the aspect ratio of the domain. 
\cite{Turner_1968} and \cite{Fernando_1989} estimated  for the average height of an intermediate layer 
that $\rm h \approx R_{\rho}^{-1}$. All simulations presented here have been performed with an aspect ratio of one, except otherwise mentioned. 
The Rayleigh number $\rm Ra_T$ is set to  $\rm Ra_T=3.5\times 10^7$ and $\rm Ra_T=5\times 10^7$, respectively.
In the following we use the related quantity $\rm Ra^*=Pr\, Ra_T$, too. It is this quantity 
which directly appears in Eqs.~(\ref{eq:nondim}) and in Eq.~(\ref{eq:shea_dim-less}), since $\rm Pr\, Ra_T\, T=Ra^*\,T$
and $\rm Pr\, Ra_S\, S = \rrho\,Ra^*\,S$ as well as $\sqrt{\rm Pr\,Ra_{\rm T} / {Ri}}=\sqrt{\rm Ra^* / {Ri}}$.

\subsection{Detection of layers}   \label{sec:layerdetect}

\begin{figure}[htb]
\begin{center}
a) \includegraphics[height=0.6\textwidth]{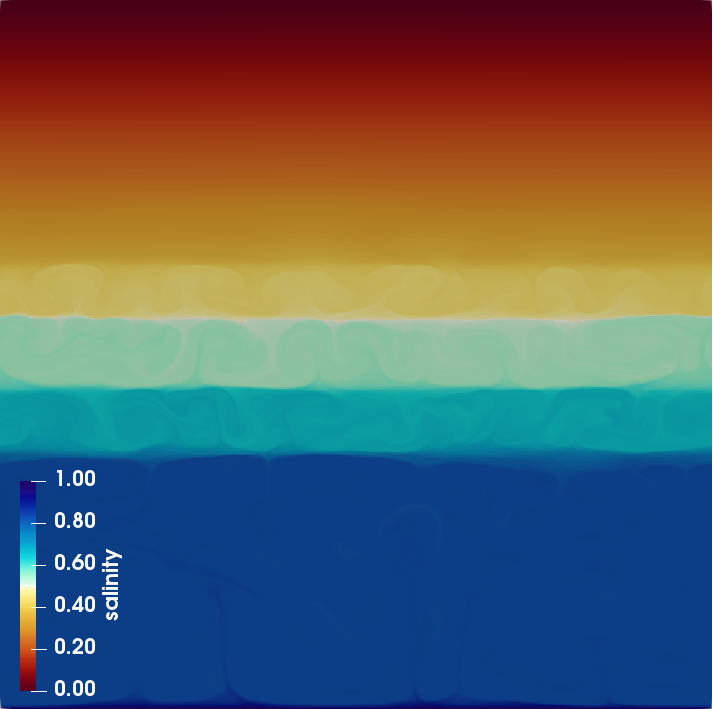} 
\end{center}
\begin{center}
b) \includegraphics[height=0.6\textwidth]{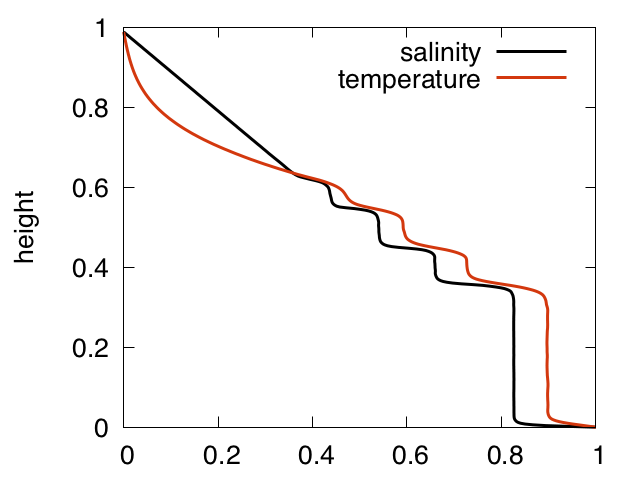}
\end{center}
\begin{center}
c) \includegraphics[height=0.5\textwidth]{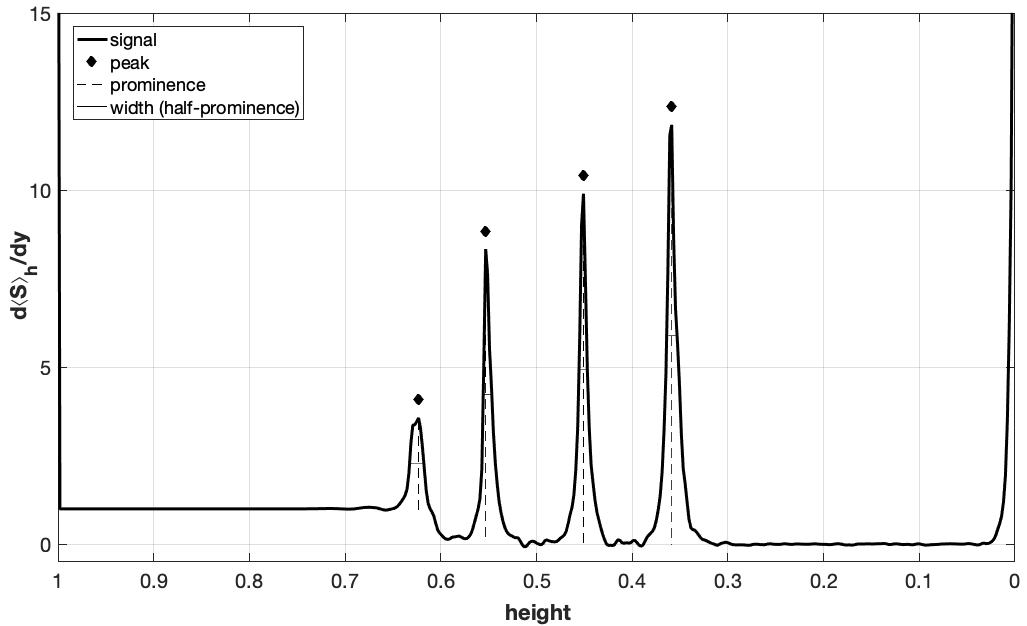}
\end{center}
\caption{(a) screenshot of the salinity for a double diffusive stack at $\tau=0.65$ for 
R$_{\rho}=3$, $\rm Pr=7$, $\rm Le=10^{-2}$ and $\rm Ra^*=3.5\times10^7$, (Case Pr7R3). 
Blue values depict high salt concentration, red ones depict low salt concentration. (b)
horizontally averaged temperature and salinity as function of height from top (y=1) to bottom (y=0).
(c) horizontally averaged saline gradient as a function of height from top (y=1) to bottom (y=0). 
Diamonds indicate located peaks with a minimum prominence of $\rm P=1$.}
\label{fig:Case_3_Rho3_Ri_inf_S_tau_065}
\end{figure}

In the following, layers are denoted as regions between interfaces. Those interfaces are free boundaries of local 
convective structures. Fig.~\ref{fig:Case_3_Rho3_Ri_inf_S_tau_065}~a) shows a stack of five layers, 
separated by four interfaces. Boundaries at the top and the bottom of the box are not interfaces in this strict 
sense, as they result from external constraints: arbitrarily close to the boundaries the fluxes of heat
and solute become purely conductive, which differs from unbounded interfaces where the double-diffusive 
instability can break a fluid column into convectively unstable layers. Hence, unbounded interfaces may
both form and merge again. Double-diffusive layers are detectable in all physical fields, but this is most easily 
done in either the temperature field or the salinity field. Due to ${\rm Le} \ll 1$ the salinity field is even more 
suitable to localize layers, since interfaces in $\rm S$ are thinner because of $\kappa_{\rm S} \ll \kappa_{\rm T}$. 
This increases the gradient in the interface and hence the flux through it. Local maxima of the horizontally averaged saline gradient $\frac{\rm d\,\langle S\rangle_h}{\rm d\,y}$ with $\rm \langle . \rangle_h$ being
the horizontal mean,
indicate 
interfaces, see Fig.~\ref{fig:Case_3_Rho3_Ri_inf_S_tau_065}~(c). This gradient 
is scanned for peaks with a minimum prominence of $\rm P=1$. 
We recall that the $y$-axis is along the direction of gravitational acceleration and points downwards
and the computation is done with the dimensionless quantities introduced further above. Four interfaces are found in this 
specific example. Fig.~\ref{fig:Case_3_Rho3_Ri_in}~(a) depicts the local maxima of the gradient as a function 
of time. Peaks of the thermal gradient (orange dots) and of the saline gradient (black dots) 
illustrate the interfaces. The mean difference between the measurements is $<0.1\%$. Clouds of points 
with vertical elongations indicate regions with strongly fluctuating interfaces. 
This occurs at the beginning of the layer formation process where thermal plumes form the first layer.

\section{Results} \label{sec:results}
\subsection{The oceanographic case without shear}

\begin{figure}[htb]
\begin{center}
a) \includegraphics[height=0.5\textwidth]{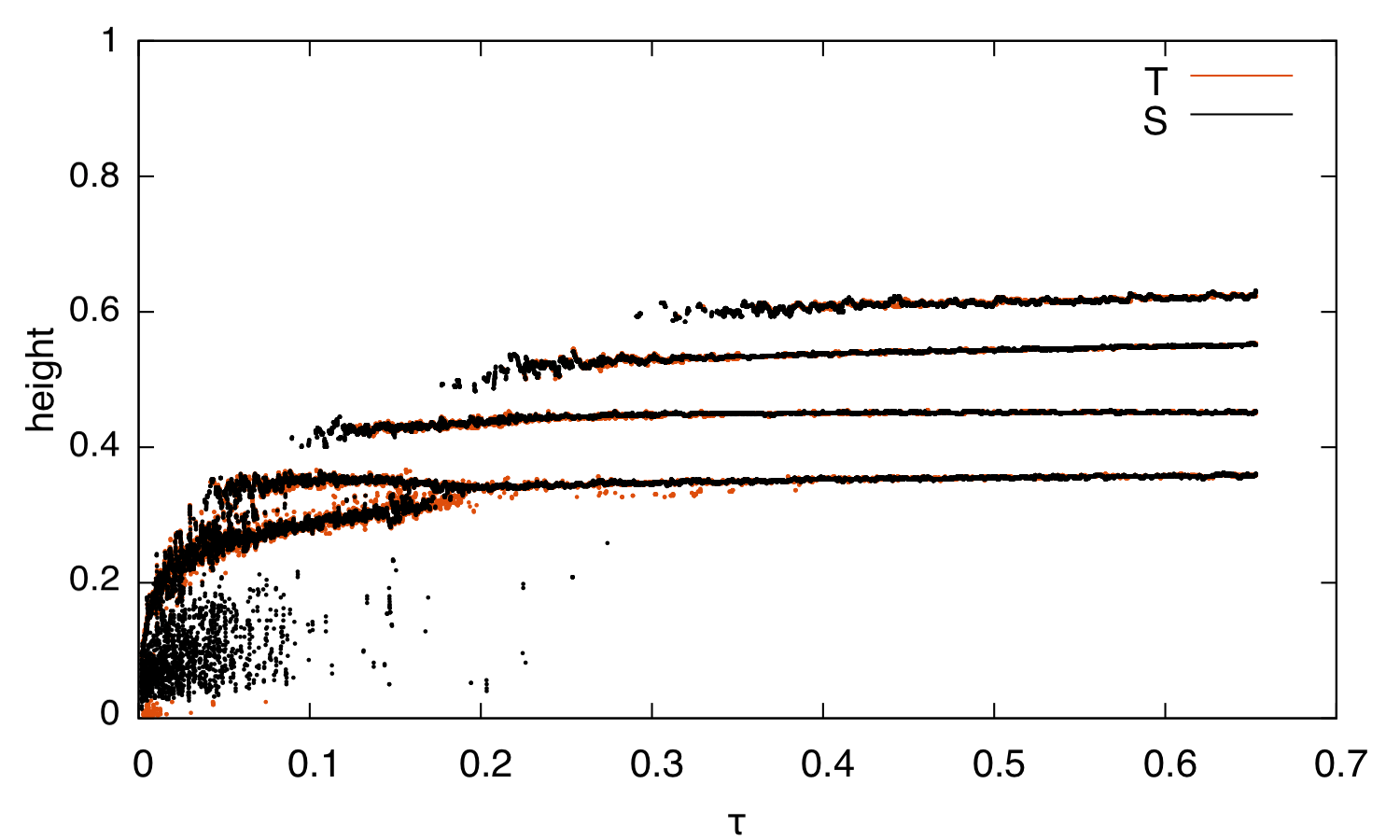} \\
\end{center}
\begin{center}
b) \includegraphics[height=0.5\textwidth]{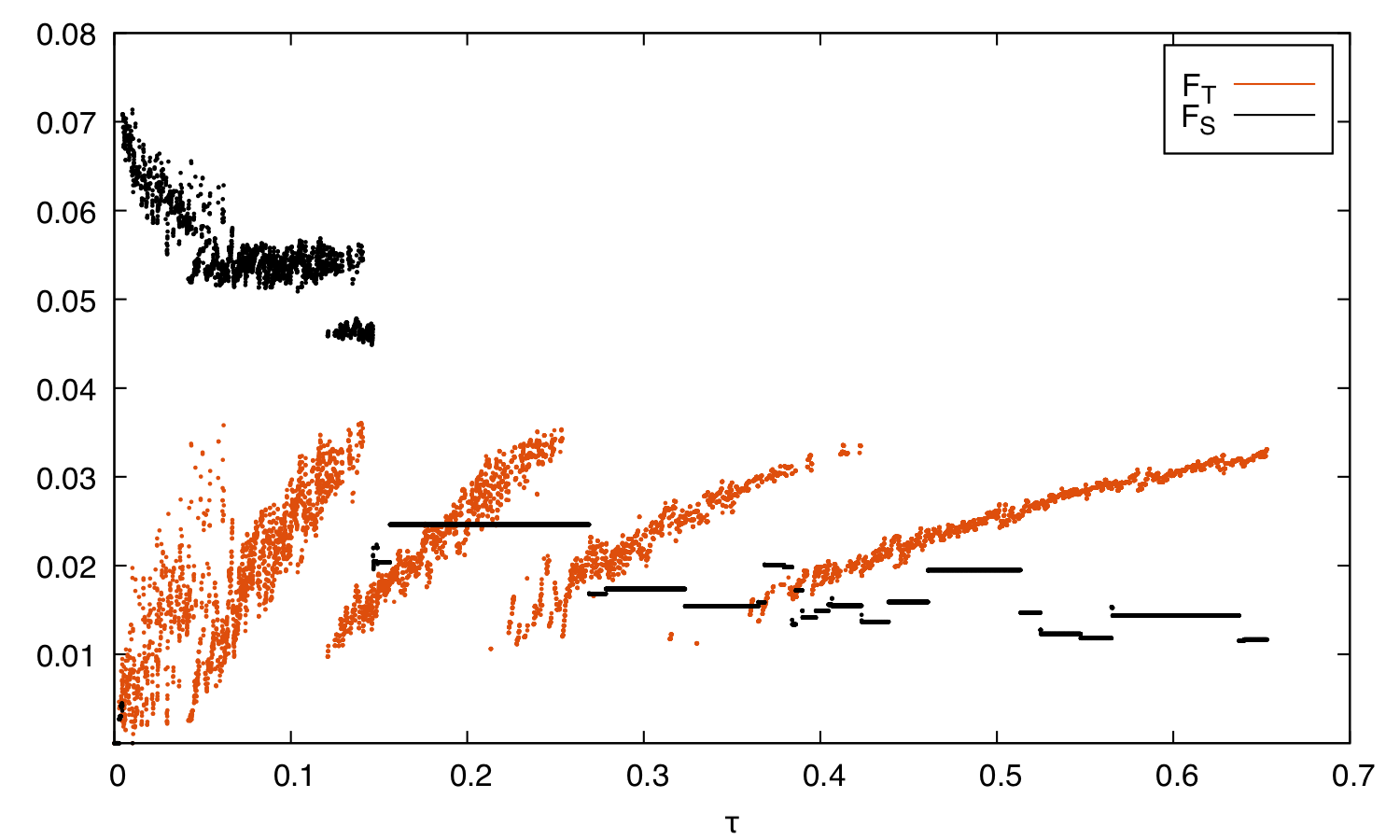}
\end{center}
\caption{Pr7R3: (a) mean location of interfaces as function of time. (b) $\rm F_T$ (orange dots) 
and $\rm F_S$ (black dots) of the first layer as function of time.}
\label{fig:Case_3_Rho3_Ri_in}
\end{figure}

\begin{figure}[htb]
\begin{center}
a) \includegraphics[height=0.5\textwidth]{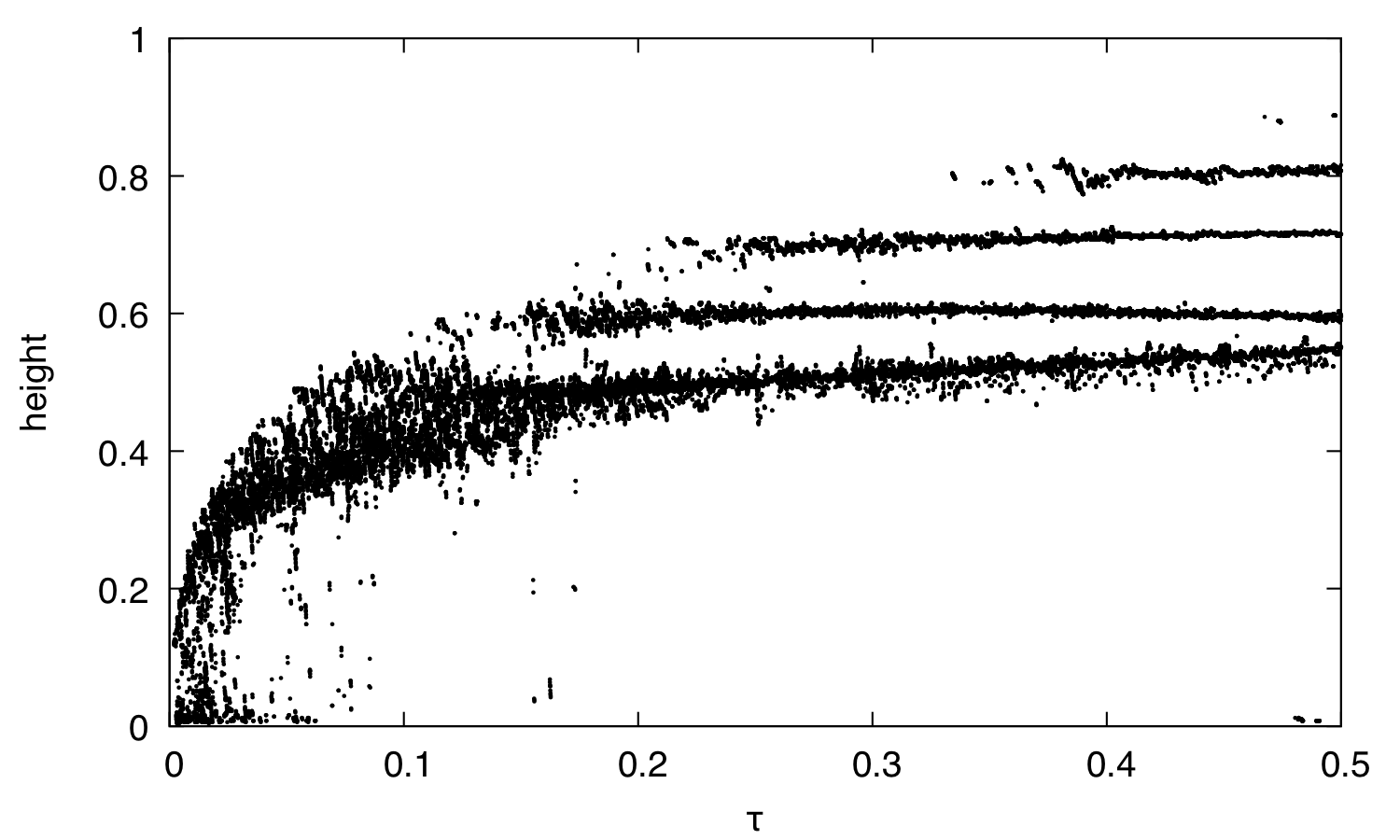} \\
\end{center}
\begin{center}
b) \includegraphics[height=0.5\textwidth]{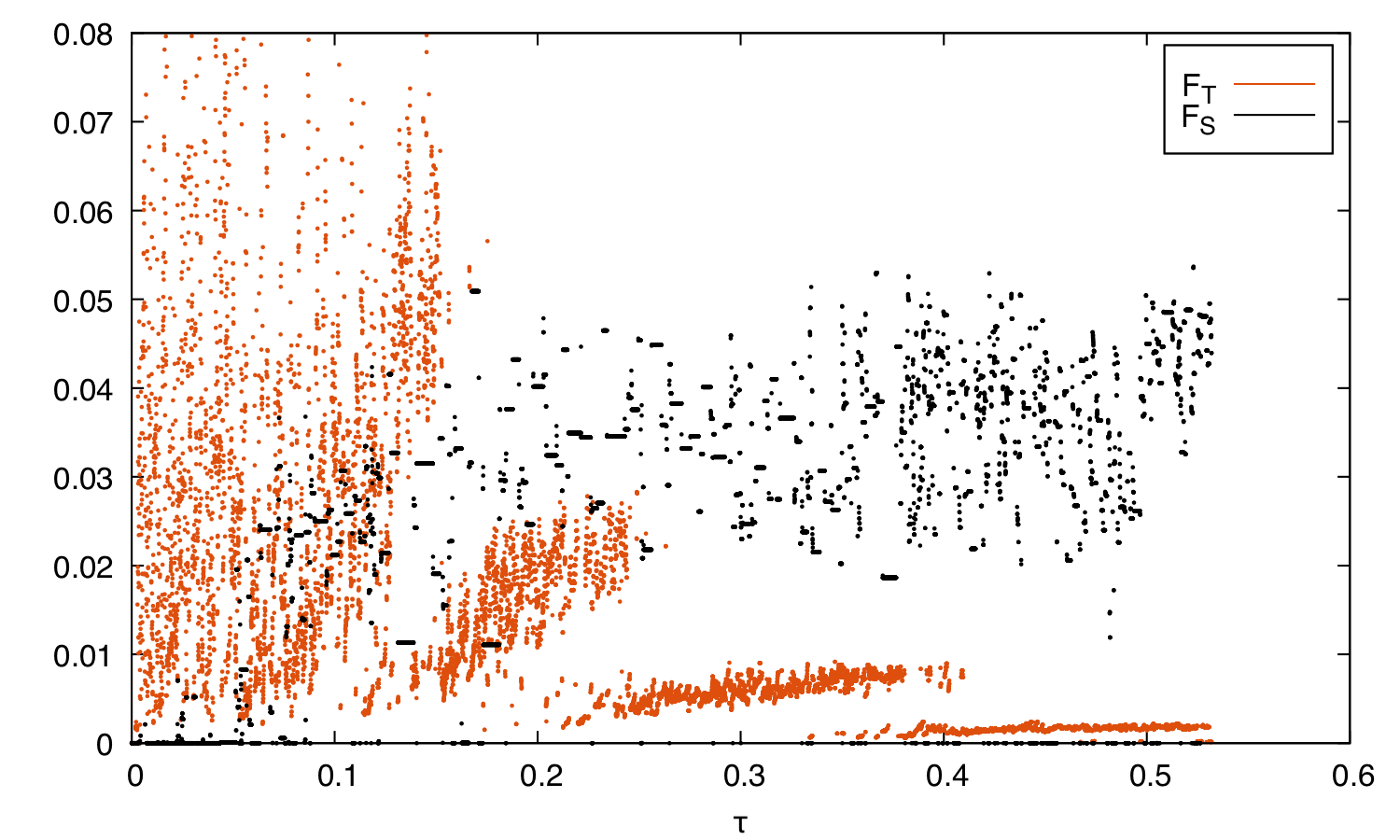}
\end{center}
\caption{Pr7R2: (a) mean location of interfaces as function of time. (b) $\rm F_T$ (orange dots) 
and $\rm F_S$ (black dots) of the first layer as function of time.}
\label{fig:Case_3_Rho2_Ri_in}
\end{figure}

Two reference cases without shear ($\rm Pr=7$, $\rm Ra^*=3.5\cdot 10^7$, $\rm Le=10^{-2}$ and $\rm R_{\rho}=\{2,3\}$) 
are used to first identify global dynamics and typical time scales. We discuss them in more detail with respect
to layer formation and the time evolution of the velocity field. Subsequently, we analyze simulations with a mean shear. 
In the following, the simulations are denoted with the abbreviations `case Pr7R2' and `case Pr7R3', respectively. 
These labels are extended through a combination of Ri and its numerical value to distinguish cases with different
mean shear. In the two cases without mean shear the formation of the first layer sets in within $\tau=0.1$. 
A merging event shortly after $\tau=0.15$ between the pre-formed initial layer and an unstable 
diffusive region on top of it ends the formation process of the first layer for both values of 
$\rm R_{\rho}$ (see Fig.~\ref{fig:Case_3_Rho3_Ri_in} (a) and Fig.~\ref{fig:Case_3_Rho2_Ri_in} (a)). Towards 
lower values of $\rm R_{\rho}$ the formation of the interface is accompanied by high fluctuations of the mean 
interface height. This arises from strong thermal updrafts. In case of Pr7R2 the height of the first layer is initially 
about $\rm y=0.4$, but increases with a lapse rate of 0.33. In comparison 
the height of the first layer is about $\rm y=0.36$, for Pr7R3 
and it remains constant for a long time. In both cases, the secondary layer formation proceeds in steps of 
$\tau=0.1-0.15$ after the initial (``seed'') layer has been formed. The subsequent formation process is quite similar 
to that one discussed in \cite{Huppert_1979}. The first layer has reached a height $\propto \rm R_{\rho}^{-1}$
when the second, independent layer begins to form around $\tau \sim 0.1$. The secondary layers are much thinner than the 
first layer and their thickness scales approximately with $\rm R_{\rho}^{-1}$. Since the mean height of the first layer increases 
with decreasing stability parameter, there are constraints for numerical simulations with $\rm 1 \lesssim R_{\rho} \lesssim 1.14$:
for $\rm R_{\rho}$ in the parameter range linearly unstable to oscillatory double-diffusive convection (for ${\rm Pr=7}$ 
and ${\rm Le=0.01}$) and for the value of ${\rm Ra_T}$ we study the interface of the first layer would have to form
in a domain which is close to the top of the simulation box.

For the reference case Pr7R3 at $\tau=0.65$ one can easily see in Fig.~\ref{fig:Case_3_Rho3_Ri_inf_S_tau_065}~(a) 
that five layers have formed which are separated by four well-defined interfaces (here, we also include the top, mostly 
diffusive ``layer'' as defined in Sect.~\ref{sec:layerdetect}). The interfaces can readily be identified 
in the horizontally averaged saline gradient (Fig.~\ref{fig:Case_3_Rho3_Ri_inf_S_tau_065}~(c)).
The prominence of the gradient in each interface exceeds the threshold value of $\rm P=1$ by a factor of at 
least two. The time series of interface locations traced for Pr7R3 is depicted in Fig.~\ref{fig:Case_3_Rho3_Ri_in}~(a). By 
the same method we have also located the interfaces formed in case of Pr7R2 (see Fig.~\ref{fig:Case_3_Rho2_Ri_in} (a)).
To this end the mean, layer-related thermal and solute Rayleigh numbers $\rm Ra_{T,L}$ and $\rm Ra_{S,L}$ 
are scaled by constant fluid properties and gravity.
This makes both expressions more comparable when applied to a single layer,
\begin{eqnarray}
\rm F_T=Ra_{T,L}/(\alpha\,g / \nu\, \kappa_T)=(T(L_{n})-T(L_{n-1}))((y(L_{n})-y(L_{n-1}))^3, \\
\rm F_S=Ra_{S,L}/(\beta \,g / \nu\, \kappa_T)=(S(L_{n})-S(L_{n-1}))((y(L_{n})-y(L_{n-1}))^3,
\end{eqnarray}
where $\rm T(L_{n})$ is the temperature measured at the layer interface L labeled with index n and $\rm y(L_{n})$ is the vertical coordinate of the interface, respectively.
The same notation is used for $\rm F_S$.
Both values are depicted along the first layer 
in Fig.~\ref{fig:Case_3_Rho3_Ri_in}~(b) and in Fig.~\ref{fig:Case_3_Rho2_Ri_in}~(b) for both stability ratios studied, respectively.
Since in the 
Boussinesq approximation the fluid properties and thus also the thermal expansion coefficient $\alpha$ and the saline 
expansion coefficient $\beta$ are constant and moreover since the gravitational acceleration is assumed to be constant 
as well, variations in temperature and solute differences for a certain layer height $\rm H$ are directly related to variations in 
the local Rayleigh numbers,
defined using temperature and solute differences between layers instead of between top and bottom,
and hence $\rm F_T$ and $\rm F_S$ quantify the local thermal and solute fluxes. 
We see from Fig.~\ref{fig:Case_3_Rho3_Ri_in}~(b) and in Fig.~\ref{fig:Case_3_Rho2_Ri_in}~(b) that for both values of 
$\rm R_{\rho}$ the thermal flux through the first interface clusters around certain values and those values change
as a function of time. Especially for Pr7R3 the time evolution of $\rm F_T$ is tightly correlated with the formation
of new layers on top of the entire stack. Each formation of a new, secondary layer is characterized by $\rm F_T$
oscillating between two clusters of large and small values until the formation of the new layer is complete 
and the larger group of values disappears. This formation process is first of all driven by an increase of the temperature
of the fluid in the first layer. But the salinity gradient through the first interface remains steep enough such that
it is essentially stable while temporarily steep temperature gradients can be sustained through which part of the heat
is released to the next layer. That eventually causes the formation of a new layer on top of the stack. The formation 
of the new secondary layer begins when lower values of $\rm F_T$ become common along high values 
($\tau=0.1$, $\tau=0.22$, and $\tau=0.32$) and ends when the larger values which are associated with the formation of the 
previous secondary layer disappear ($\tau=0.11$, $\tau=0.26$, and $\tau=0.42$). 
Since for Pr7R3 the size of the first layer barely changes from $\tau \sim 0.1$ onwards, the formation of
new secondary layers eventually allows a reduction of $\rm F_T$ across the first interface. The drop of $\rm F_T$ 
during a formation event results from the increasing convective heat transport in the newly formed top layer 
once a critical gradient had been reached in the first interface. This reduces the storage of heat in the first layer 
and its interface and thus stabilizes the latter for a more extended period of time: the height of the layers with 
respect to $\rm T$ and $\rm S$ is very similar and $\rm F_S$ just slightly drops on average between 
$\tau=0.15$ and $\tau=0.65$. The situation is more complex for Pr7R2. It shows a similar evolution of 
$\rm F_T$ with time, but the variations in $\rm S(L_{n})-S(L_{n-1})$ are much larger. Indeed, the 
second layer appears to merge with the first one probably before $\tau=0.6$ and the height of the first layer 
keeps growing linearly with time. 

Clearly though, each newly formed secondary layer is coupled closely to the thermal flux through the first 
interface and hence the `seed' layer. This drives the long-term evolution of the layers on thermal time scales.

\subsection{The oceanographic case with shear}
\begin{figure}[htb]
\begin{center}
a) \includegraphics[height=0.3\textwidth]{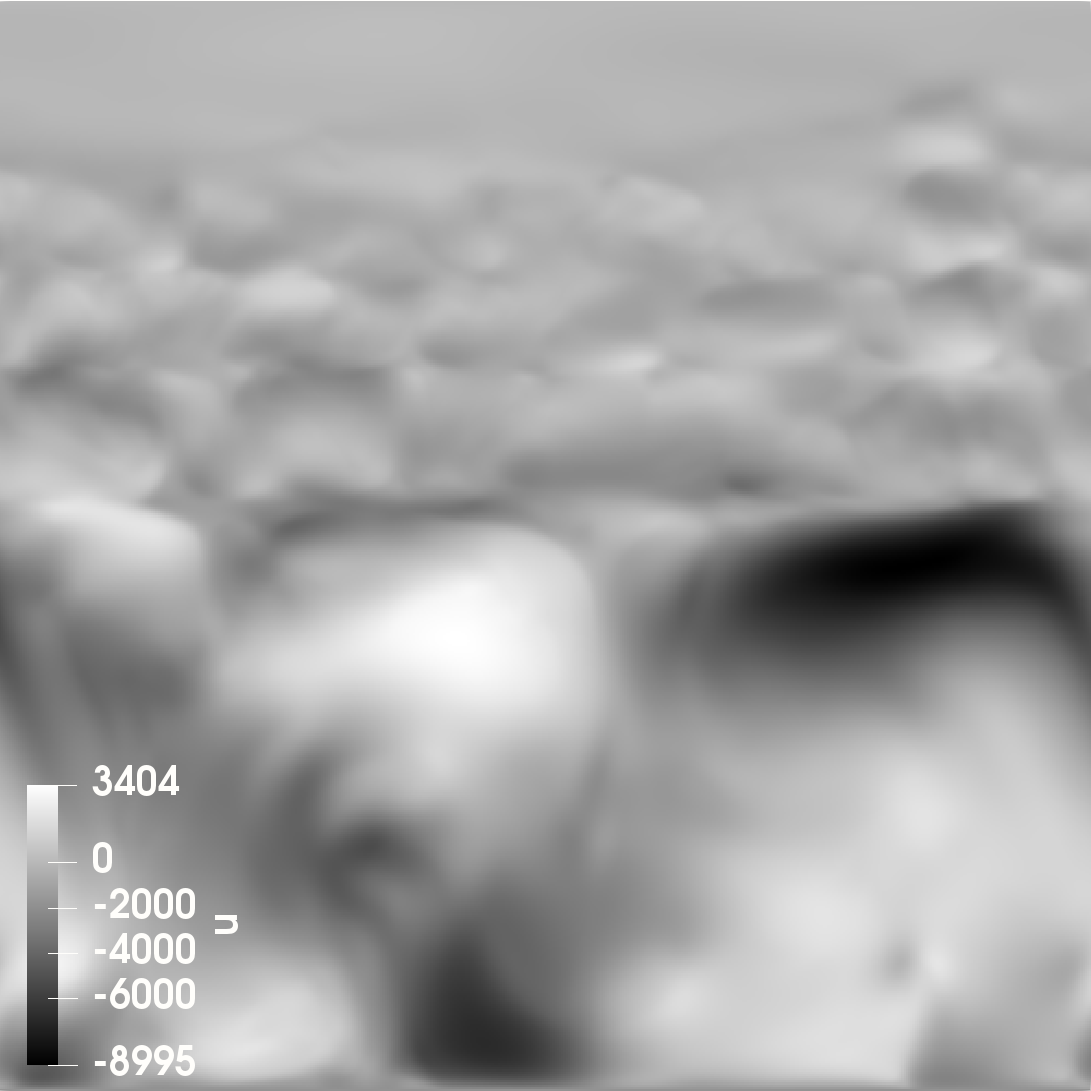}
b) \includegraphics[height=0.3\textwidth]{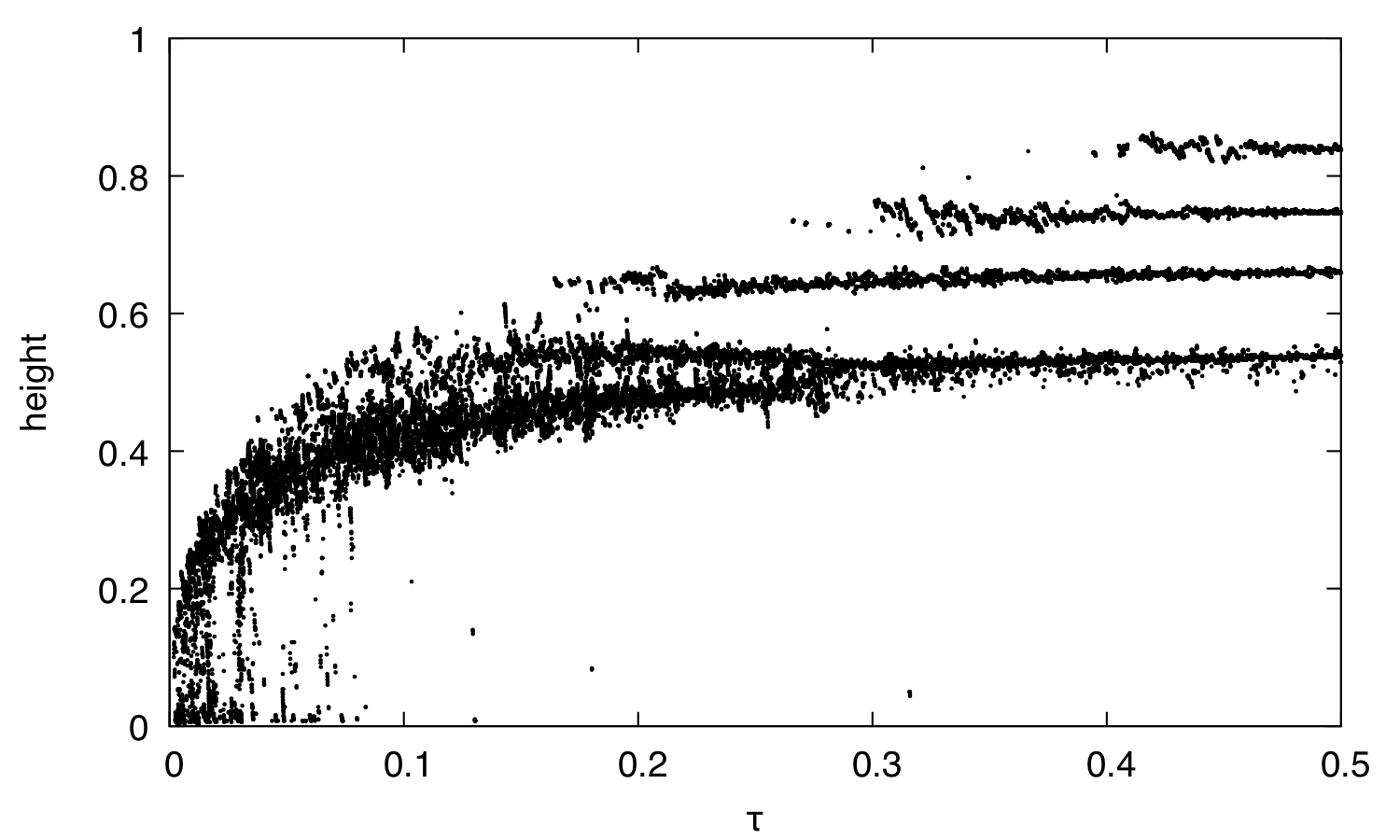} \\
c) \includegraphics[height=0.3\textwidth]{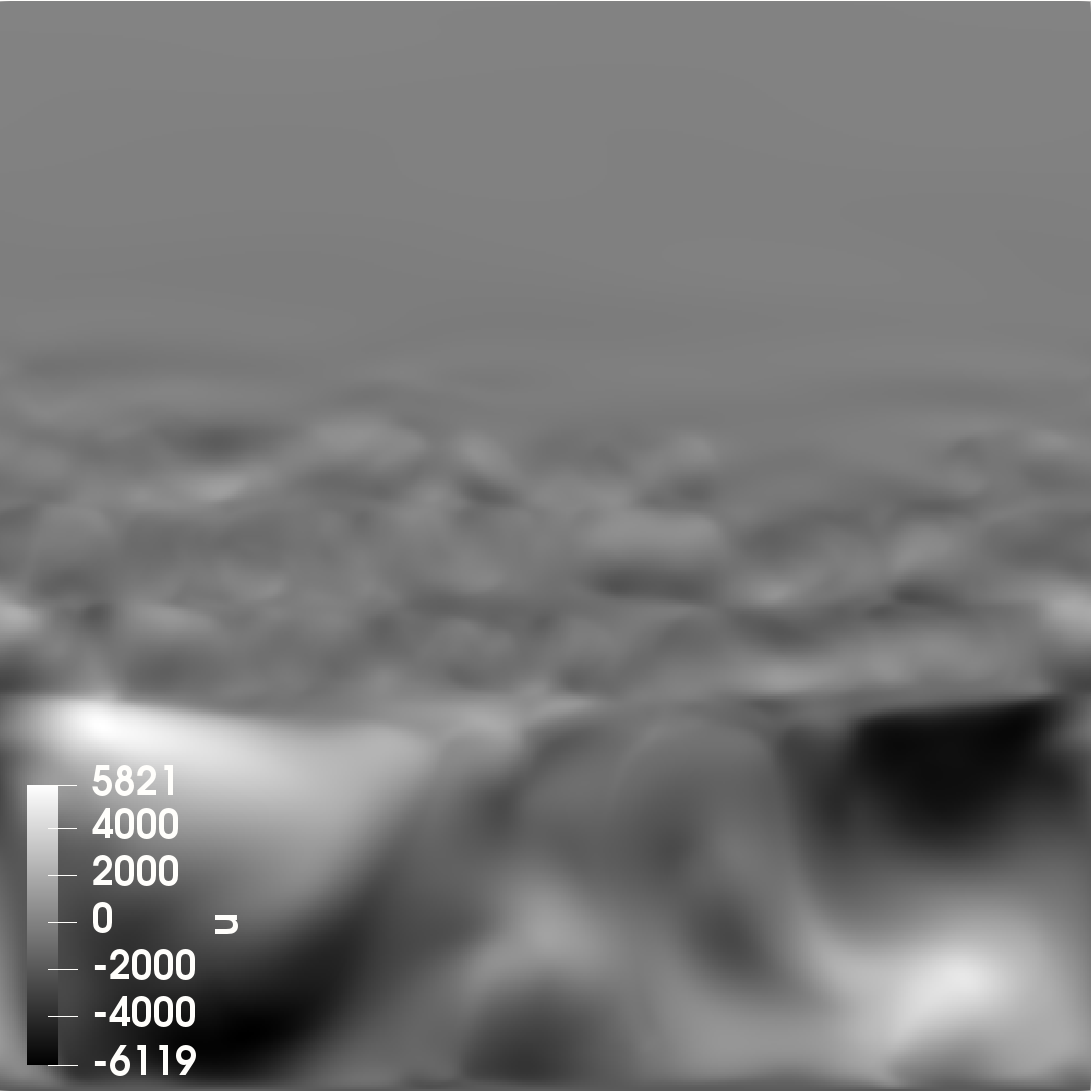}
d) \includegraphics[height=0.3\textwidth]{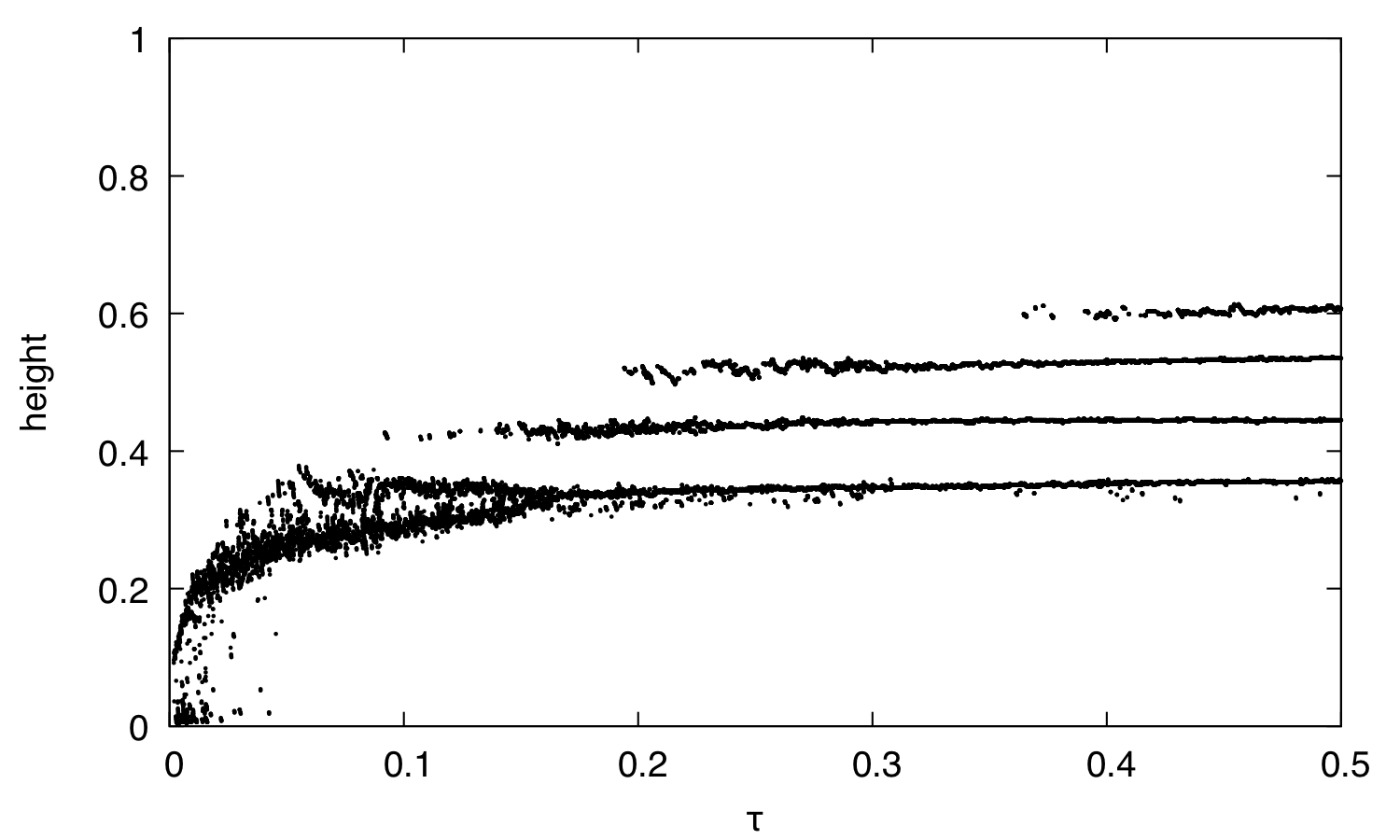}
\end{center}
\caption{Simulations with $\rm Pr=7$ and $\rm Ri=10$ for $\rm R_{\rho}=2$ (top row) and  $\rm R_{\rho}=3$ (bottom row). The left 
column depicts the vertical velocity at $\tau=0.5$ and the right column depicts positions of interfaces.}
\label{fig:Pr7Ri10-overview}
\end{figure}
\begin{figure}[htb]
\begin{center}
a) \includegraphics[height=0.3\textwidth]{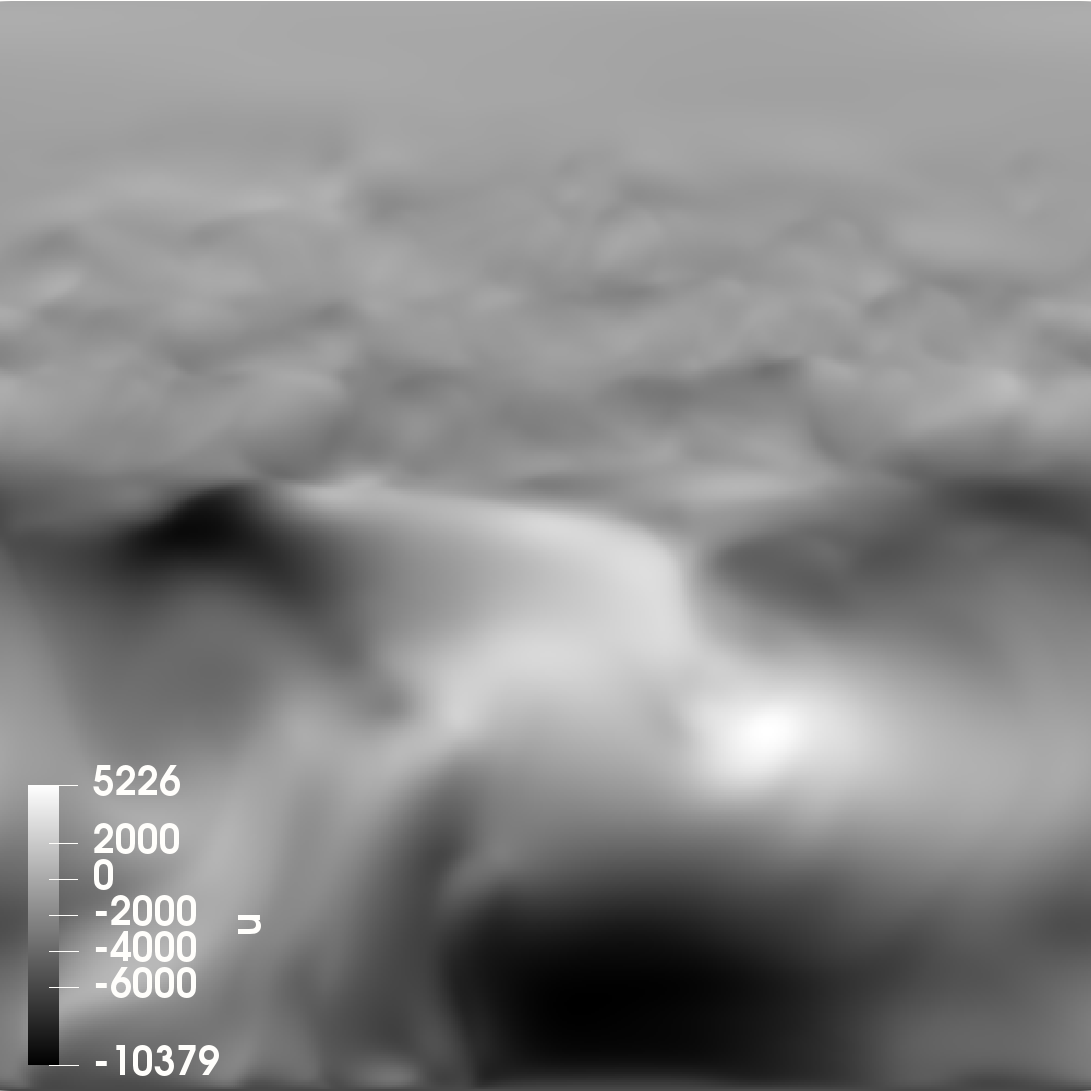}
b) \includegraphics[height=0.3\textwidth]{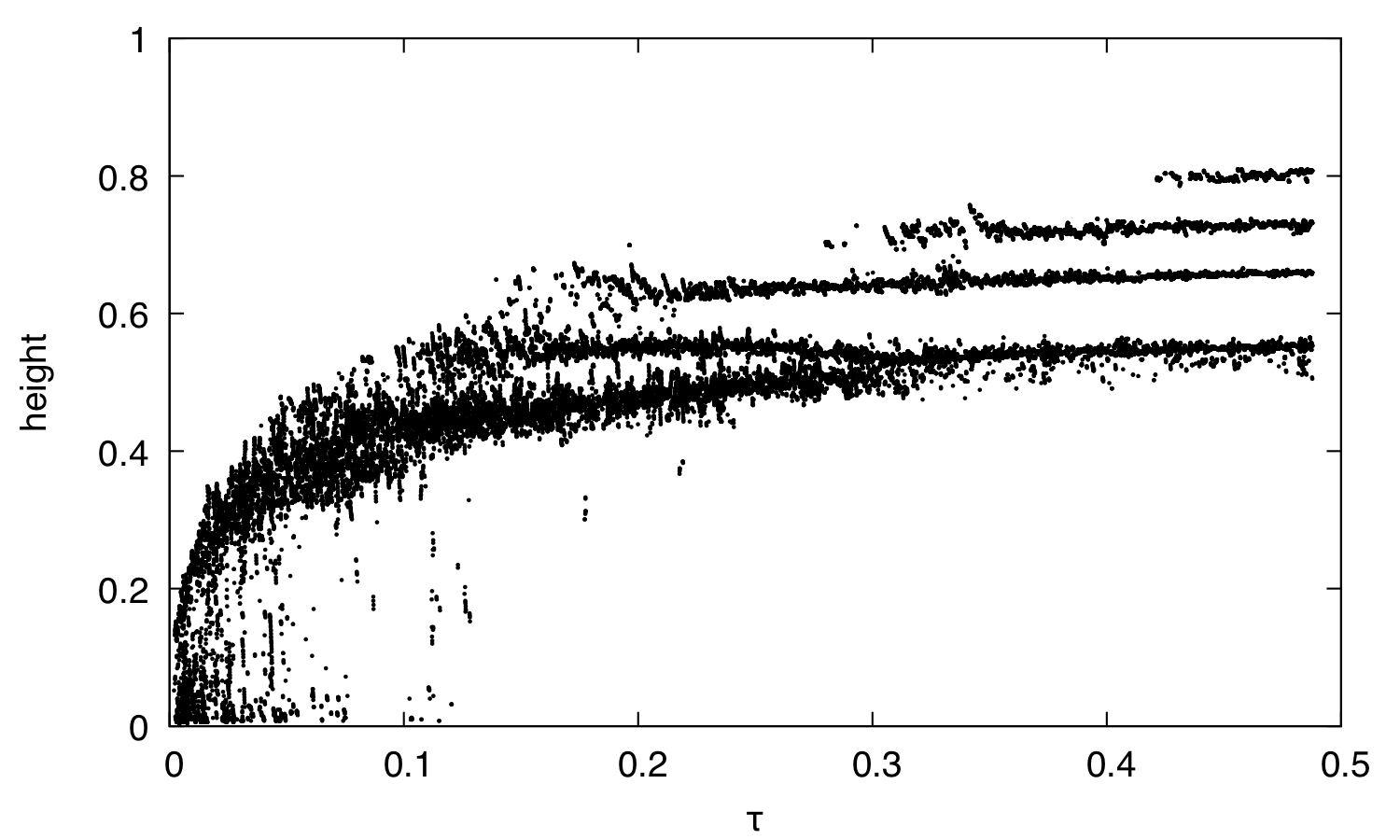}  \\
c) \includegraphics[height=0.3\textwidth]{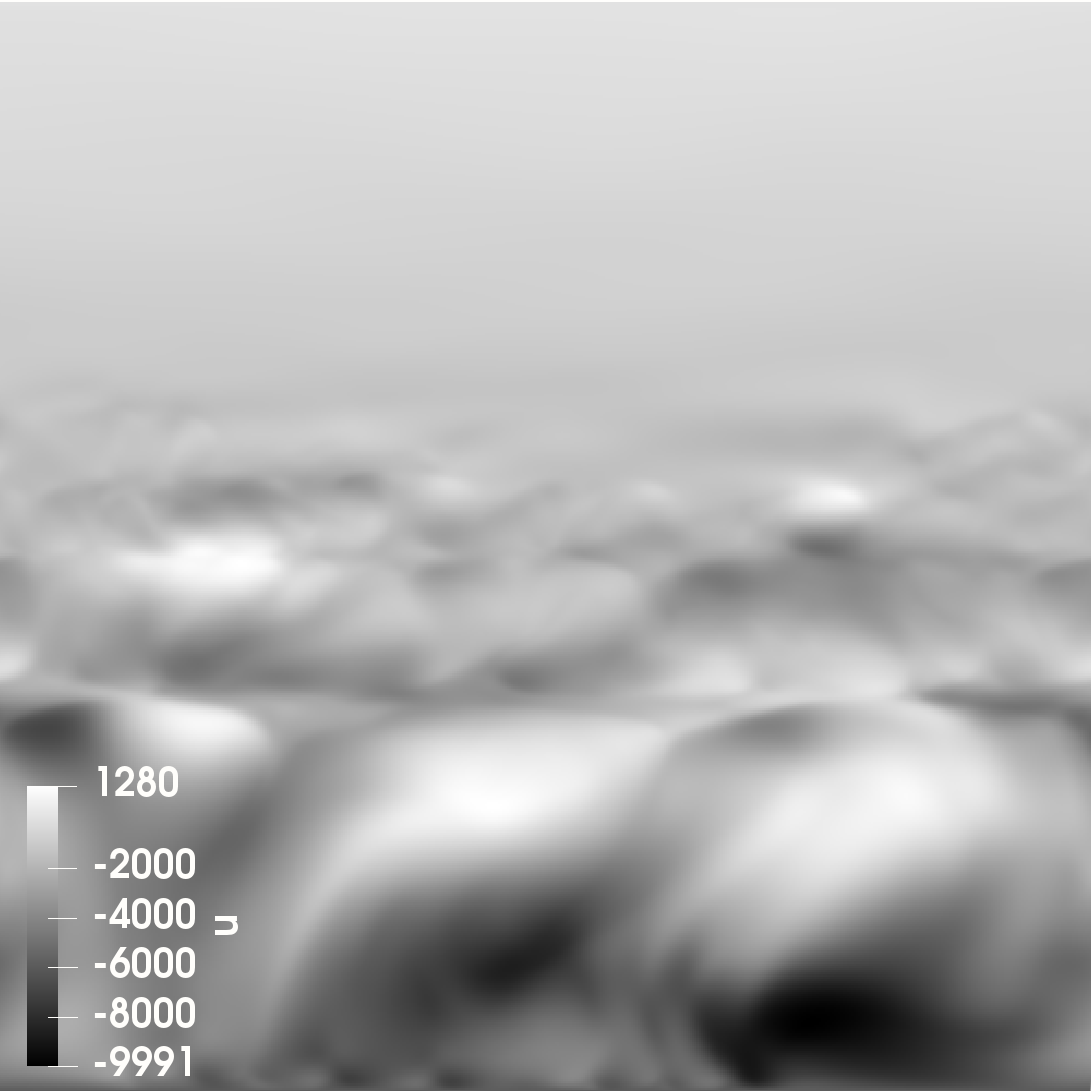}
d) \includegraphics[height=0.3\textwidth]{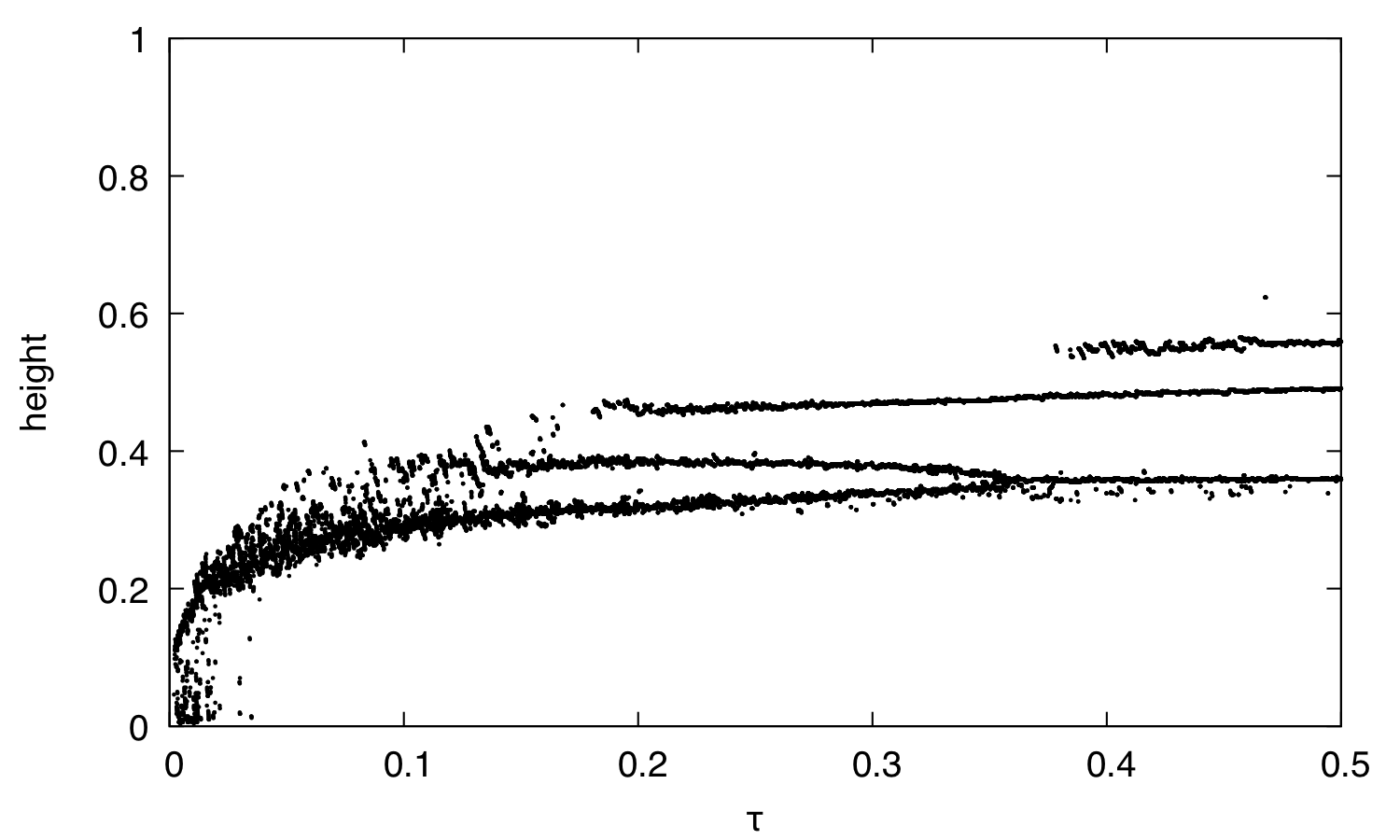}
\end{center}
\caption{Simulations with $\rm Pr=7$ and $\rm Ri=1$ for $\rm R_{\rho}=2$ (top row) and  $\rm R_{\rho}=3$ (bottom row). The left 
column depicts the vertical velocity at $\tau=0.5$ and the right column depicts positions of interfaces.}
\label{fig:Pr7Ri1-overview}
\end{figure}
\begin{figure}[htb]
\begin{center}
a) \includegraphics[height=0.3\textwidth]{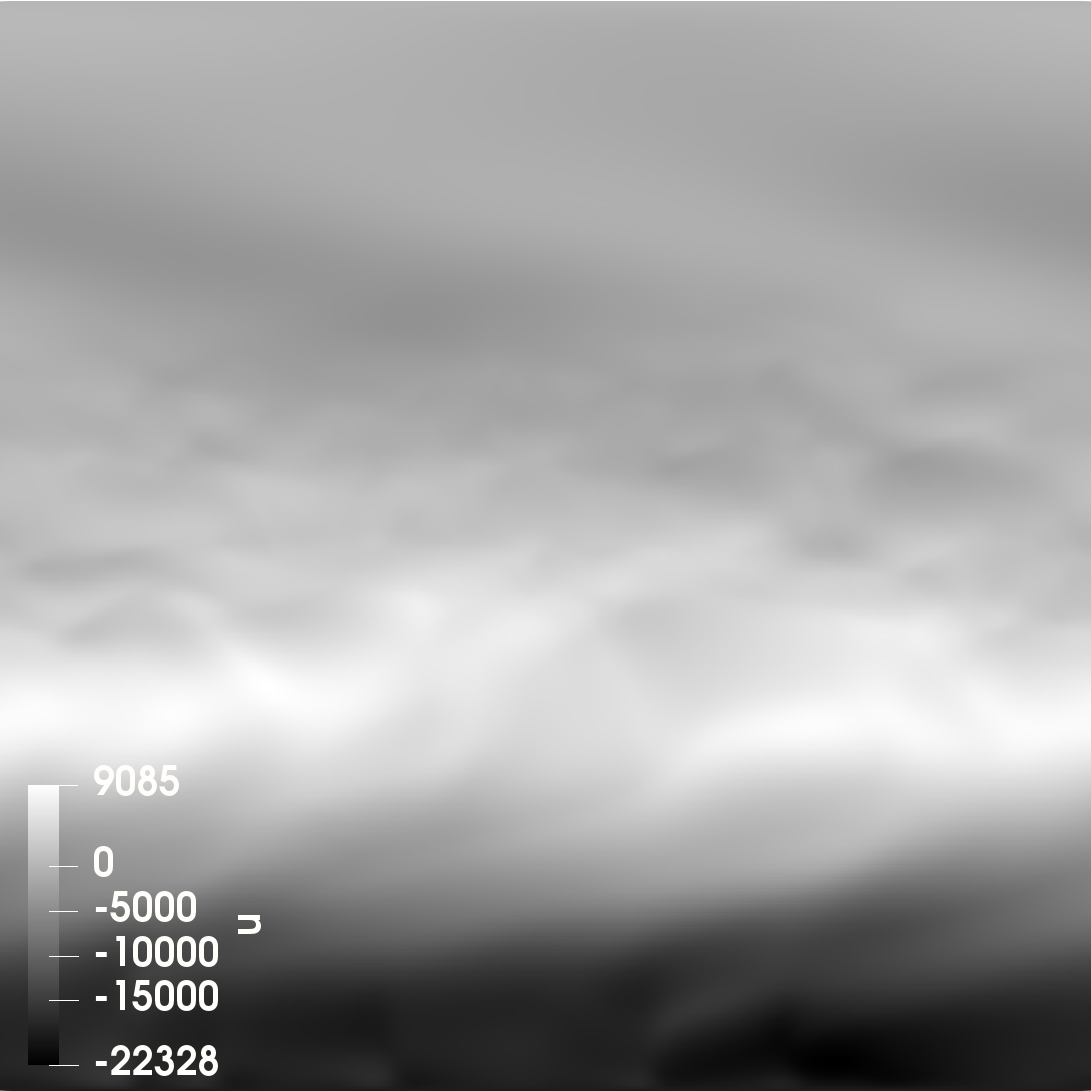}
c) \includegraphics[height=0.3\textwidth]{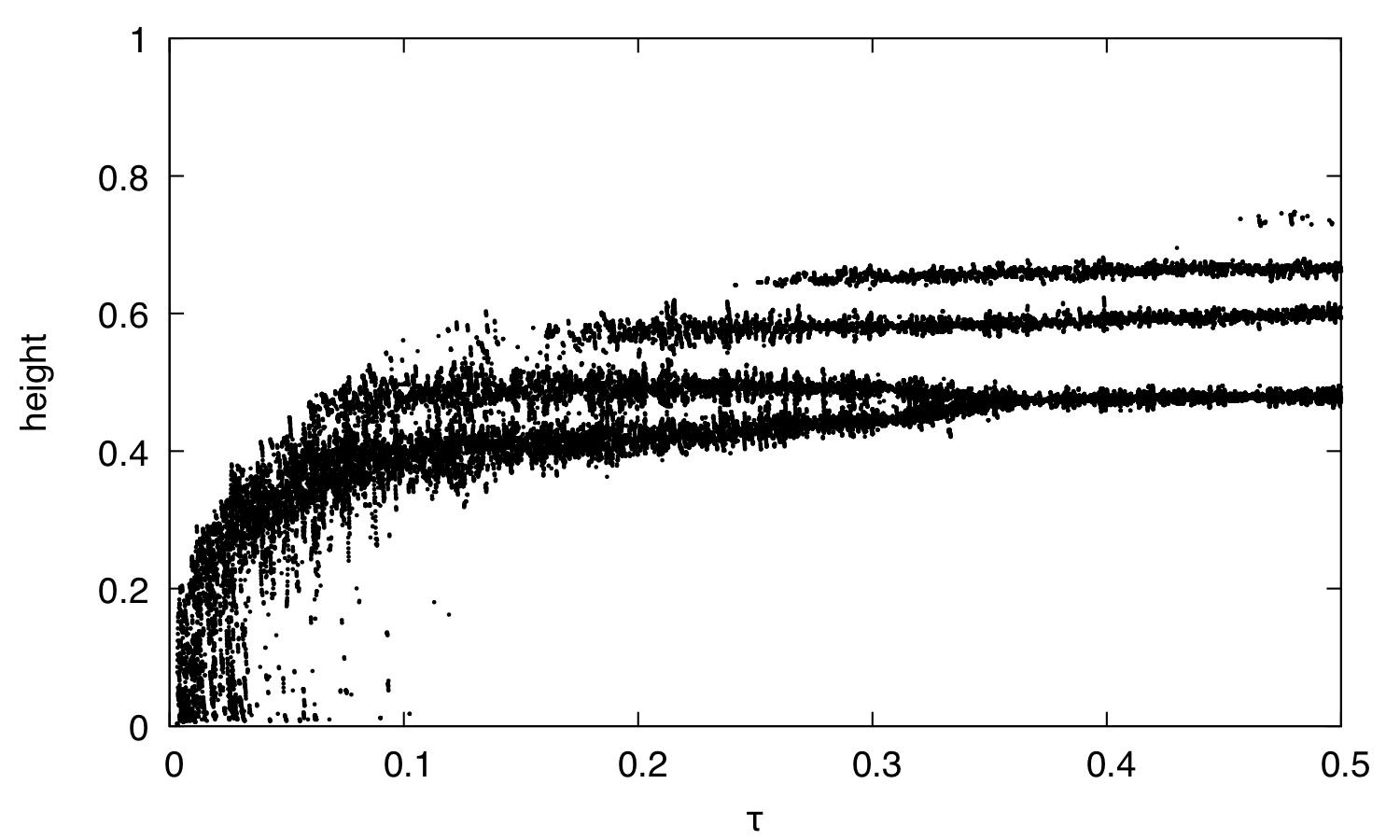}  \\
a) \includegraphics[height=0.3\textwidth]{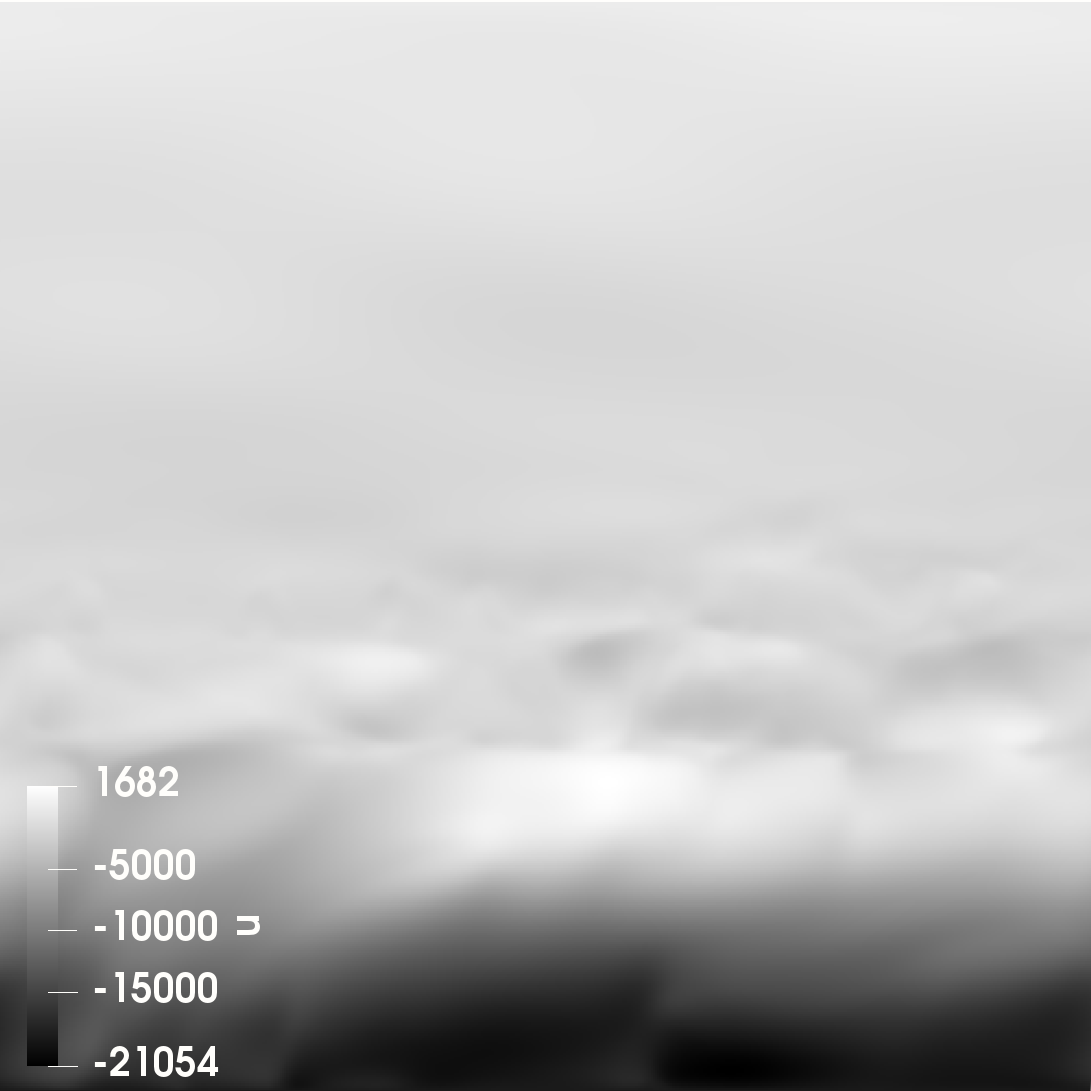}
d) \includegraphics[height=0.3\textwidth]{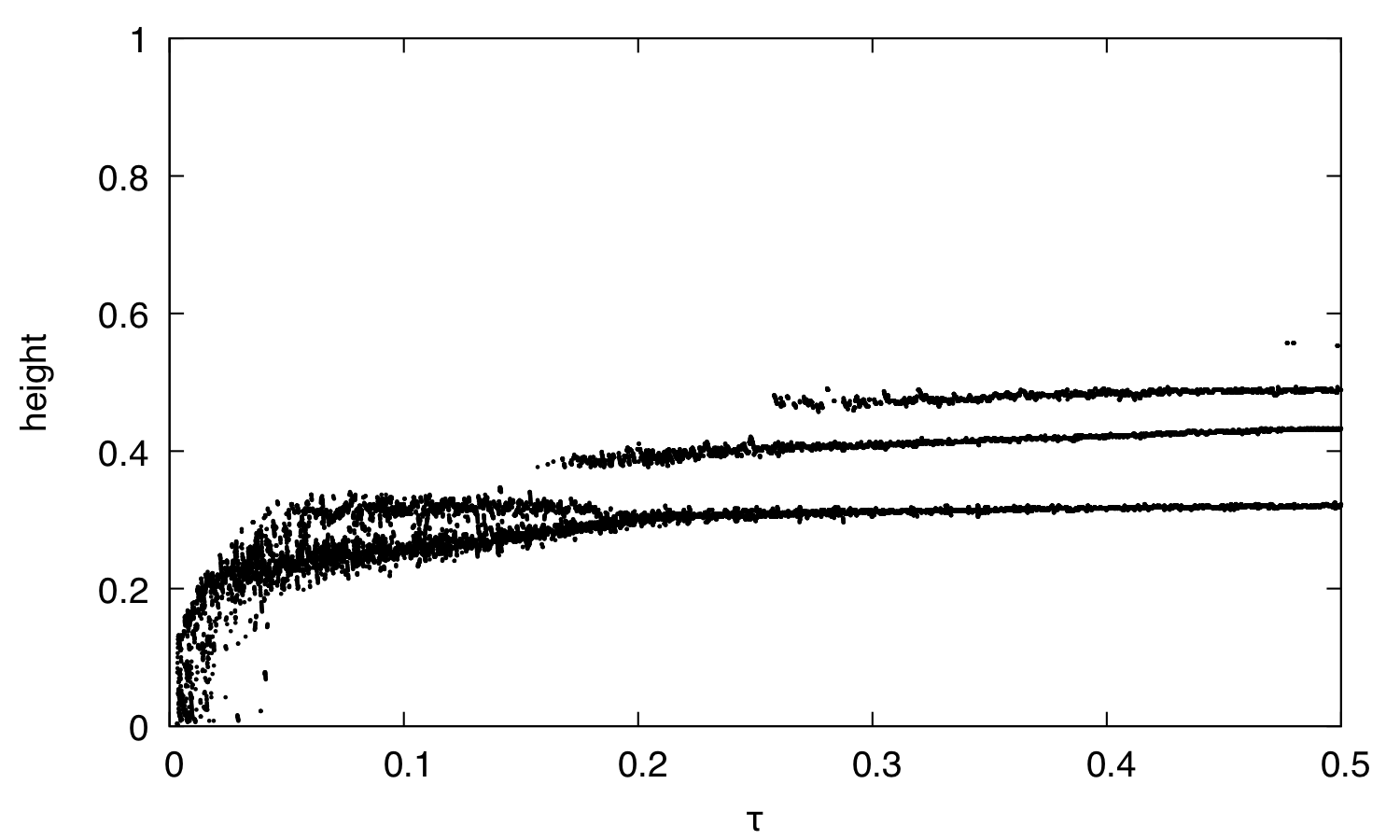}
\end{center}
\caption{Simulations with $\rm Pr=7$ and $\rm Ri=10^{-1}$ for $\rm R_{\rho}=2$ (top row) and  $\rm R_{\rho}=3$ (bottom row).
The left column depicts the vertical velocity at $\tau=0.3$ and $\tau=0.35$, respectively, and the 
right column depicts positions of interfaces.}
\label{fig:Pr7Ri01-overview}
\end{figure}

For the oceanographic case six simulations have been performed with shear for values of ${\rm Ri=10}$, 
${\rm Ri=1}$, and ${\rm Ri=10^{-1}}$ for $\rm R_{\rho}=\{2,3\}$. Fig.~\ref{fig:Pr7Ri10-overview}, Fig.~\ref{fig:Pr7Ri1-overview},
and Fig.~\ref{fig:Pr7Ri01-overview} summarize the main results. Layer formation is found in all these cases. Again, for 
$\rm R_{\rho}=2$ significantly higher first layer heights are found than for $\rm R_{\rho}=3$. The secondary layer formation process 
acts on the same time scales ($\tau \sim 0.1$) as for ${\rm Ri=\infty}$. However, the formation of the first layer 
is slowed down and is completed only around $\tau \sim 0.3$ ($\rm R_{\rho}=2)$ and $\tau \sim 0.2$ ($\rm R_{\rho}=3)$, 
respectively. This is about twice as long as for ${\rm Ri=\infty}$. Moreover, the first merger with the initially formed 
second layer is delayed strongly as a function of Ri$^{-1}$. The height of the first layer itself though is not influenced 
significantly by Ri. However, shear influences the global velocity field. 

\begin{figure}[htb]
\begin{center}
a) \includegraphics[height=0.6\textwidth]{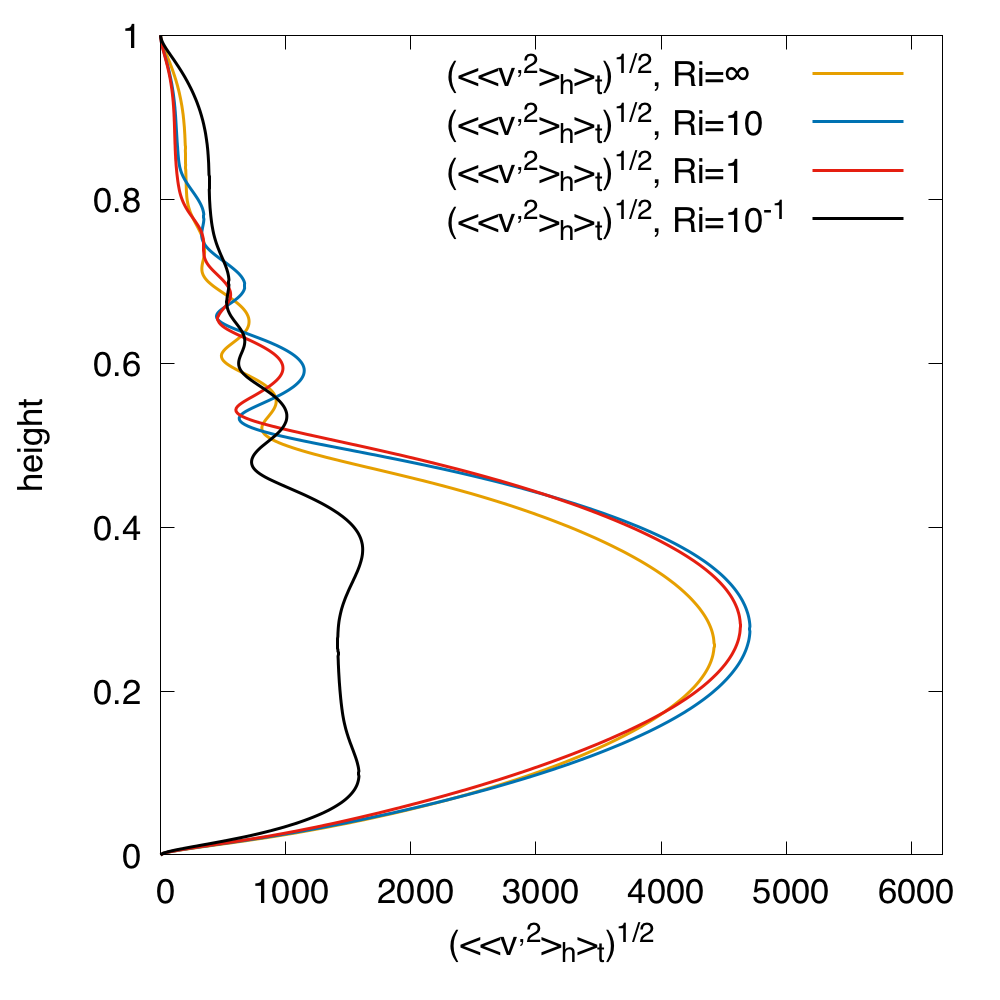} 

b) \includegraphics[height=0.6\textwidth]{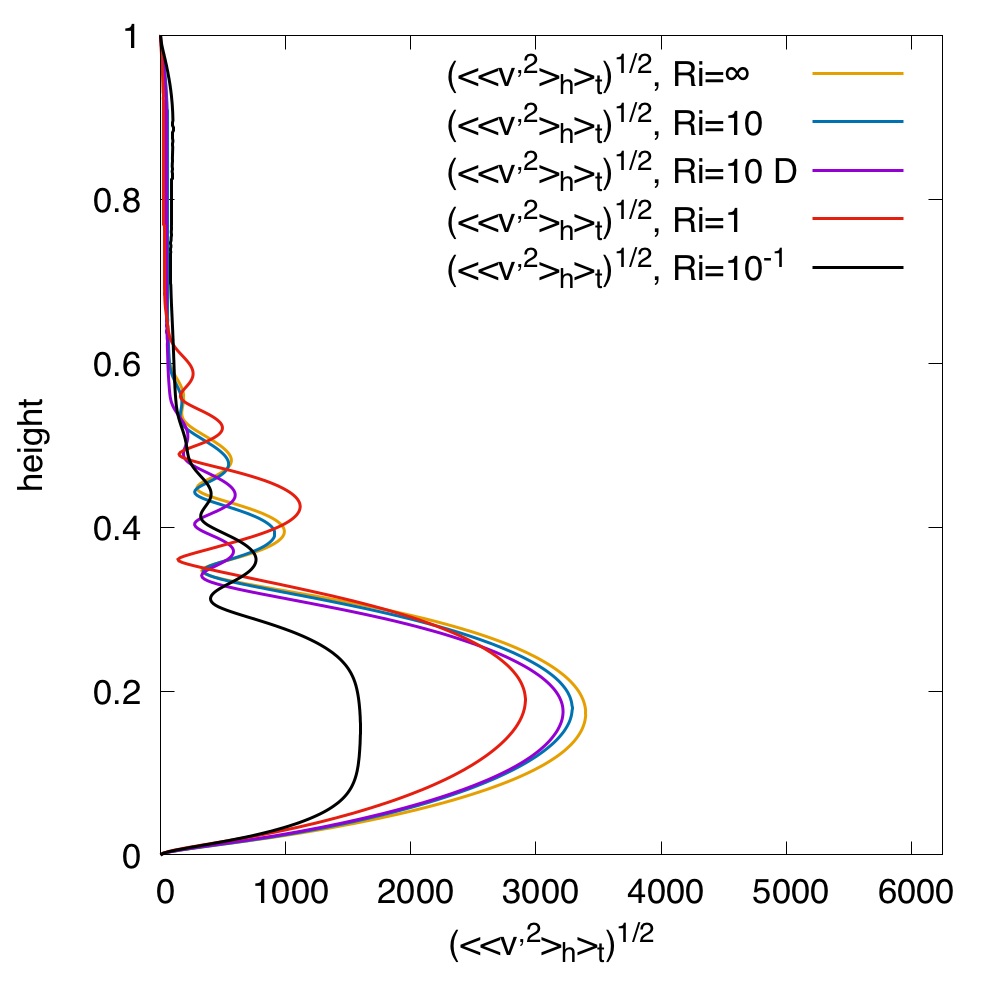}
\end{center}
\caption{Fluctuations of the vertical velocity component v averaged over time 
$\rm \sqrt{\langle\langle(v')^2\rangle_h\,\rangle_{t}}$ with $\rm v'=v-\langle v\rangle_h$ for (a) $\rm R_{\rho}=2$ 
and (b)  $\rm R_{\rho}=3$ for the oceanographic case. 
The case of $\rm Ri=10 D$ (purple line) has moving plates at the top and the bottom.}
\label{fig:v_Pr7Rrho}
\end{figure}

\begin{figure}[htb]
\begin{center}
a) \includegraphics[height=0.6\textwidth]{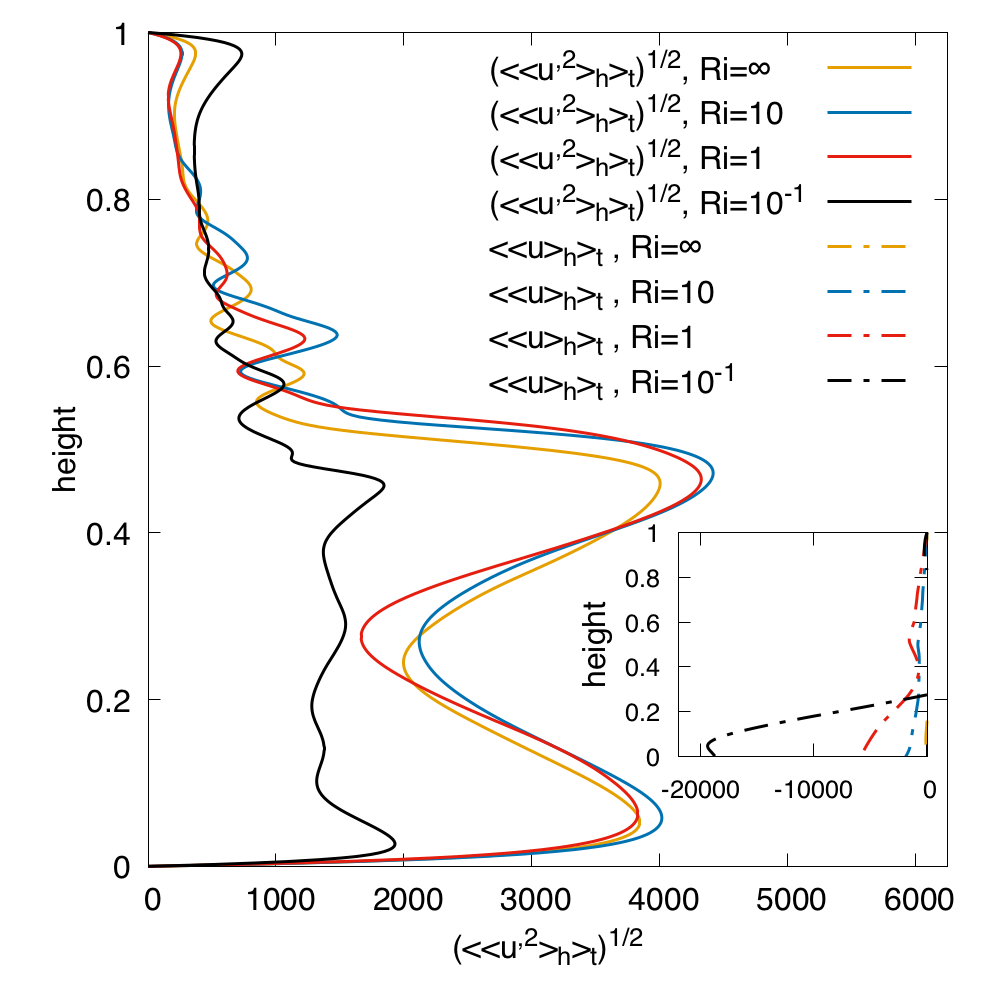} 

b) \includegraphics[height=0.6\textwidth]{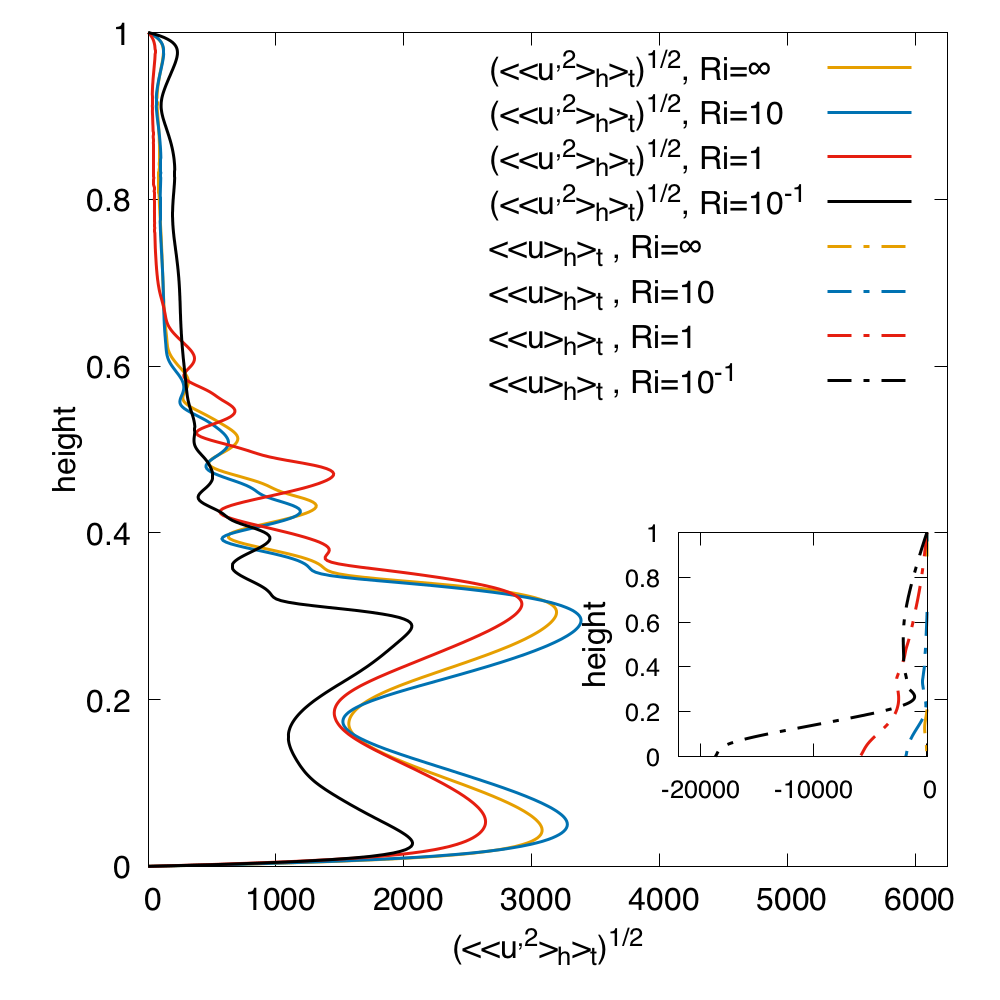}
\end{center}
\caption{Fluctuations of the horizontal velocity component u averaged over time 
$\rm \sqrt{\langle\langle(u')^2\rangle_h\,\rangle_{t}}$ with $\rm u'=u-\langle u\rangle_h$ for (a) $\rm R_{\rho}=2$ 
and (b)  $\rm R_{\rho}=3$ for the oceanographic case.
Insets: mean horizontal velocity component averaged over time $\rm \langle{\langle u\rangle}_h\,\rangle_{t}$. The case of $\rm Ri=10 D$ (purple line) has moving plates at the top and the bottom. }
\label{fig:u_Pr7Rrho}
\end{figure}

The influence of shear is depicted in Fig.~\ref{fig:Pr7Ri01-overview}~(a) as 2D velocity field 
and as mean value in  Fig.~\ref{fig:v_Pr7Rrho}~(a) and Fig.~\ref{fig:u_Pr7Rrho}~(a) for $\rm R_{\rho}=2$ 
as well as in Fig.~\ref{fig:v_Pr7Rrho}~(b) and Fig.~\ref{fig:u_Pr7Rrho}~(b)
for $\rm R_{\rho}=3$. The vertical and horizontal velocity components are analyzed 
using the first statistical moment, i.e., the horizontally and temporally
averaged values $\rm \langle{\langle v\rangle}_h\,\rangle_{t}$ and $\rm \langle{\langle u\rangle}_h\,\rangle_{t}$.
In the following, $\rm \langle .\rangle_h$ is the horizontal mean and $\rm \langle .\rangle_{t}$ is the mean in time.
Fluctuations are analysed with the second statistical moments 
$\rm \sqrt{\langle\langle(v-\langle v\rangle_h)^2\rangle_h\,\rangle_{t}}$ and
$\rm \sqrt{\langle\langle(u-\langle u\rangle_h)^2\rangle_h\,\rangle_{t}}$.
The velocity profiles are averaged over $\tau \sim 0.15$, where two secondary layers are detected in all simulations, cf. 
Fig.~\ref{fig:Case_3_Rho3_Ri_in}~(a) between $0.18\leq\tau\leq0.33$. The mean vertical velocities show 
values of $\rm \langle{\langle v\rangle}_h\,\rangle_{t} \lesssim 1$ for the analyzed time span and converge 
towards zero when averaging them over sufficiently long time scales.

The temporally averaged mean horizontal velocity differs significantly from the mean vertical velocity field.
Fig.~\ref{fig:v_Pr7Rrho} (insets of (a) and (b)) shows the latter for varying stability parameter $\rm R_{\rho}$. 
As for vertical velocities the horizontal components of cases with ${\rm Ri}=\infty$ are of ${\bf o}(1)$, i.e., of the
order of the diffusion velocity (see insets of Fig.~\ref{fig:v_Pr7Rrho}~(b) and Fig.~\ref{fig:u_Pr7Rrho}~(b)). 
Even for the highest shear rate, at ${\rm Ri}=10^{-1}$, the formation of secondary layers always occurs despite
the formation of the first layer is delayed by a factor of two in time.
However, the case of Pr7R2Ri01 shows a strong counter-drift in the upper part of the first layer (for comparison 
see also Fig.~\ref{fig:v_Pr7Rrho}~(a), inset). 
It is found that the oppositely directed shear flow in the upper part of the first layer vanishes after the formation of the 
first layer. Thus, a horizontal flow in the first layer directed against the mean shear indicates that the simulation
is not yet dynamically relaxed.

A simulation with moving plates at the bottom and the top has been performed, too. 
It is represented by the label $\rm Ri=10 D$ and depicted in Figs. \ref{fig:v_Pr7Rrho} and \ref{fig:u_Pr7Rrho} (purple lines). 
In contrast with $\rm Ri=10$ it distributes the shear by letting the plates move at half the speed of
the bottom plate for $\rm Ri=10$, as discussed in Sect.~\ref{sec:bc_and_shear}.
As can be noted from comparing in particular with the case $\rm Ri=10$ there is no difference
visible for the diffusive region on top, neither for vertical nor for horizontal root mean square velocities.
Evidently, this shear rate at the given local stratification as characterized by the quantity $\rm R_{\rho, \rm loc}$ 
introduced and discussed in the next subsection is too small to generate a velocity field which could interact 
with the oscillatory double-diffusive convective instability to create layers,
a mechanism discussed by \cite{Radko_2016}. Also for the stack of layers the differences are either 
small anyway or in a range that can be anticipated from the different time evolution for both cases.

\subsection{The giant planet case without shear}  

\begin{figure}[htb]
\begin{center}
a) \includegraphics[height=0.5\textwidth]{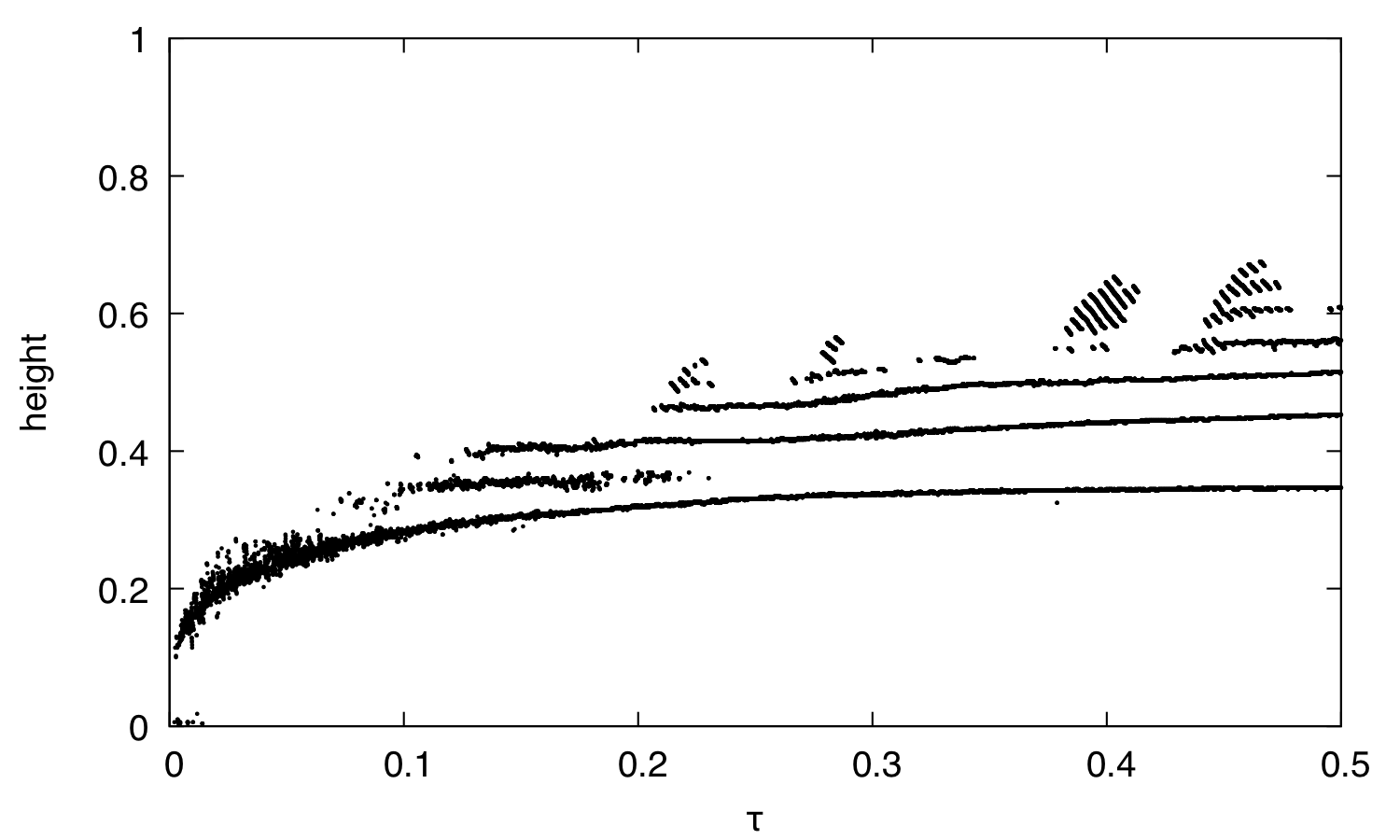} \\
\end{center}
\begin{center}
b) \includegraphics[height=0.5\textwidth]{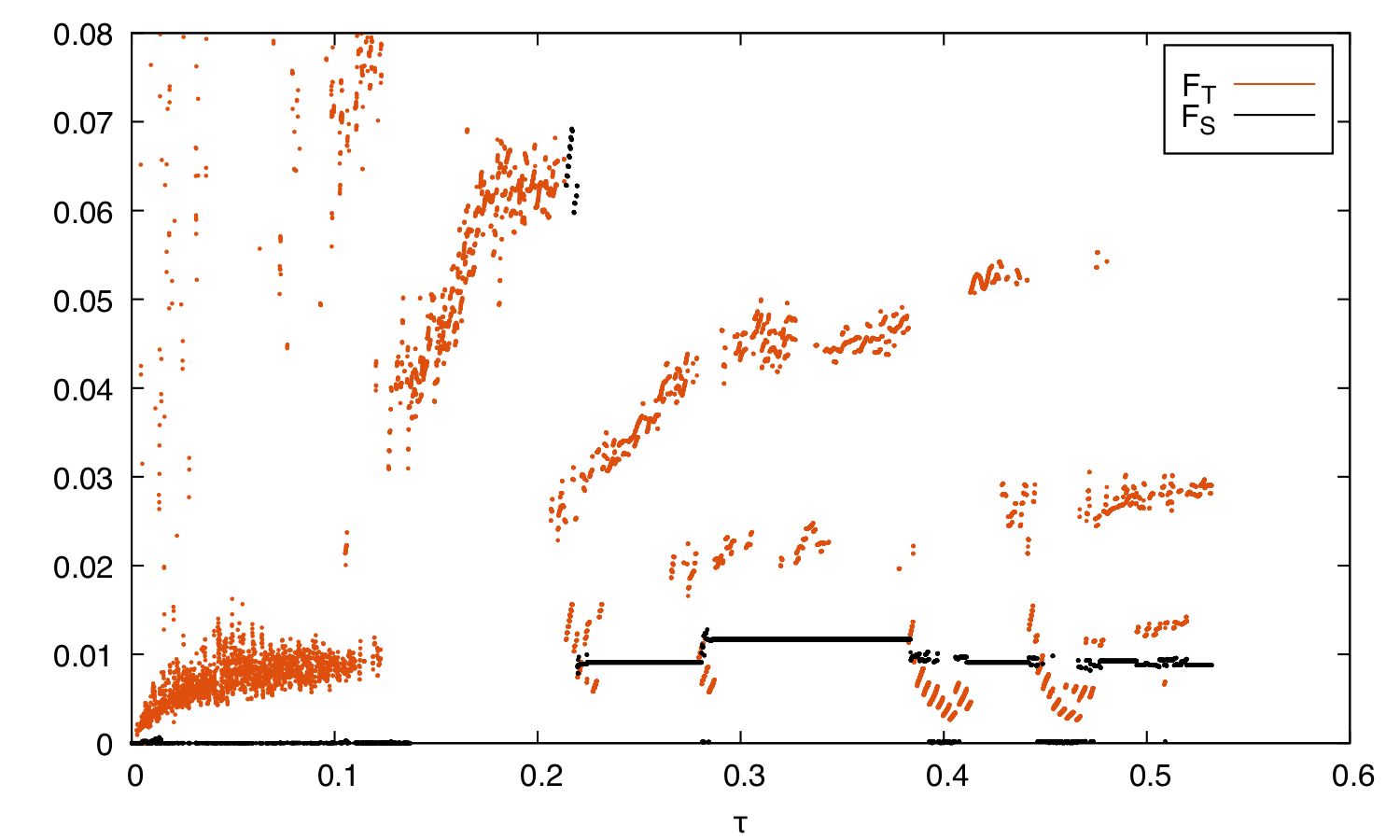}
\end{center}
\caption{Pr01R4: (a) mean location of interfaces as function of time. (b) $\rm F_T$ (orange dots) 
and $\rm F_S$ (black dots) of the first layer as function of time.}
\label{fig:Case_2_Rho4_Ri_in}
\end{figure}

\begin{figure}[htb]
\begin{center}
a) \includegraphics[height=0.5\textwidth]{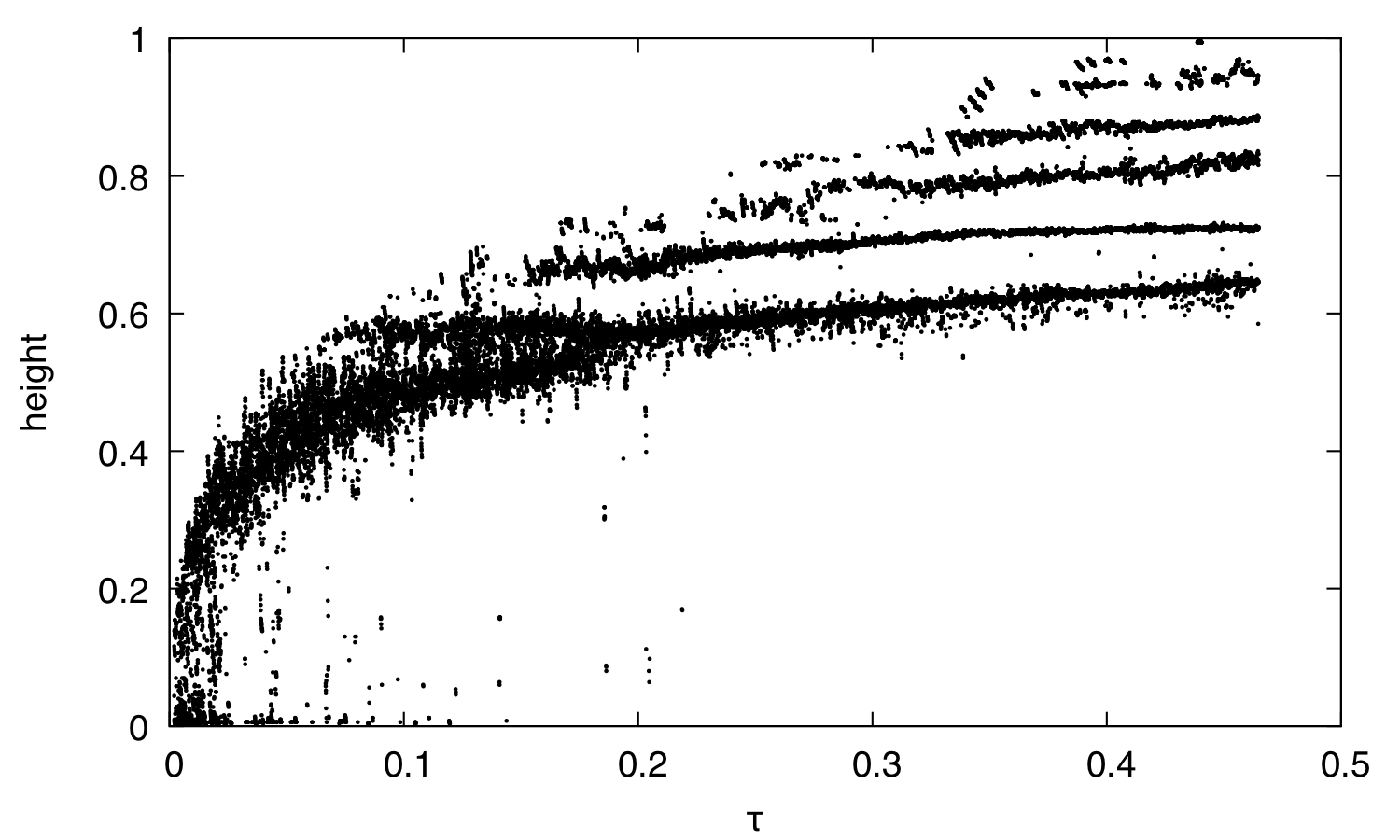} \\
\end{center}
\begin{center}
b) \includegraphics[height=0.5\textwidth]{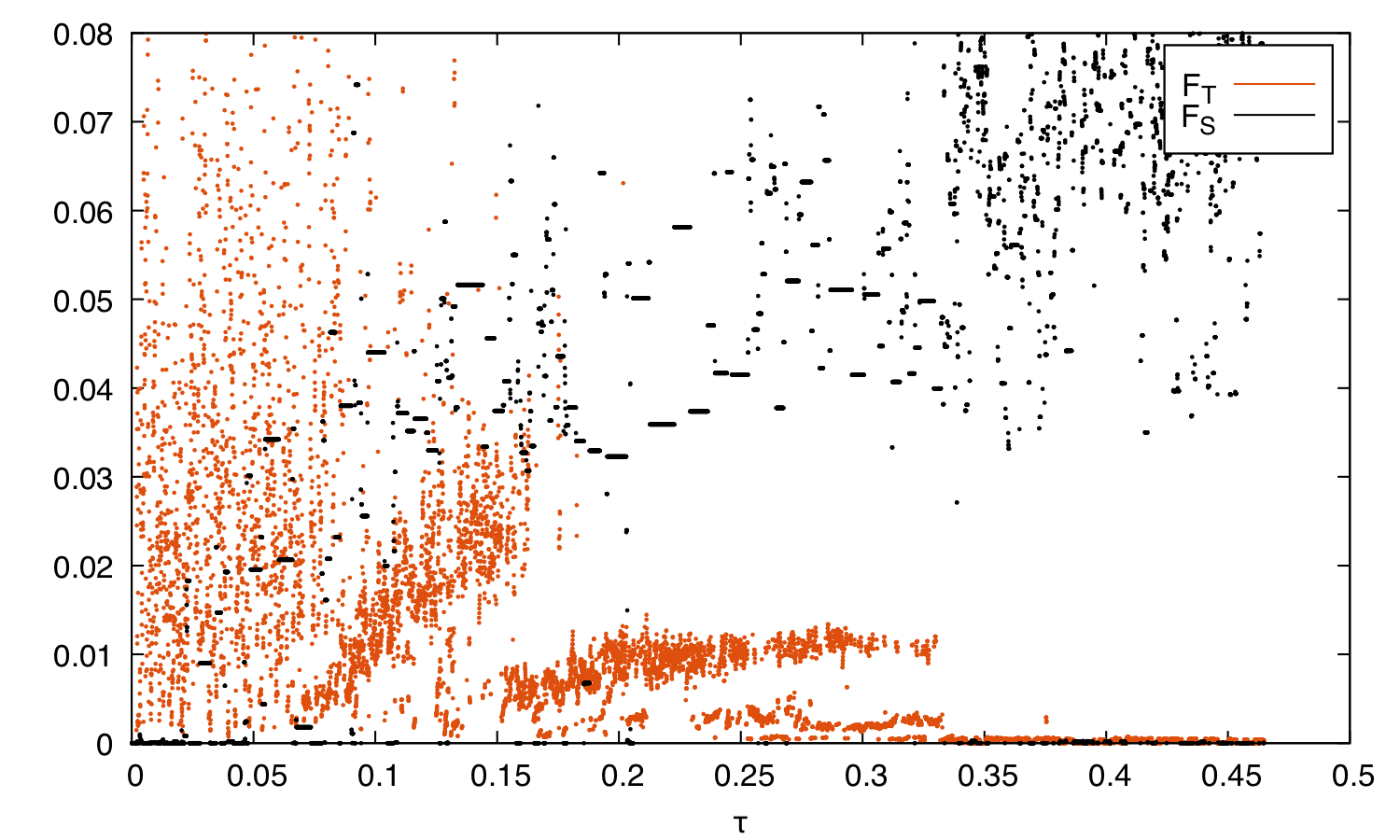}
\end{center}
\caption{Pr01R2: (a) mean location of interfaces as function of time. (b) $\rm F_T$ (orange dots) 
and $\rm F_S$ (black dots) of the first layer as function of time.}
\label{fig:Case_2_Rho2_Ri_in}
\end{figure}

\begin{figure}[htb]
\begin{center}
(a) \includegraphics[height=0.5\textwidth]{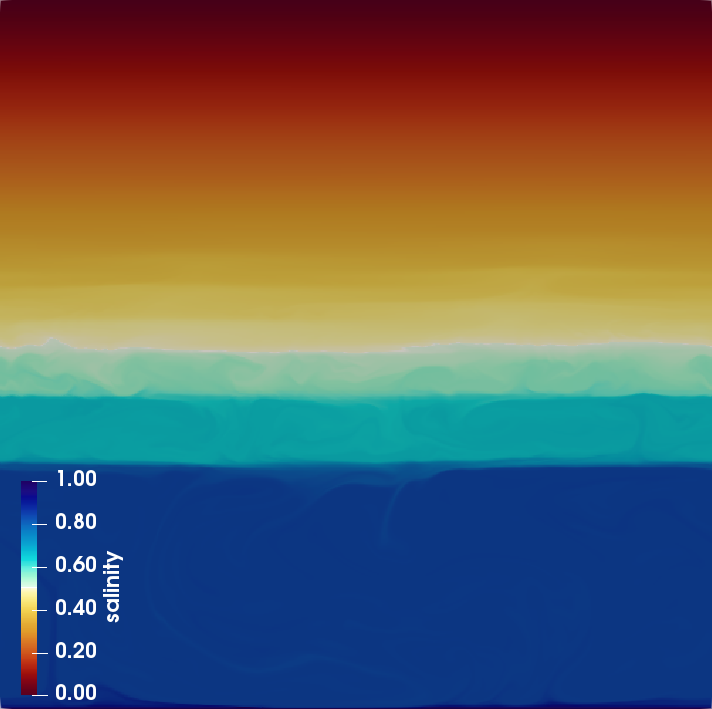}

 (b) \includegraphics[height=0.6\textwidth]{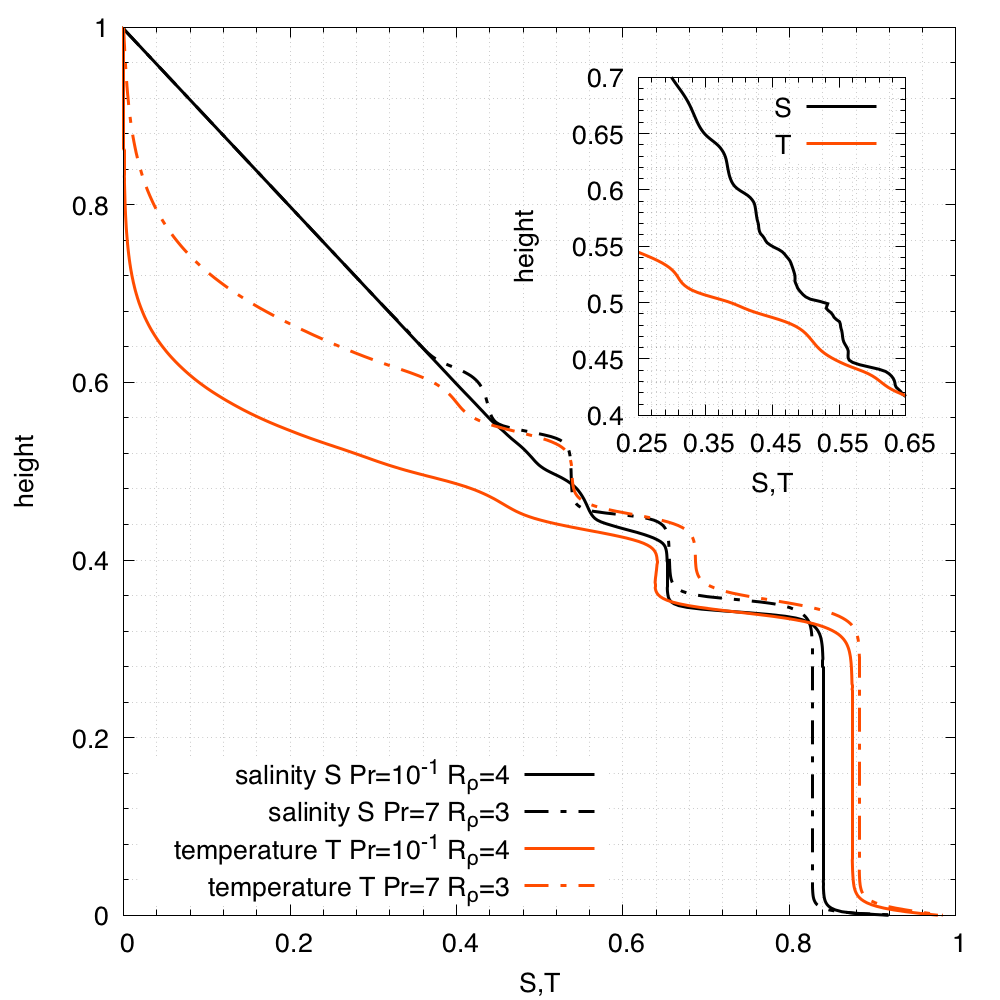}

(c) \includegraphics[height=0.6\textwidth]{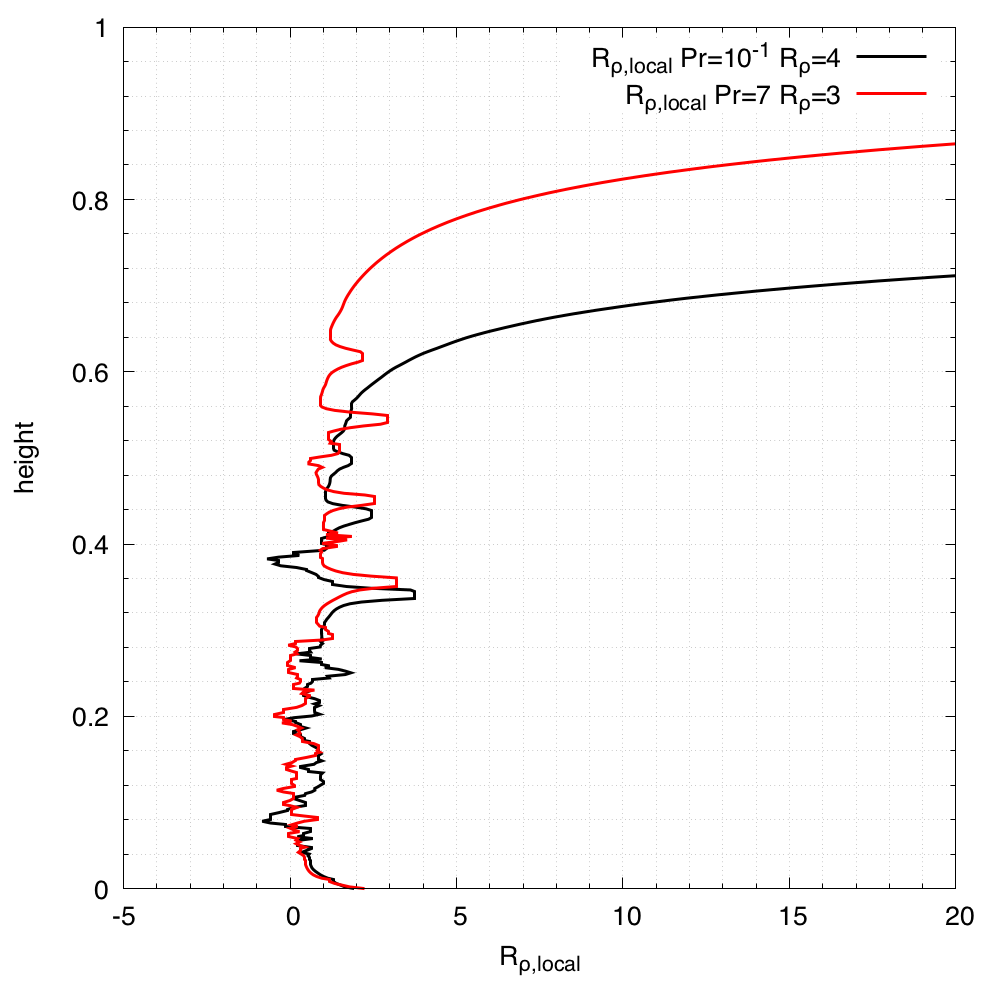}
\end{center}
\caption{(a) salinity field at $\tau=0.39$ for Pr01R4. Multiple, interconnected, ODDC layers with height $y\sim 0.2$ 
             above the first secondary layer. (b) horizontally averaged salinity and temperature for Pr01R4 with zoom in the region of ODDC layers
             above the first layer. (c) horizontally averaged local stability parameter for the case of $\rm Pr=7$ and $\rm Pr=10^{-1}$.}
\label{fig:Case_2_Rho3_Ri_inf_S_tau_039}
\end{figure}

Two reference cases without shear ($\rm Pr=10^{-1}$, $\rm Ra^*=5.0\cdot 10^7$, $\rm Le=10^{-2}$, 
and $\rm R_{\rho}=\{2,4\}$) have been investigated. They are analyzed here in the same way as the oceanographic 
ones. In principle, for comparable $\rm R_{\rho}$ and $\rm Ra^*$ the time scales and global dynamics 
are identical to cases where $\rm Pr=7$. The height of the first layer increases for decreasing stability 
parameter $\rm R_{\rho}$ and the heights of secondary layers are of the order of $\approx \rrho^{-1}$.
The formation of the first layer is complete within $\tau=0.2$. However, the formation of secondary layers  
differs significantly from the oceanographic case. Consecutive formation of layers in time intervals of $\tau=0.1$ 
is found as in the case of ${\rm Pr}=7$. Additionally, we find the simultaneous formation of multiple layers.
Here, two scenarios are observed: firstly, two layers may form at nearly the same time. These layers
are stable with respect to their height. Secondly, multiple but unstable layers form. They survive only for 
a fraction of the thermal time scale before they dissolve or merge to form a single, larger layer. 
Both types of layers fulfill the prominence criteria ($\rm P>1$) and are hence detected by the 
algorithm. They can readily be found in Fig.~\ref{fig:Case_2_Rho4_Ri_in} (a) and Fig.~\ref{fig:Case_2_Rho2_Ri_in}~(a). 
Especially Pr01R4 shows the formation of multiple, unstable layers at $\tau=0.4$ and $\tau=0.45$. Here, up to four 
layers form instantaneously. The unstable layers are best visualized in the two-dimensional field of salinity,
shown here at $\tau=0.4$ for Pr01R4 in Fig.~\ref{fig:Case_2_Rho3_Ri_inf_S_tau_039} (b), inset.

We recall that contrary to the oceanographic case, the diffusive region above the initially formed secondary 
layers can gradually become unstable to (oscillatory) double-diffusive convection (ODDC) for the seemingly
more stable linear stratification which begins to form on top of the stack of secondary layers, because for 
$\rm Pr=10^{-1}$ and $\rm Le=10^{-2}$ we have $1 < \rrho < \rrhomax < 10$ for ODDC to occur, which for
a mean $\rm R_{\rho}$ of 2 and 4 is easily fulfilled. Moreover, we can expect that these conditions are met 
locally above the top interface once diffusion has operated for sufficiently long time to evolve the temperature 
distribution into a linear one in that region (cf.\ Fig.~\ref{fig:Case_2_Rho3_Ri_inf_S_tau_039}~(b)). The 
establishment of a doubly linear stratification in a purely diffusive region occurs independently of the specific 
value of $\rm Pr$ (cf.\ Fig.~\ref{fig:Case_3_Rho3_Ri_inf_S_tau_065}(b)) in our normalized, dimensionless unit 
of time, Eq.~(\ref{eq:entdim}). In the end for $\rm Pr=10^{-1}$ layer formation similar to that one described
in \citet{Radko_2003} and \citet{Mirouh_2012} can occur in that region and as described in the latter, these
layers merge again to form larger entities. For $\rm Pr=7$ this process is unlikely to occur since without shear
\citep{Radko_2016} the allowed range for $\rm R_{\rho}$ is extremely narrow (between 1 and 1.14 for the
instability in first place and even smaller for actual layer formation through this mechanism, since this requires 
$\rm 1 < \rrho < \rrhomax < 1.14$). On the other hand, for the `giant planet scenario' layers formed by ODDC
appear alongside the layers created due to heat release from the primary layer at the bottom. 

The thermal and solute fluxes across the interfaces parameterized by $\rm F_T$ and $\rm F_S$ are depicted for the 
first layer in Fig.~\ref{fig:Case_2_Rho4_Ri_in}~(b) and in Fig.~\ref{fig:Case_2_Rho2_Ri_in}~(b). Again, each formation
of a new secondary layer is characterized by a drop in the thermal gradient through the first layer and its adjacent
interface. Simultaneous formation of two secondary layers is characterized by overlapping clusters, cf.\
Fig.~\ref{fig:Case_2_Rho4_Ri_in}~(b) for $\tau=0.2-0.5$. The same situation is found in 
Fig.~\ref{fig:Case_2_Rho2_Ri_in}~(b) between $\tau=0.15-0.25$. The formation of the unstable layers is 
also visible in Fig.~\ref{fig:Case_2_Rho4_Ri_in}~(b): they are related to filament clusters found between 
$\tau=0.4-0.5$. The saline flux of the first layer is not correlated with the formation of secondary or 
unstable layers. Black dots, e.g., in Fig.~\ref{fig:Case_2_Rho4_Ri_in}~(b) are not clustered nor do they show 
any drops except for a few events during the formation of unstable layers.

To highlight the difference in layer formation for the `oceanographic' scenario (Pr7R3) and the `giant planet'
case (Pr01R4) in Fig.~\ref{fig:Case_2_Rho3_Ri_inf_S_tau_039}~(b) we compare the dependence of $\rm T$ and $\rm S$
on height after $\tau=0.39$. For Pr7R3 a later state is shown in 
Fig.~\ref{fig:Case_3_Rho3_Ri_inf_S_tau_065}~(b): at $\tau=0.65$, the temperature in the third and fourth layer 
has increased and the gradients in $\rm S$ and $\rm T$ at the top interface now coincide. But already at $\tau=0.39$ 
the latter is stable (recall $\rrho=3$). For the case of Pr01R4 we observe that only the lowermost 
two layers are fully developed ($\rm T$ and $\rm S$ independent of height within the convectively mixed layer). 
The third layer features transport by both diffusion and convection, as there is clearly a non-zero gradient throughout
the layer. Above $\rm y \sim 0.44$ 
many small layers appear, particularly in $\rm S$. The inset highlights those little staircases
which are the unstable `ODDC' layers discussed just above and visible also both in Fig.~\ref{fig:Case_2_Rho4_Ri_in}~(a)
and, indirectly, in Fig.~\ref{fig:Case_2_Rho4_Ri_in}~(b). 
In Fig.~\ref{fig:Case_2_Rho3_Ri_inf_S_tau_039}~(c) we 
show the local stability parameter $\rm R_{\rho, \rm loc}$ computed from averaging over the last interval of
$\tau=0.1$ available for each simulation, i.e., for $\tau=0.4-0.5$ for Pr01R4 and for $\tau=0.55-0.65$ for Pr7R3.
Reinterpreting $\rm \Delta S$ and $\rm \Delta T$ as temperature differences 
between adjacent 
vertical grid points, $\rm \delta S$ and $\rm \delta T$, we define $\rm R_{\rho, \rm loc}  = \rrho \delta S / \delta T$
(cf.\ the quantity $\rm \tilde{R}_{\rho}$ in \citealt{Spruit_2013} which was introduced 
to develop a more detailed model of ODDC interfaces). For pure diffusion, $\rm R_{\rho, \rm loc}=\rrho$. Inside 
a convectively unstable region, this ratio is poorly defined, since both $\rm \delta S$ and $\rm \delta T$ can be very small,
but due to higher temperature diffusivity than salt diffusivity and diffusion still contributing to the transport,
we can expect $\rm R_{\rho, \rm loc} \lesssim 1$, i.e., the solute differences get mixed more efficiently than temperature
differences. This is what we observe for the lowermost, primary layer in Fig.~\ref{fig:Case_2_Rho3_Ri_inf_S_tau_039}~(c).
More importantly, we can clearly depict the average vertical position of the layer interfaces: with values
$\rm R_{\rho, \rm loc} \gtrsim 1.2$ all four interfaces clearly stick out for Pr7R3. Between them 
$\rm R_{\rho, \rm loc} \lesssim 1$ as expected for the convectively mixed part of the layers. Remarkably, the minimum 
on top of the fourth interface, at $\rm y \sim 0.64$, has $\rm R_{\rho, \rm loc} \sim 1.2$ with $\rm R_{\rho, \rm loc}$ 
increasing further above. Due to diffusion $\rm R_{\rho, \rm loc} \rightarrow \rrho$
for those layers, but since $\rm R_{\rho, \rm loc} \gtrsim 1.14$, no ODDC instability occurs. For Pr01R4 we find
$\rm R_{\rho, \rm loc} \lesssim 10$ up to $\rm y \sim 0.68$ and $\rm R_{\rho, \rm loc} \lesssim 4$ up to $\rm y \sim 0.61$.
Indeed, this is where the small staircases show up in Fig.~\ref{fig:Case_2_Rho3_Ri_inf_S_tau_039}~(b) thus corroborating
layer formation being caused by the double-diffusive instability in that region.

Recalling Fig.~\ref{fig:Case_2_Rho3_Ri_inf_S_tau_039}~(c), particularly for the case Pr01R4, one might wonder why 
the interfaces between the layers are not prone to the double-diffusive instability as well, since this is where values 
of $\rm R_{\rho, \rm loc}$ just slightly larger than 1 are readily observed. But those interfaces are exposed to the strong 
mixing induced by convection which occurs on a much shorter time scale than the local thermal time scale governing 
the double-diffusive instability. Hence, already existing layers may grow and thereby merge or their 
interfaces may erode and thereby also lead to layer merging (e.g., \citealt{Spruit_2013}) instead of forming new, small 
staircases. Favorable conditions for the double-diffusive instability can only be found on the top of the stack of 
secondary layers.

\subsection{The giant planet case with shear}

\begin{figure}[htb]
\begin{center}
 \includegraphics[height=0.27\textwidth]{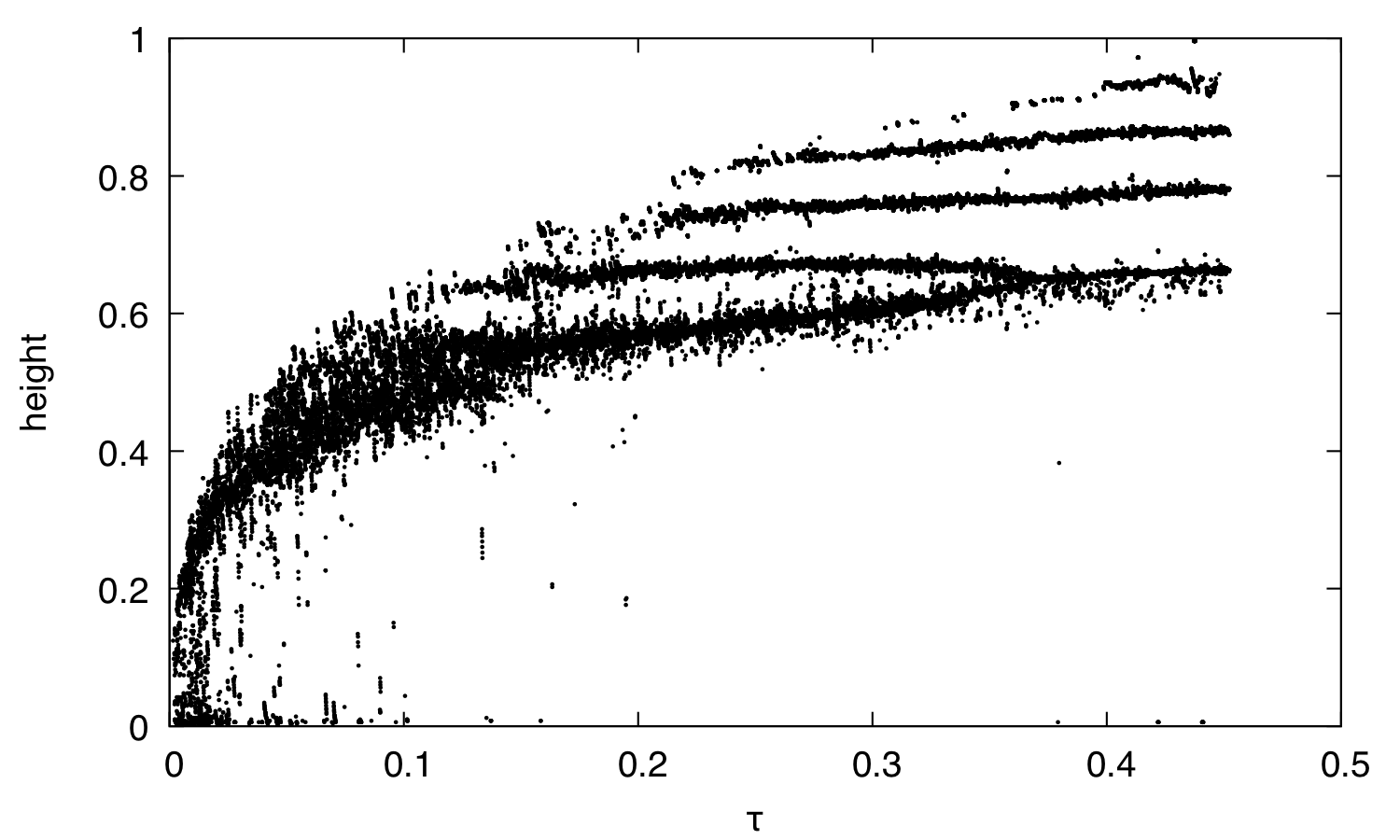}
 \includegraphics[height=0.27\textwidth]{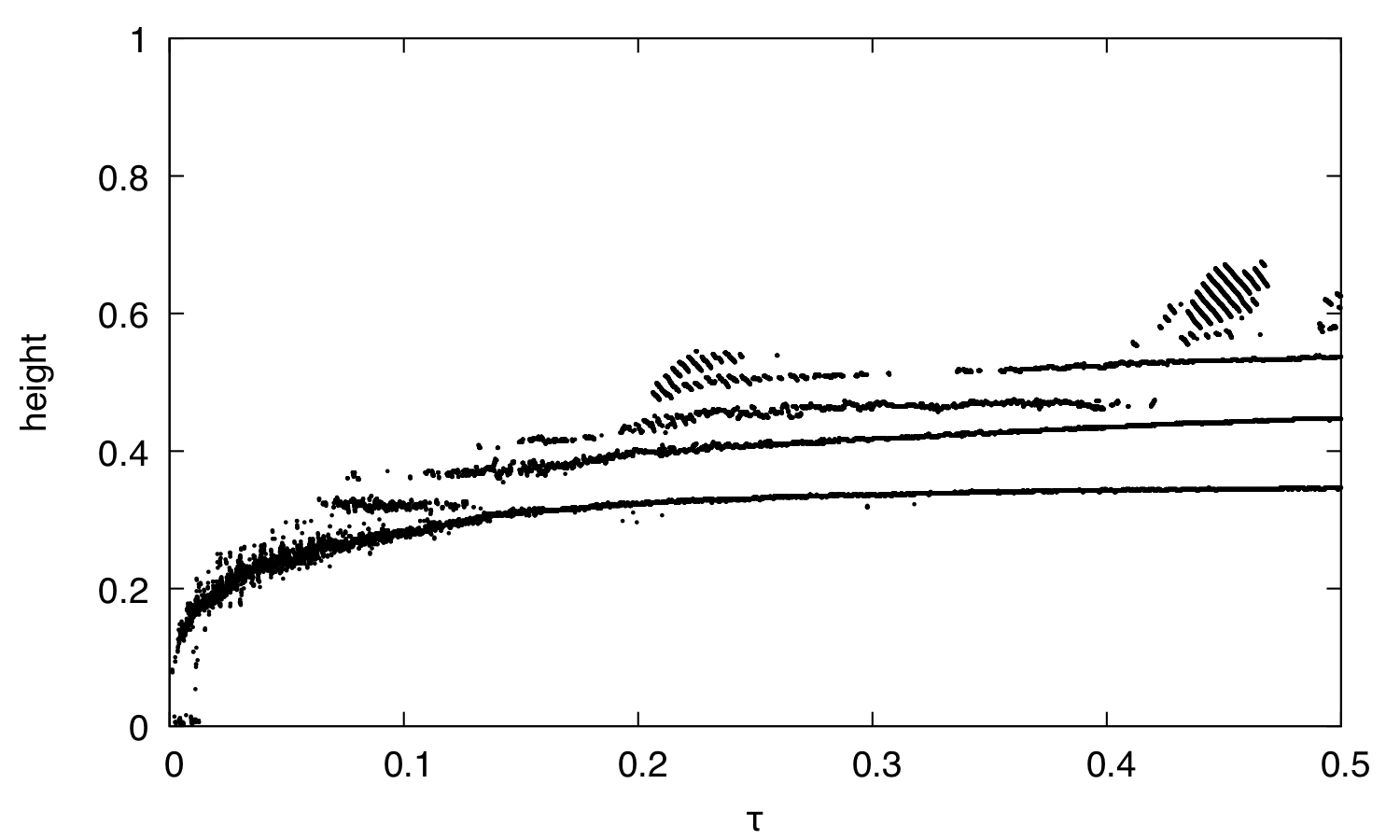}
\end{center}
\caption{Positions of interfaces for $\rm Pr=10^{-1}$ and $\rm Ri=10$ for $\rm R_{\rho}=2$ (left) and  $\rm R_{\rho}=4$ (right).}
\label{fig:Pr01Ri10-overview}
\end{figure}

\begin{figure}[htb]
\begin{center}
 \includegraphics[height=0.27\textwidth]{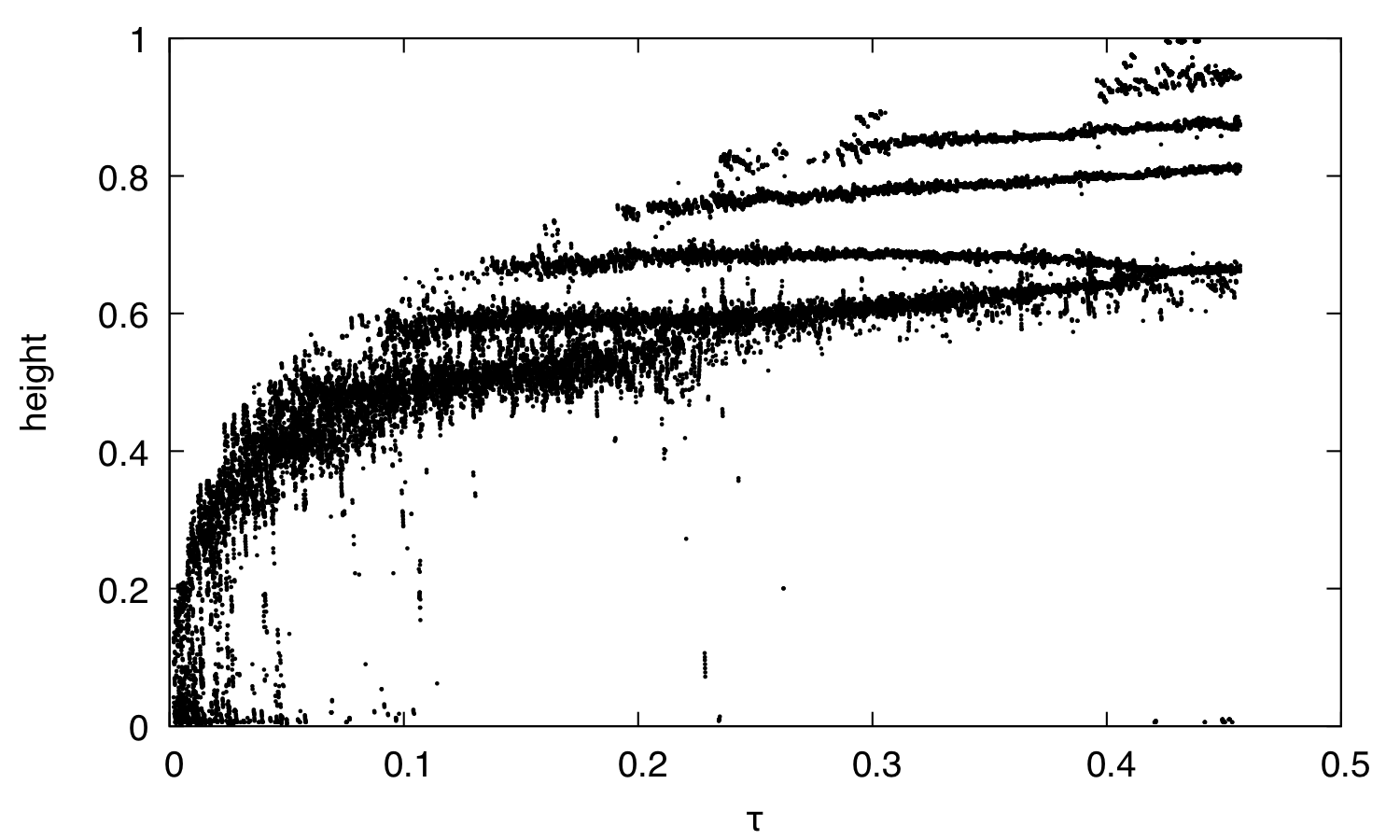}
 \includegraphics[height=0.27\textwidth]{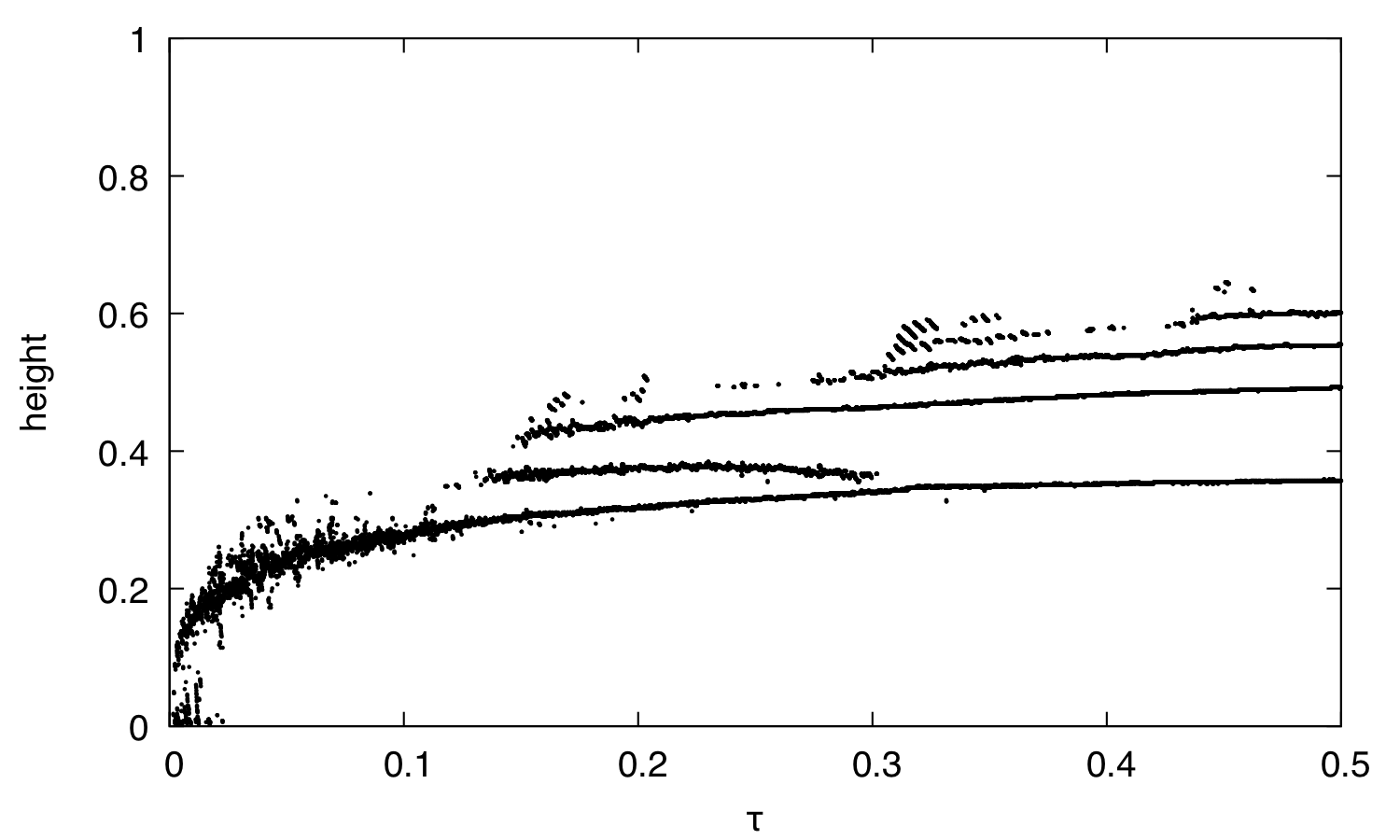}
\end{center}
\caption{Positions of interfaces for $\rm Pr=10^{-1}$ and $\rm Ri=1$ for $\rm R_{\rho}=2$ (left) and  $\rm R_{\rho}=4$ (right).}
\label{fig:Pr01Ri1-overview}
\end{figure}

\begin{figure}[htb]
\begin{center}
 \includegraphics[height=0.27\textwidth]{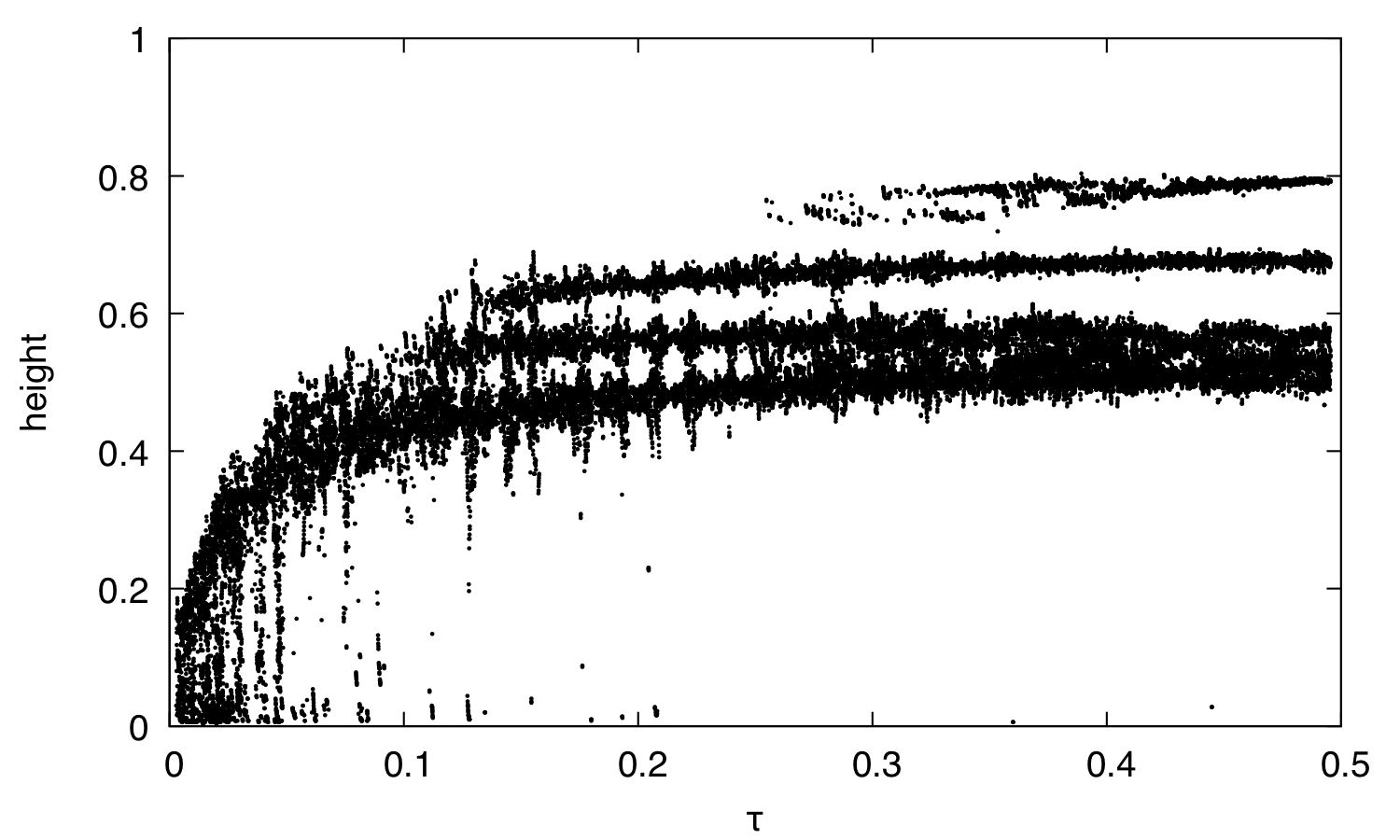}
 \includegraphics[height=0.27\textwidth]{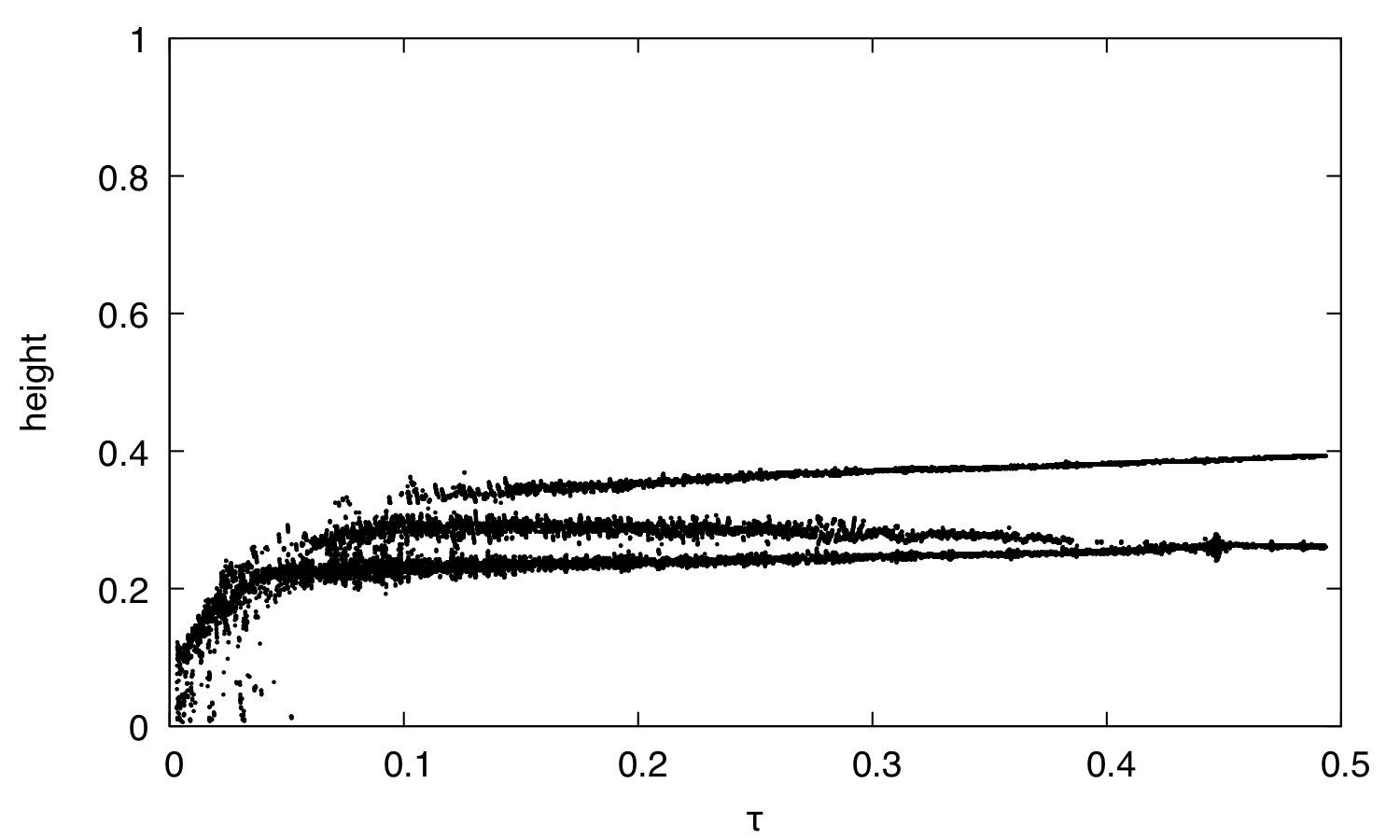}
\end{center}
\caption{Positions of interfaces for $\rm Pr=10^{-1}$ and $\rm Ri=10^{-1}$ for $\rm R_{\rho}=2$ (left) and  $\rm R_{\rho}=4$ (right).}
\label{fig:Pr01Ri01-overview}
\end{figure}

For the giant planet case six simulations have been performed with shear for values of ${\rm Ri=10}$, 
${\rm Ri=1}$, and ${\rm Ri=10^{-1}}$ for $\rrho=\{2,4\}$. Again, layer formation is found in all six cases and for
$\rm R_{\rho}=2$ we find significantly higher first layer heights than for $\rm R_{\rho}=4$ similar to the case of $\rm Pr=7$
discussed above. The secondary layer formation process acts on similar time scales as for ${\rm Ri=\infty}$,
namely $\tau \sim 0.1$. This is different from the oceanographic case where the formation of the first layer
was slowed down while the formation of secondary layers acted on the same time scales as in cases
without shear. Simultaneous secondary layer formation and the formation of unstable layers is found in all cases.
Merging processes between secondary layers are found in, e.g., Fig.~\ref{fig:Pr01Ri10-overview} (a) at $\tau=0.3$ and 
Fig.~\ref{fig:Pr01Ri1-overview} (a) at $\tau=0.4$  and Fig.~\ref{fig:Pr01Ri1-overview} (b) at $\tau=0.3$. 
This appears to be either more frequent or it happens at least faster for cases without shear.
If we compare scenarios where $\Pr=7$ with those where $\Pr=0.1$, each of them subject to shear,  
we notice that these merging events also occur for `true' secondary layers rather than just for the structures
formed on top of the `seed' layer during the first $\tau=0.05$.
The resulting layer height is the sum of the heights of its precursors and does not influence the stack of 
layers above. We also find layer formation driven by the ODDC instability at all values of $\rm Ri$
investigated here, most prominently in Fig.~\ref{fig:Pr01Ri10-overview}~(b) at $\tau=0.22$ and $\tau=0.43$, in
Fig.~\ref{fig:Pr01Ri1-overview}~(b) basically from $\tau=0.15$ till $\tau=0.45$, and in 
Fig.~\ref{fig:Pr01Ri01-overview}~(a) for $\tau \sim 0.28$. Some hints of ODDC induced layer formation may 
also be found in 
Fig.~\ref{fig:Pr01Ri10-overview}~(a) at $\tau=0.3$ and in Fig.~\ref{fig:Pr01Ri1-overview}~(a) from
$\tau=0.3$ till $\tau=0.4$. The ODDC layer formation on top of the stack, largely independent of $\rm Ri$,
is the second important difference to the case of $\rm Pr=7$.

\begin{figure}[htb]
\begin{center}
a) \includegraphics[height=0.6\textwidth]{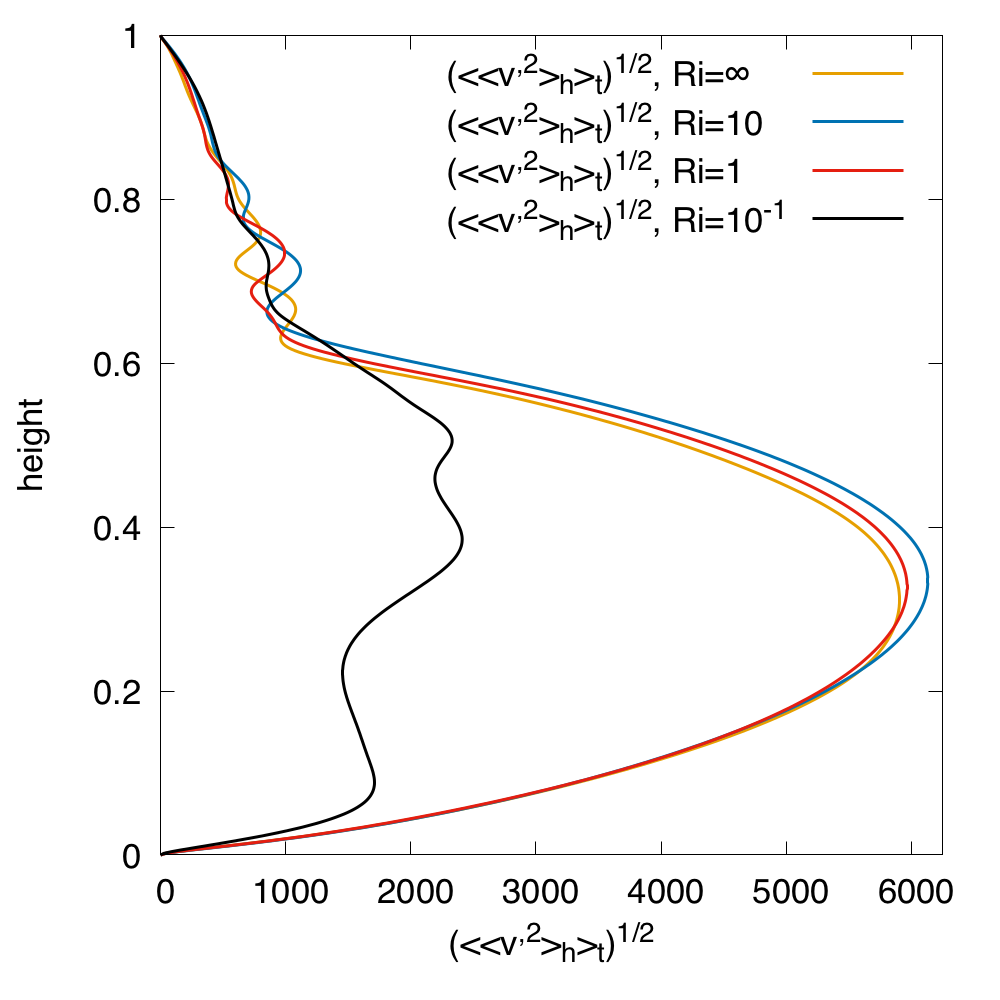} 

b) \includegraphics[height=0.6\textwidth]{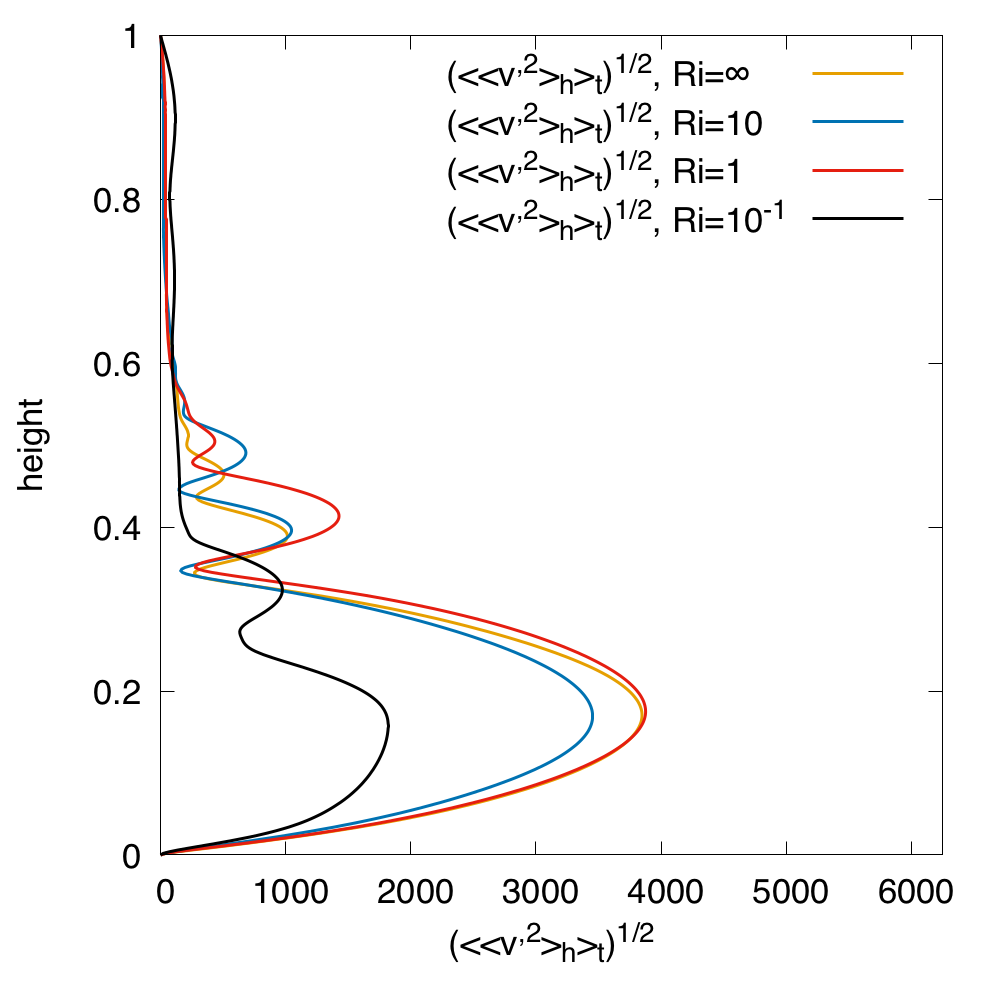}
\end{center}
\caption{Fluctuations of the vertical velocity component v averaged over time 
$\rm \sqrt{\langle\langle(v')^2\rangle_h\,\rangle_{t}}$ with $\rm v'=v-\langle v\rangle_h$ for (a) $\rm R_{\rho}=2$ 
and (b)  $\rm R_{\rho}=4$ for the giant planet case.}
\label{fig:v_Pr01Rrho}
\end{figure}

\begin{figure}[htb]
\begin{center}
a) \includegraphics[height=0.6\textwidth]{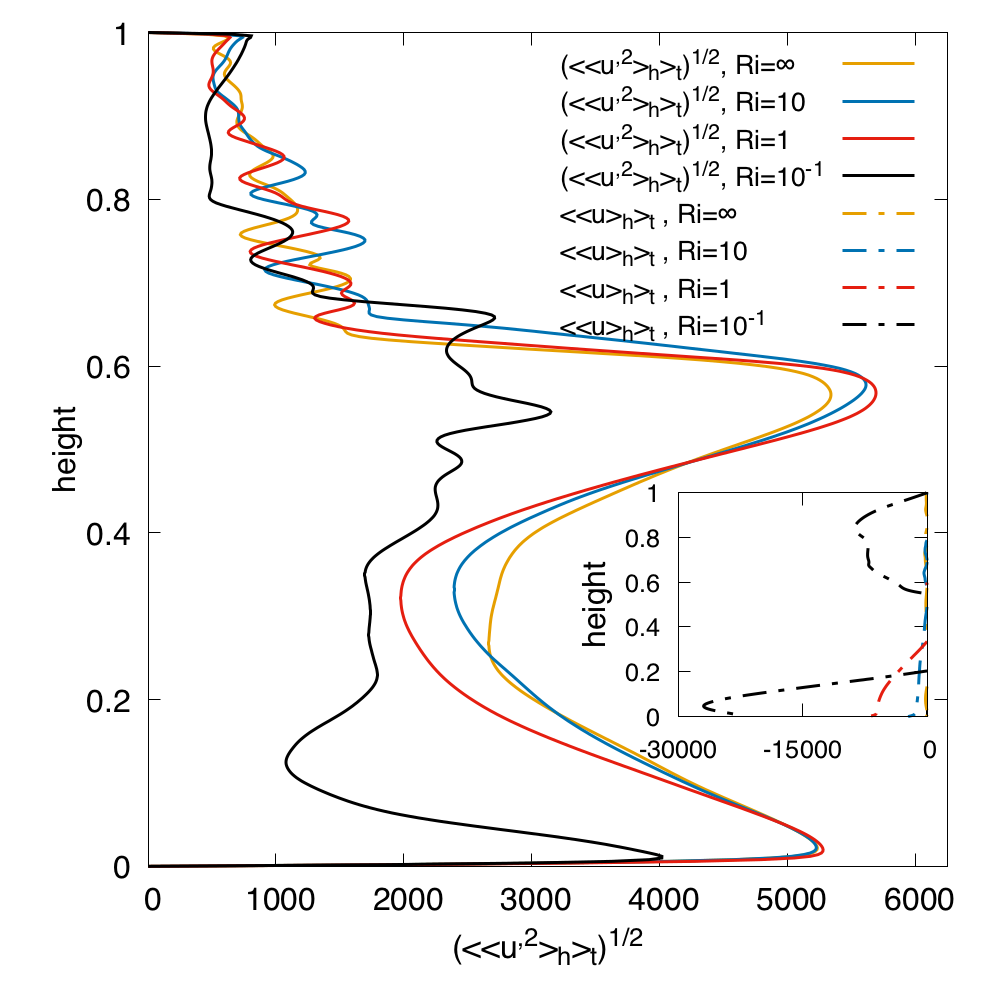} 

b) \includegraphics[height=0.6\textwidth]{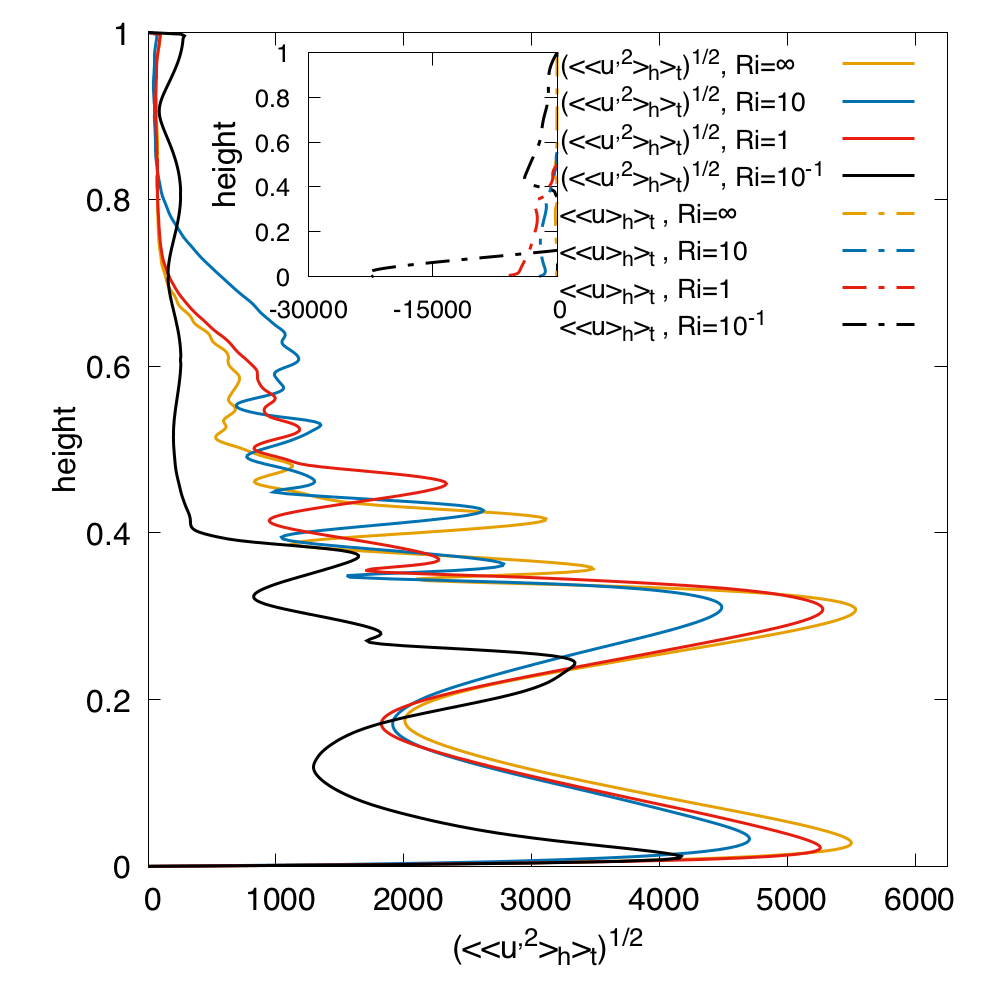}
\end{center}
\caption{Fluctuations of the horizontal velocity component u averaged over time 
$\rm \sqrt{\langle\langle(u')^2\rangle_h\,\rangle_{t}}$ with $\rm u'=u-\langle u\rangle_h$ for (a) $\rm R_{\rho}=2$ 
and (b)  $\rm R_{\rho}=4$ for the giant planet case.
Insets: mean horizontal velocity component averaged over time $\rm \langle{\langle u\rangle}_h\,\rangle_{t}$.}
\label{fig:u_Pr01Rrho}
\end{figure}

In Fig.~\ref{fig:v_Pr01Rrho} and Fig.~\ref{fig:u_Pr01Rrho} we compare the ensemble averaged mean and root mean square 
vertical and horizontal velocities, respectively, for the three different shear rates and the corresponding reference case
without shear. As for the case with $\rm Pr=7$ (see Fig.~\ref{fig:v_Pr7Rrho} and Fig.~\ref{fig:u_Pr7Rrho}) the mean vertical
velocity is below the diffusion velocity for all shear rates and for cases without shear, as to be expected for (kinetically)
relaxed simulations and a spatial discretization respecting conservation of transported quantities. The root mean square
vertical velocities for intermediate, moderate, and no shear are larger in the primary layer for the lower Prandtl number
of $\rm Pr=0.1$ than for  $\rm Pr=7$, as expected for a less viscous fluid and slightly larger $\rm Ra^{*}$. 
Likewise, convective velocities are considerably slowed down particularly in the first layer for the case of a high shear rate. 
An important difference though are the higher vertical velocities in the diffusive region at the top of the stack 
for $\rrho=2$: this is not only the case for the highest shear rate, but also for lower or even no mean 
shear. For enhanced stratification ($\rrho=3$ for $\rm Pr=7$ and $\rrho=4$ for $\rm Pr=0.1$) the case with 
$\rm Pr=0.1$ shows vertical root mean square velocities notably higher than the diffusion velocity only for $\rm Ri=0.1$
whereas they are negligible for lower shear. In comparison, for $\rm Pr=7$ this region is characterized essentially by 
pure diffusion. This holds for all shear rates as well as for no shear. Comparing Fig.~\ref{fig:v_Pr7Rrho} with 
Fig.~\ref{fig:v_Pr01Rrho} it is evident that the interfaces between layers are characterized by low 
vertical velocities. One reason that they are not even smaller stems from the interfaces not being static entities but 
features which move upwards and downwards as well, at least within a limited region. For large shear the observed
vertical root mean square velocities, particularly for $\rrho=2$, are clearly a consequence of the dominant shear 
generating horizontal flow and eventually also vertical motions. 

Comparing the mean horizontal flow depicted in Fig.~\ref{fig:u_Pr7Rrho} and in Fig.~\ref{fig:u_Pr01Rrho} we note that
the case of $\rm Ri=0.1$ appears to be not yet fully relaxed for $\rrho=2$. Otherwise we observe, as expected, essentially
no mean motion for the case without shear and mean flow increasing towards the bottom, as expected from the lower
boundary condition introducing the shear in first place. For horizontal root mean square velocities we can notice a prominent 
minimum in the first layer for all simulations. 

This is due to convective rolls forming in the layer which necessarily leads to higher horizontal convective motions 
near the interface and the lower boundary compared to the middle of the layer. Considering the cases with $\rm Pr=7$ 
and $\rrho=3$ and those with $\rm Pr=0.1$ and $\rrho=4$ we note that only for $\rm Ri=0.1$ we find horizontal root
mean square velocity fluctuations above the diffusion velocity in the upper, diffusive region.
Thus, in the diffusive region the velocity field generated by the mean shear is small. Together with the still large
values of $\rm R_{\rho, \rm loc}$ in that region this might explain why we do not find ODDC layer formation for $\rm Pr=7$ 
and $\rrho=3$ for the case of non-zero shear. This is contrary to what is expected from the work of \cite{Radko_2016} 
who found this to be the case for $\rm Pr=10$ and a linear stratification in S and T as well as a sinusoidal shear  
that is compatible with the horizontally periodic boundary condition assumed in his work.

From the scenarios studied here we cannot confirm that shear enhances ODDC layer formation as discussed
in his work. However, this might well originate from the stratifications in our simulations which have not been
set up to detect the mechanism discussed in \cite{Radko_2016}.

\section{Discussion and concluding remarks} \label{sec:discussion}

Direct numerical simulations of double diffusive layer formation are still a challenging task.
Resolving all relevant spatial and temporal scales is complicated by low values of molecular 
diffusivities and the fact that layer formation evolves on the thermal time scale. 
We present the first study of bottom heated double diffusive layer formation with and without shear
for the oceanographically relevant case of $\rm Pr=7$ and the astrophysically relevant (giant planet)  
case of $\rm Pr=10^{-1}$. The dependency of layer formation on  $\rm Pr$, $\rrho$, and $\rm Ri$ is 
studied. For this purpose two numerical codes are used, the ANTARES code suite and the open source 
software OpenFoam. The latter code is used to validate the Boussinesq approximation based model on 
Turner's water tank experiment (\cite{Turner_1964}) and analytical models thereof (e.g., \citealt{Fernando_1987}).

Our simulations with ANTARES reveal the successive formation of layers stable throughout a large
fraction of the thermal time scale. Their stability is determined by the conditions within the interfaces
and characterized by $\rm R_{\rho, \rm loc}$ and layers form despite $\rrho > \rm (1+Pr)/(Pr+Le)$. This also
corroborates the theory and numerical modelling of layer formation in volcanic rift lakes 
(\citealt{Carpenter_2012a,Carpenter_2012b} and \citealt{Sommer_2014}).
We also confirm classical results from local ODDC stability analysis as
reviewed in Sect.~\ref{sec:introduction}.

Three different types of layers have appeared in our numerical simulations. 
The initial `seed' or first layer results from a sudden heating source being activated, 
hence the stratification above the source (the bottom plate) has a stratification unstable
in the sense of the Ledoux criterion. Convection sets in, but cannot
spread to the top of the simulation box due to the counteracting solute
gradient. Instead a layer is formed with an initially rather unstable
interface. This is exactly the laboratory setup of \cite{Turner_1964},
but is much less likely to occur in nature. We might think of a sudden
heating due to volcanic activity underneath a rift lake or a helium
core or shell flash in an evolved star. But that stellar scenario is
related to a dynamically highly unstable case and the present, idealized
calculation probably has only very limited implications on such processes, if any.

Interestingly, secondary layer formation is found for all cases we have studied.
The secondary layers form as a result of heat transport from the bottom which
increases the gradients in the interfaces in the stack so that eventually the top 
interface becomes Ledoux unstable and a new layer forms which stabilizes the 
underlying layers. This is a scenario which can well be expected to occur in 
oceanographic cases, in giant planets, or in stars (for the latter in regions on
top of nuclear burning zones), since only an existing
convective layer with an interface that can become unstable in the sense of the 
Ledoux criterion is required to trigger the formation of further layers.
Thereby, we found a connection between the new layers 
on top with the bottom region which occurs on the convective time scale. This interaction 
is highlighted in the thermal flux of the first layer, where the formation of each additional 
secondary layer leaves a unique footprint.
The secondary layers form even in ODDC stable stratifications, i.e.,  
if $\rrho \geqslant 2$ and $\rm Pr=7$. 
Additionally, recent results by \cite{Radko_2016} on shear furthering layer formation in
stratifications stable to ODDC (i.e., for the oceanographic case) could not be confirmed. 
Probably, this regime was not covered by the regions on top of the layer stacks in the 
simulations presented here.
On the other hand, we have found the third type of layers, those originating
from the ODDC instability, essentially for all cases where $\rm Pr=0.1$, located always
on top of the pre-existing stack of secondary layers, for zero mean shear and for all 
non-zero shear rates we have considered (its presence cannot be confirmed for the case
of $\rrho = 4$ and $\rm Ri = 0.1$). Interestingly, the ODDC layers are not perturbed even 
by very large shear.

For the giant planet case of a small Prandtl number the layer 
formation due to the oscillatory double-diffusive convective instability occurs 
only once the conditions on top of the stack, described by the local stability
parameter $\rm R_{\rho, \rm loc}$, become favorable for ODDC
to set in and the time scale for destabilizing the top interface through heat 
transfer from the interfaces underneath has become longer than the local 
ODDC evolution time. Thus, we have found those types of layers only once 
typically about three layers have been formed by the two other mechanisms
described further above.

Shear causes that the formation of the initial, convective
layer is slowed down. However, this happens only on fractions of the thermal time scale. 
Furthermore, shear dilutes the interface boundaries, which are less turbulent 
and hence more pronounced without shear.

We conclude that to understand convective layer formation under
conditions found in the ocean or also in giant planets (and in principle also in
various types of stars, see Sect.~\ref{sec:introduction}) it is important
to study the consequences of a whole range of possible initial stratifications.
A linear initial stratification in temperature and solute appears only realistic 
if the system could evolve and relax unperturbed for more than 
the global thermal time scale.
Hence, layer formation in astro- and geophysical systems might appear 
as a combination of heating induced and ODDC induced processes depending on 
the history and initial stratification of the system
as well as on the competition between global thermal evolution, evolution
of the layers themselves, or external forcings induced on the system, such as seasonal
variations or the like. This may determine later evolutionary stages, since one
might end up with a large mixed bottom layer more quickly than if the ODDC
instability were the only mechanism of convective layer formation.
Shear does not change this process in a fundamental way, but accurate predictive models
are more complex than concluded from models developed for cases without shear and
stratifications which are just linear in S and T.

We hence suggest that future research in this field should also
include the study of a wider scenario of initial conditions and performing flux 
measurements and mixing time calculations which are based on numerical simulations
similar to those we have done here and comparisons of those with semi-analytical
models that are used for global models required in those research fields.

\section*{Acknowledgements}

F.~Kupka gratefully acknowledges financial support through Austrian
Science Fund (FWF) projects P~25229-N27 and P~29172-N27. The numerical
simulations have been performed on the Vienna Scientific Cluster VSC
(project 70708), resources dedicated to the Faculty of Mathematics at VSC-3. 
We thank M.H.~Montgomery for providing us with computational 
resources at the TACC Stampede2 cluster (University of Texas, Austin).
\newline \\
Compliance with ethical standards: The authors declare
 that they have no conflict of interest.

\bibliographystyle{spbasic} 
\bibliography{bib_layer}

\begin{thebibliography}{63}
\providecommand{\natexlab}[1]{#1}
\providecommand{\url}[1]{{#1}}
\providecommand{\urlprefix}{URL }
\expandafter\ifx\csname urlstyle\endcsname\relax
  \providecommand{\doi}[1]{DOI~\discretionary{}{}{}#1}\else
  \providecommand{\doi}{DOI~\discretionary{}{}{}\begingroup
  \urlstyle{rm}\Url}\fi
\providecommand{\eprint}[2][]{\url{#2}}

\bibitem[{Armitage and House(1962)}]{Armitage_1962}
Armitage KB, House HB (1962) A limnological reconnaissance in the area of
  me-murdo sound, antarctica. Limnol Oceanog 7:36--41

\bibitem[{Baines and Gill(1969)}]{Baines_1969}
Baines PG, Gill AE (1969) On thermohaline convection with linear gradients.
  Journal of Fluid Mechanics 37(2):289--306

\bibitem[{{Bascoul}(2007)}]{Bascoul_2007}
{Bascoul} GP (2007) {Numerical simulations of semiconvection}. In: {Kupka} F,
  {Roxburgh} I, {Chan} KL (eds) Convection in Astrophysics, IAU Symposium, vol
  239, pp 317--319, \doi{10.1017/S1743921307000658}

\bibitem[{Batchelor(2000)}]{Batchelor_2000}
Batchelor GK (2000) An {I}ntroduction to {F}luid {D}ynamics. Cambridge
  Mathematical Library, Cambridge University Press, Cambridge

\bibitem[{Beckermann et~al(1991)Beckermann, Fan, and
  Mihailovic}]{Beckermann_1991}
Beckermann C, Fan C, Mihailovic J (1991) Numerical simulations of
  double-diffusive convection in a hele-shaw cell. International Video Journal
  of Engineering Research 1:71--82

\bibitem[{{Biello}(2001)}]{Biello_2001}
{Biello} JA (2001) {Layer formation in semiconvection}. PhD thesis, THE
  UNIVERSITY OF CHICAGO

\bibitem[{Canuto(1999)}]{Canuto_1999}
Canuto VM (1999) Turbulence in stars. {III}. unified treatment of diffusion,
  convection, semiconvection, salt fingers, and differential rotation.
  Astrophys J 524:311--340

\bibitem[{{Canuto}(2011{\natexlab{a}})}]{Canuto_2011a}
{Canuto} VM (2011{\natexlab{a}}) {Stellar mixing. II. Double diffusion
  processes}. Astronomy and Astrophysics 528:A77,
  \doi{10.1051/0004-6361/201014448}

\bibitem[{{Canuto}(2011{\natexlab{b}})}]{Canuto_2011b}
{Canuto} VM (2011{\natexlab{b}}) {Stellar mixing. III. The case of a passive
  tracer}. Astronomy and Astrophysics 528:A78,
  \doi{10.1051/0004-6361/201015372},
  \urlprefix\url{https://doi.org/10.1051/0004-6361/201015372}

\bibitem[{Carpenter et~al(2012{\natexlab{a}})Carpenter, Sommer, and
  W{\"u}est}]{Carpenter_2012a}
Carpenter JR, Sommer T, W{\"u}est A (2012{\natexlab{a}}) Simulations of a
  double-diffusive interface in the diffusive convection regime. Journal of
  Fluid Mechanics 711:411--436, \doi{10.1017/jfm.2012.399},
  \urlprefix\url{http://dx.doi.org/10.1017/jfm.2012.399}

\bibitem[{Carpenter et~al(2012{\natexlab{b}})Carpenter, Sommer, and
  W{\"u}est}]{Carpenter_2012b}
Carpenter JR, Sommer T, W{\"u}est A (2012{\natexlab{b}}) Stability of a
  double-diffusive interface in the diffusive convection regime. Journal of
  Physical Oceanography 42(5):840--854, \doi{10.1175/jpo-d-11-0118.1},
  \urlprefix\url{http://dx.doi.org/10.1175/JPO-D-11-0118.1}

\bibitem[{Chabrier and Baraffe(2007)}]{Chabrier_2007}
Chabrier G, Baraffe I (2007) Heat transport in giant (exo)planets: A new
  perspective. The Astrophysical Journal Letters 661(1):L81,
  \urlprefix\url{http://stacks.iop.org/1538-4357/661/i=1/a=L81}

\bibitem[{Descy et~al(2012)Descy, Darchambeau, and Schmid}]{Descy_2012}
Descy JP, Darchambeau F, Schmid M (2012) Lake Kivu: Limnology and
  biogeochemistry of a tropical great lake. Aquatic Ecology Series, Springer

\bibitem[{Ding and Li(2014)}]{Ding_2014}
Ding CY, Li Y (2014) Properties of semi-convection and convective overshooting
  for massive stars. Monthly Notices of the Royal Astronomical Society
  438(2):1137--1148, \doi{10.1093/mnras/stt2262}

\bibitem[{Fernando(1987)}]{Fernando_1987}
Fernando HJS (1987) The formation of a layered structure when a stable salinity
  gradient is heated from below. Journal of Fluid Mechanics 182:525--541

\bibitem[{Fernando(1989)}]{Fernando_1989}
Fernando HJS (1989) Buoyancy transfer across a diffusive interface. Journal of
  Fluid Mechanics 209:1--34

\bibitem[{{Flanagan} et~al(2013){Flanagan}, {Lefler}, and
  {Radko}}]{Flanagan_2013}
{Flanagan} JD, {Lefler} AS, {Radko} T (2013) {Heat transport through diffusive
  interfaces}. Geophys Res Lett 40:2466--2470, \doi{10.1002/grl.50440}

\bibitem[{Garaud(2018)}]{Garaud_2018}
Garaud P (2018) Double-diffusive convection at low prandtl number. Annual
  Review of Fluid Mechanics 50(1):275--298,
  \doi{10.1146/annurev-fluid-122316-045234},
  \urlprefix\url{https://doi.org/10.1146/annurev-fluid-122316-045234}

\bibitem[{{Grossman} and {Taam}(1996)}]{Grossman_1996}
{Grossman} SA, {Taam} RE (1996) {Double-Diffusive Mixing-Length Theory,
  Semiconvection and Massive Star Evolution}. Monthly Notices of the Royal
  Astronomical Society 283:1165--1178, \doi{10.1093/mnras/283.4.1165}

\bibitem[{Happenhofer et~al(2013)Happenhofer, Grimm-Strele, Kupka,
  L{\"o}w-Baselli, and Muthsam}]{Happenhofer_2013}
Happenhofer N, Grimm-Strele H, Kupka F, L{\"o}w-Baselli B, Muthsam H (2013) A
  low mach number solver: Enhancing applicability. Journal of Computational
  Physics 236:96 -- 118, \doi{https://doi.org/10.1016/j.jcp.2012.11.002},
  \urlprefix\url{http://www.sciencedirect.com/science/article/pii/
  S0021999112006572}

\bibitem[{Huppert and Moore(1976)}]{Huppert_1976}
Huppert H, Moore DR (1976) Nonlinear double-diffusive convection. Journal of
  Fluid Mechanics 78(4):821--854

\bibitem[{{Huppert} and {Linden}(1979)}]{Huppert_1979}
{Huppert} HE, {Linden} PF (1979) {On heating a stable salinity gradient from
  below}. Journal of Fluid Mechanics 95:431--464,
  \doi{10.1017/S0022112079001543}

\bibitem[{{Huppert} and {Turner}(1981)}]{Huppert_1981}
{Huppert} HE, {Turner} JS (1981) {Double-diffusive convection}. Journal of
  Fluid Mechanics 106:299--329, \doi{10.1017/S0022112081001614}

\bibitem[{{Kato}(1966)}]{Kato_1966}
{Kato} S (1966) {Overstable Convection in a Medium Stratified in Mean Molecular
  Weight}. Publ Astron Soc Japan 18:374

\bibitem[{{Kupka} and {Muthsam}(2017)}]{Kupka_2017}
{Kupka} F, {Muthsam} HJ (2017) {Modelling of stellar convection}. Living
  Reviews in Computational Astrophysics 3:1, \doi{10.1007/s41115-017-0001-9}

\bibitem[{Kupka et~al(2015)Kupka, Losch, Zaussinger, and Zweigle}]{Kupka_2015}
Kupka F, Losch M, Zaussinger F, Zweigle T (2015) Semi-convection in the ocean
  and in stars: A multi-scale analysis. Meteorologische {Z}eitschrift
  24(3):343--358, \doi{10.1127/metz/2015/0643}

\bibitem[{{Langer} et~al(1985){Langer}, {El Eid}, and {Fricke}}]{Langer_1985}
{Langer} N, {El Eid} MF, {Fricke} KJ (1985) {Evolution of massive stars with
  semiconvective diffusion}. Astronomy and Astrophysics 145:179--191

\bibitem[{{Leconte} and {Chabrier}(2012)}]{Leconte_2012}
{Leconte} J, {Chabrier} G (2012) A new vision of giant planet interiors: Impact
  of double diffusive convection. Astronomy and Astrophysics 540:A20,
  \doi{10.1051/0004-6361/201117595},
  \urlprefix\url{https://doi.org/10.1051/0004-6361/201117595}

\bibitem[{Leconte and Chabrier(2013)}]{Leconte_2013}
Leconte J, Chabrier G (2013) Layered convection as the origin of saturn/'s
  luminosity anomaly. Nature Geosci 6(5):347--350,
  \urlprefix\url{http://dx.doi.org/10.1038/ngeo1791}

\bibitem[{{Ledoux}(1947)}]{Ledoux_1947}
{Ledoux} P (1947) Stellar models with convection and with discontinuity of the
  mean molecular weight. Astrophys J 105:305--321, \doi{10.1086/144905}

\bibitem[{Lesieur(2008)}]{Lesieur_2008}
Lesieur M (2008) Turbulence in Fluids. Fluid Mechanics and Its Applications,
  Springer Netherlands,
  \urlprefix\url{https://books.google.de/books?id=xKUDN22Y7OYC}

\bibitem[{{Maeder} et~al(2013){Maeder}, {Meynet}, {Lagarde}, and
  {Charbonnel}}]{Maeder_2013}
{Maeder} A, {Meynet} G, {Lagarde} N, {Charbonnel} C (2013) The thermohaline,
  richardson, rayleigh-taylor, solberg-h\o{}iland, and gsf criteria in rotating
  stars. A\&A 553:A1, \doi{10.1051/0004-6361/201220936},
  \urlprefix\url{https://doi.org/10.1051/0004-6361/201220936}

\bibitem[{{Merryfield}(1995)}]{Merryfield_1995}
{Merryfield} WJ (1995) {Hydrodynamics of semiconvection}. Astrophysical Journal
  444:318--337, \doi{10.1086/175607}

\bibitem[{{Mirouh} et~al(2012){Mirouh}, {Garaud}, {Stellmach}, {Traxler}, and
  {Wood}}]{Mirouh_2012}
{Mirouh} GM, {Garaud} P, {Stellmach} S, {Traxler} AL, {Wood} TS (2012) {A New
  Model for Mixing by Double-diffusive Convection (Semi-convection). I. The
  Conditions for Layer Formation}. Astrophysical Journal 750:61,
  \doi{10.1088/0004-637X/750/1/61}

\bibitem[{Moore and Garaud(2016)}]{Moore_2016}
Moore K, Garaud P (2016) Main sequence evolution with layered semiconvection.
  Astrophysical Journal 817:54 (12pp.)

\bibitem[{Mutabazi et~al(2016)Mutabazi, Yoshikawa, Fogaing, Travnikov,
  Crumeyrolle, Futterer, and Egbers}]{Mutabazi_2016}
Mutabazi I, Yoshikawa HN, Fogaing MT, Travnikov V, Crumeyrolle O, Futterer B,
  Egbers C (2016) Thermo-electro-hydrodynamic convection under microgravity: a
  review. Fluid Dynamics Research 48(6):061413

\bibitem[{Muthsam et~al(2010)Muthsam, Kupka, L{\"o}w-Baselli, Obertscheider,
  Langer, and Lenz}]{Muthsam_2010}
Muthsam H, Kupka F, L{\"o}w-Baselli B, Obertscheider C, Langer M, Lenz P (2010)
  Antares -- a numerical tool for astrophysical research with applications to
  solar granulation. New Astronomy 15(5):460--475,
  \doi{10.1016/j.newast.2009.12.005},
  \urlprefix\url{http://dx.doi.org/10.1016/j.newast.2009.12.005}

\bibitem[{{Radko}(2003)}]{Radko_2003}
{Radko} T (2003) {A mechanism for layer formation in a double-diffusive fluid}.
  Journal of Fluid Mechanics 497:365--380, \doi{10.1017/S0022112003006785}

\bibitem[{{Radko}(2010)}]{Radko_2010}
{Radko} T (2010) {Equilibration of weakly nonlinear salt fingers}. Journal of
  Fluid Mechanics 645:121, \doi{10.1017/S0022112009992552}

\bibitem[{{Radko}(2013)}]{Radko_2013}
{Radko} T (2013) Double-Diffusive Convection. Cambridge University Press

\bibitem[{{Radko}(2016)}]{Radko_2016}
{Radko} T (2016) Thermohaline layering in dynamically and diffusively stable
  shear flows. Journal of Fluid Mechanics 805:147--170,
  \doi{10.1017/jfm.2016.547},
  \urlprefix\url{http://dx.doi.org/10.1017/jfm.2016.547}

\bibitem[{{Rosenblum} et~al(2011){Rosenblum}, {Garaud}, {Traxler}, and
  {Stellmach}}]{Rosenblum_2011}
{Rosenblum} E, {Garaud} P, {Traxler} A, {Stellmach} S (2011) {Turbulent Mixing
  and Layer Formation in Double-diffusive Convection: Three-dimensional
  Numerical Simulations and Theory}. Astrophysical Journal 731:66,
  \doi{10.1088/0004-637X/731/1/66}

\bibitem[{{Silva Aguirre} et~al(2011){Silva Aguirre}, {Ballot}, {Serenelli},
  and {Weiss}}]{Aguirre_2011}
{Silva Aguirre} V, {Ballot} J, {Serenelli} AM, {Weiss} A (2011) Constraining
  mixing processes in stellar cores using asteroseismology - impact of
  semiconvection in low-mass stars. Astronomy and Astrophysics 529:A63,
  \doi{10.1051/0004-6361/201015847},
  \urlprefix\url{https://doi.org/10.1051/0004-6361/201015847}

\bibitem[{{Sommer} et~al(2014){Sommer}, {Carpenter}, and
  {W{\"u}est}}]{Sommer_2014}
{Sommer} T, {Carpenter} JR, {W{\"u}est} A (2014) {Double-diffusive interfaces
  in Lake Kivu reproduced by direct numerical simulations}. Geophys Res Lett
  41:5114--5121, \doi{10.1002/2014GL060716}

\bibitem[{{Spiegel}(1969)}]{Spiegel_1969}
{Spiegel} EA (1969) {Semiconvection}. Comments on Astrophysics and Space
  Physics 1:57

\bibitem[{Spruit(1992)}]{Spruit_1992}
Spruit H (1992) The rate of mixing in semiconvective zones. Astronomy and
  Astrophysics 253:131--138

\bibitem[{{Spruit}(2013)}]{Spruit_2013}
{Spruit} HC (2013) {Semiconvection: theory}. Astronomy and Astrophysics
  552:A76, \doi{10.1051/0004-6361/201220575}

\bibitem[{Stern(1960)}]{Stern_1960}
Stern ME (1960) The ``salt-fountain''and thermohaline convection. Tellus
  12(2):172--175

\bibitem[{Stevenson(1985)}]{Stevenson_1985}
Stevenson DJ (1985) Cosmochemistry and structure of the giant planets and their
  satellites. Icarus 62(1):4 -- 15,
  \doi{http://dx.doi.org/10.1016/0019-1035(85)90168-X},
  \urlprefix\url{http://www.sciencedirect.com/science/article/pii/
  001910358590168X}

\bibitem[{Suarez et~al(2010)Suarez, Childress, and Tyler}]{Suarez_2010}
Suarez F, Childress AE, Tyler SW (2010) Temperature evolution of an
  experimental salt-gradient solar pond. Journal of Water and Climate Change
  1(4):246--250, \doi{10.2166/wcc.2010.101}

\bibitem[{{Tayler}(1956)}]{Tayler_1956}
{Tayler} RJ (1956) {The evolution of unmixed stars}. Monthly Notices of the
  Royal Astronomical Society 116:25, \doi{10.1093/mnras/116.1.25}

\bibitem[{Turner(1968)}]{Turner_1968}
Turner JS (1968) The behaviour of a stable salinity gradient heated from below.
  Journal of Fluid Mechanics 33:183--200

\bibitem[{Turner and Stommel(1964)}]{Turner_1964}
Turner JS, Stommel H (1964) A new case of convection in the presence of
  combined vertical salinity and temperature gradients. Proceedings of the
  National Academy of Sciences of the United States of America 52(1):49--53,
  \urlprefix\url{https://doi.org/10.1073/pnas.52.1.49}

\bibitem[{Veronis(1965)}]{Veronis_1965}
Veronis G (1965) On finite amplitude instability in thermohaline convection. J
  Mar Res 23:1--17

\bibitem[{{Walin}(1964)}]{Walin_1964}
{Walin} G (1964) {Note on the stability of water stratified by both salt and
  heat}. Tellus 16:389

\bibitem[{{Weller} et~al(1998){Weller}, {Tabor}, {Jasak}, and
  {Fureby}}]{Weller_1998}
{Weller} HG, {Tabor} G, {Jasak} H, {Fureby} C (1998) {A tensorial approach to
  computational continuum mechanics using object-oriented techniques}.
  Computers in Physics 12:620--631, \doi{10.1063/1.168744}

\bibitem[{{Wood} et~al(2013){Wood}, {Garaud}, and {Stellmach}}]{Wood_2013}
{Wood} TS, {Garaud} P, {Stellmach} S (2013) {A New Model for Mixing by
  Double-diffusive Convection (Semi-convection). II. The Transport of Heat and
  Composition through Layers}. Astrophysical Journal 768:157,
  \doi{10.1088/0004-637X/768/2/157}

\bibitem[{{Xiong}(1986)}]{Xiong_1986}
{Xiong} DR (1986) {The evolution of massive stars using a non-local theory of
  convection}. Astronomy and Astrophysics 167:239--246

\bibitem[{Young and Rosner(2000)}]{Young_2000}
Young Y, Rosner R (2000) Numerical simulation of double-diffusive convection in
  a rectangular box. Phys Rev E 61:2676--2694, \doi{10.1103/PhysRevE.61.2676},
  \urlprefix\url{http://link.aps.org/doi/10.1103/PhysRevE.61.2676}

\bibitem[{Zaussinger(2011)}]{Zaussinger_2011}
Zaussinger F (2011) Numerical simulation of double-diffusive convection. PhD
  thesis, {Univ. Vienna}, \urlprefix\url{https://othes.univie.ac.at/13172/}

\bibitem[{Zaussinger and Spruit(2013)}]{Zaussinger_2013a}
Zaussinger F, Spruit HC (2013) Semiconvection: numerical simulations. Astronomy
  and {A}strophysics 554:A119, \doi{10.1051/0004-6361/201220573},
  \urlprefix\url{http://dx.doi.org/10.1051/0004-6361/201220573}

\bibitem[{{Zaussinger} et~al(2013){Zaussinger}, {Kupka}, and
  {Muthsam}}]{Zaussinger_2013b}
{Zaussinger} F, {Kupka} F, {Muthsam} HJ (2013) {Semi-convection}. In: {Goupil}
  M, {Belkacem} K, {Neiner} C, {Ligni{\`e}res} F, {Green} JJ (eds) Lecture
  Notes in Physics, Berlin Springer Verlag, Lecture Notes in Physics, Berlin
  Springer Verlag, vol 865, p 219, \doi{10.1007/978-3-642-33380-4\_11}

\bibitem[{Zaussinger et~al(2017)Zaussinger, Kupka, Egbers, Neben, H{\"u}cker,
  Bahr, and Schmitt}]{Zaussinger_2017}
Zaussinger F, Kupka F, Egbers C, Neben M, H{\"u}cker S, Bahr C, Schmitt M
  (2017) Semi-convective layer formation. In: Pogorelov N, Pogorelov E, Zank G
  (eds) 11th International Conference on Numerical Modeling of Space Plasma
  Flows: ASTRONUM-2016, vol 837, p 012012, \doi{doi
  :10.1088/1742-6596/837/1/012012},
  \urlprefix\url{http://stacks.iop.org/1742-6596/837/i=1/a=012012}

\end{thebibliography}

\appendix
\section{Model validation}  \label{append:lab_validation}

The numerical description and the physical model (Boussinesq approximation) are now validated against experimental 
and theoretical results. This way we show that our 2D simulation setup used in Sect.~\ref{sec:methods} 
and~\ref{sec:results} provides a detailed physical description of layer formation in double-diffusive convection over 
heated bottom boundaries, apart from the differences that have to be accounted  for to exactly reproduce the laboratory 
system. Since we are interested in a proof of concept, we have used a different simulation code which was easier 
to adapt for the laboratory comparison but much less suitable for the study we present in Sect.~\ref{sec:methods} 
and~\ref{sec:results} (dimensionless units, need for excellent scaling in parallel computing, high approximation order, etc.).

First experiments of double-diffusive layering over a heated plate with a stable salinity gradient have been 
performed by \cite{Turner_1964}. The experimental setup in \cite{Turner_1964} consists of a tank with 
dimensions of ${\rm 0.25^3\,m^3}$. The initially stable salt water gradient measured $1\%$ in density. The tank 
was heated from below at a heating rate of ${\rm 2\,cals\, cm^{-2} min^{-1}}$, which gives ${\rm 1395.57 W/m^2}$ in 
Si units, respectively. After 10 minutes the first layer formed at ${\rm H=0.1\,m}$. The final height of this first layer 
was reached after an hour at ${\rm H=0.125\,m}$, whereas a second layer also formed above the first one during this hour.
After ${\rm 95\, min}$ three layers are visible, where the height of the second and the third are about ${\rm 0.037\,m}$
(see Fig.~2 in \citealt{Turner_1964}, the heating rate at the bottom was about ${\rm 1400\,W m^{-2}}$, the density 
gradient due to salinity measured 1\%). The same experiment was repeated with doubled heating rates and steeper 
salinity gradients. Doubling the density gradient increased the amount of layers (Fig.~3a in \citealt{Turner_1964},
${\rm 1400\,W m^{-2}}$ heat rate and a salinity gradient of 2\%), whereas the doubled heating rate at the bottom 
increased the heights of the layers (Fig.~3b in \citealt{Turner_1964}, ${\rm 2800\,W m^{-2}}$ heating rate and a salinity 
gradient of 1\%). However, the reader needs to keep in mind that the evolution of the layer formation is time dependent. 
This makes it necessary to have time stamps available. In summary, only the first experiment can be used to validate 
a numerical code. Here, four time stamps and all physical properties are available.

The numerical setup follows the laboratory experiment where heat is imposed at the bottom (denoted with `bot') by 
the heat flux $\bf F=\rm k \nabla T$, which yields a Neumann boundary condition on temperature $\rm T$,
\begin{equation}
\left (\frac{\partial \rm T}{\partial \rm y}\right )_{\rm bot} = \rm F_y/k,
\end{equation}
with $\rm F_y$ being the vertical component of $\rm \bf F$ and $\rm k$ the thermal conductivity. The salinity gradient 
$\rm \Delta S$ is defined by the density difference $\rm \Delta \rho$ of $1\%$ between the bottom and the top of the tank.
The salinity is calculated by $\rm \Delta S=\Delta \rho / (\rho \beta)$, cf. \cite{Turner_1968}, Eq.~7 and~18.
All fluid properties are assumed to be constant at reference temperature $\rm T_{ref}=293\,K$,  cf. Tab.~\ref{tab:parameters}. 
\begin{table}
\begin{center}
\caption{Parameters used for the validation simulation.}
\begin{tabular}{c c c c }
\hline
\hline
variable name & unit                             & parameter                               & value/range  \\
\hline
H                & m                                  & height                                     & 0.25 \\
L, D            & m                                  & length, depth                           & 0.25 \\
$\nu$          & ${\rm m^2 s^{-1}}$          & kinematic viscosity                  & $10^{-6}$ \\
${\rm\kappa_T}$ & ${\rm m^2 s^{-1}}$   & thermal diffusivity                    & $1.42 \cdot 10^{-7}$  \\
k                & ${\rm W\,m^{-1}K^{-1}}$   & thermal conductivity                & $0.6$  \\
${\rm\kappa_S}$ & ${\rm m^2 s^{-1}}$   & saline diffusivity                      & $1.42 \cdot 10^{-9}$  \\
$\alpha$      & ${\rm K^{-1}}$                 & thermal expansion coefficient  & $2.3 \cdot 10^{-4}$  \\
$\beta$       & $1$                                & saline expansion coefficient   & $7.6 \cdot 10^{-4}$  \\
$\rho$         & ${\rm kg\, m^{-3}}$          & density                                  & 1025  \\
T                & ${\rm K}$                       & ref. temperature                      & 293.15  \\
S               & ${\rm g\, kg^{-1}}$           & salinity                                   & 12.7836  \\
$\rm F_z$        & ${\rm W/m^2}$                       & heat flux                            & 1395  \\
$\rm c_p$        & ${\rm m^2 K^{-1} s^{-2}}$ & specific heat capacity              & 4182  \\
\hline
\hline
\label{tab:parameters}
\end{tabular}
\end{center}
\end{table}%

The top of the tank (denoted with `top') is closed against the atmosphere fulfilling an insulator boundary condition. In the following, 
$\rm (\frac{\partial T}{\partial y})_{\rm top} = 0$. This is justified by the following two premises: 
a) the radiative heat flux at the top $\rm F_{\rm top}$ is three orders of magnitudes lower than $\rm F_{\rm bot}$. 
b) the layering occurs only in the lower $2/3$ of the tank, where the thermal influence of the upper boundary 
is negligible on the short time scales of the experiment. The side walls are insulated, too and fulfill 
$\rm (\frac{\partial T}{\partial \bf n})_{\rm x}= 0$. The solute boundaries are assumed to be impermeable for the solute at 
all boundaries, whence $\rm (\frac{\partial S}{\partial \bf n})=0$. Impenetrable, stress free boundary conditions 
are considered for the velocity field.

The numerical solution of the governing equations are calculated with the finite volume based software OpenFOAM~5.x, \cite{Weller_1998}. 
More precisely, the set of equations given by Eq.~(\ref{eq:BA_extended}) and the Boussinesq approximation 
Eq.~(\ref{eq:basic}) are solved with the PISO algorithm. Here, the momentum equation is solved by neglecting the 
pressure term. This results in a velocity field unfulfilling the incompressibility constraint. Subsequently, a Poisson 
equation is formulated to estimate the pressure. This pressure is used to correct the momentum field to fulfill 
$\nabla\cdot \bf u=\rm 0$. This is the same numerical approach as used in ANTARES, however, the equations 
are treated in physical units. The entire simulation time of 5700 sec is identical with the last experimental snap shot.

\begin{figure}
\begin{center}
\includegraphics[width=0.7\textwidth]{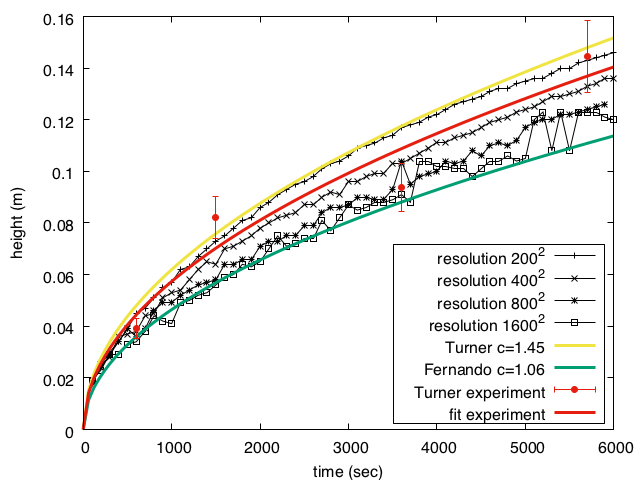}
\end{center}
\caption{Height of the first layer as function of time for various resolutions. 
            Theoretical and experimental results are depicted as straight lines and
            single full dots, respectively.}
\label{fig:validation}
\end{figure}

The influence of the resolution is tested with a grid study. More precisely, rectangular grids with 
$200 \times 200$, $400 \times 400$, $800 \times 800$, and $1600 \times 1600$ are compared using parameters 
of the standard case. The main testing parameter is the height of the first layer, which is plotted for all four cases 
in Fig.~\ref{fig:validation}. Only a small difference 
in the layer height of approx.\ 3\% is found in comparison with selected 3D simulations. As the overall layer 
formation process is not influenced by the third spatial dimension all simulations are conducted in 2D. 
The highest resolution resolves the saline boundary layer with two points and the thermal 
boundary layer with 20 points. This estimate is based on Eq.~(\ref{eq:Sscale}) and (\ref{eq:Tscale})
(see also \citealt{Zaussinger_2011}, Appendix~B.1). Obviously, the layer height decreases with higher resolutions.
Additionally, the diffusive interface shows differences. In case of $200 \times 200$ and $400 \times 400$ 
cells the interface is very flat and the shape is not much influenced by the convective zone. This changes 
for $800 \times 800$ cells, where the interface gets turbulent and a unique layer height is not given any more. 
The interface shows structural changes in case of $1600 \times 1600$ cells.
On top of the actual interface a small mixing region is established. This additional layer is separated from 
the underlying zone, but important for the evolution of the second layer. It is neither purely diffusive nor 
purely convective, but stable enough not to be mixed with other layers. However, this \textit{connection layer} 
is only visible at highest resolutions. In case of $800 \times 800$ the connection layer is merged  into the 
first layer due to higher numerical diffusivity. The height of the first layer changes only by $<5\%$ for the 
highest resolutions. The grid study on the height of the first layer reveals a minimum resolution of 
$800 \times 800$ cells for simulating this experiment with OpenFoam. This is two times higher than our simulations 
performed with ANTARES. We suggest this to be a consequence of the second order scheme used in OpenFoam.

In \cite{Turner_1968} the authors described the growth of the first layer with an equilibrium model. They estimated
the height by $\rm h=C\cdot t^{1/2}S_*^{-1/2}H_*^{1/2}$, where $\rm S_*=- g \beta (\frac{d\,S}{d\,y})$,
$\rm H_*=-\frac{g\alpha H}{\rho c}$ and $\rm c=\sqrt{2}$. For known $\rm H_*$ and $\rm S_*$ the growth rate 
can be easily obtained and used to validate numerical models. \citet{Fernando_1987} extended the range 
of the constant value c to $\rm 1.06<C<1.63$. The height of the first layer for both bounding values for parameters 
as given in Tab.~\ref{tab:parameters} is depicted in Fig.~\ref{fig:validation}. The height varies between 0.11--0.15~m 
after $\rm t=6000\,{\rm s}$. Besides theoretical models we considered an experiment (cf.\ Fig.~2 in \citealt{Turner_1964}), 
too. The layer heights are measured by hand and depicted as full dots in Fig.~\ref{fig:validation}. However, 
only four snapshots are available in the original work. Further experiments are available from other authors, 
but the lack of experimental parameters makes it difficult to recreate them by numerical simulations.
The height of layers in the experiment of \cite{Turner_1964} are as follows:
0.041\,m after 600\,s, 0.093\,m after 1500\,s, 0.112\,m after 3600\,s, and 0.142\,m after 5700\,s.
Three of four values are within the plane spanned by the theoretical estimates (blue-yellow) for the first layer height. 
All tested resolutions are with in this plane, too. A fitting curve of the four experimental points ranges within 
the two lower resolutions $200 \times 200$ and $400 \times 400$. This curve has an exponential growth rate of 0.5 and 
a constant $\rm c=1.34$ for given values of $\rm S_*$ and $\rm H_*$. However, fluctuations of the layer height are assumed 
to be within 10\%, which is based on uncertainties of the fluid properties and the thermal properties of the tank material. 
The numerical simulations act on the same time scales as the experiment. This means, that the same amount of secondary 
layers are visible at comparable time stamps. Based on the comparison between theoretical estimates for the first layer 
height (\cite{Turner_1968} and \cite{Fernando_1987}) and four experimental snap shots published in \cite{Turner_1964}, 
we conclude that the governing equations given by Eq.~(\ref{eq:basic}) and Eq.~(\ref{eq:BA_extended})
are valid for simulations in the context of layer formation in double-diffusive convection. 

\end{document}